\pdfoutput=1 

\documentclass[preprint,pre,showpacs,preprintnumbers,amsmath,amssymb]{revtex4}

\usepackage{graphicx}
\usepackage{dcolumn}
\usepackage{bm}
\usepackage{braket}
\usepackage{subfigure} 
\usepackage{booktabs}
\usepackage{tabularx}
\usepackage{multirow}

\begin{document}

\preprint{PREPRINT}

\title{Numerical Study of the Properties of the Central Moment Lattice Boltzmann Method}

\author{Yang Ning}
\email{yning@uwyo.edu}

\affiliation{Department of Mechanical Engineering, University of
Wyoming, Laramie, WY 82071\\}

\author{Kannan N. Premnath}
\email{knandhap@uwyo.edu}

\affiliation{Department of Mechanical Engineering, University of
Wyoming, Laramie, WY 82071\\}

\date{\today}

\begin{abstract}
Central moment lattice Boltzmann method (LBM) is one of the more recent developments among the lattice kinetic schemes for computational fluid dynamics. A key element in this approach is the use of \emph{central} moments to specify collision process and forcing, and thereby naturally maintaining Galilean invariance, an important characteristic of fluid flows. When the different central moments are relaxed at different rates like in a standard multiple relaxation time (MRT) formulation based on \emph{raw} moments, it is endowed with a number of desirable physical and numerical features. Since the collision operator exhibits a cascaded structure, this approach is also known as the cascaded LBM. While the cascaded LBM has been developed sometime ago, a systematic study of its numerical properties, such as accuracy, grid convergence and stability for well defined canonical problems is lacking and the present work is intended to fulfill this need. We perform a quantitative study of the performance of the cascaded LBM for a set of benchmark problems of differing complexity, viz., Poiseuille flow, decaying Taylor-Green vortex flow and lid-driven cavity flow. We first establish its grid convergence and demonstrate second order accuracy under diffusive scaling for both the velocity field and its derivatives, i.e. components of the strain rate tensor, as well. The method is shown to quantitatively reproduce steady/unsteady analytical solutions or other numerical results with excellent accuracy. The cascaded MRT LBM based on central moments is found to be of similar accuracy when compared with the standard MRT LBM based on raw moments, when detailed comparison of the flow fields are made, with both well reproducing even small scale vortical features. Numerical experiments further demonstrate that the central moment MRT LBM results in significant stability improvements when compared with certain existing collision models at moderate additional computational cost.\\
\end{abstract}

\pacs{47.11.Qr,05.20.Dd,47.27.-i}
\maketitle

\section{\label{sec:intro}Introduction}
Early developments in the area of computational fluid dynamics (CFD) have focused on the solution of the classical discretizations of the
continuum description of fluid motion. During the last two decades, there has been much interest and effort in the development of schemes
that derive their basis on a more smaller scale picture involving particle motion, which may be classified as mesoscopic methods.
One of the most promising of such approaches is the lattice Boltzmann method (LBM)~\cite{chen98,succi01,luo10}. Based on kinetic theory,
it involves the solution of the lattice Boltzmann equation (LBE), which specifies the evolution of the particle populations along discrete
directions, which comprise the lattice. This evolution involves a Lagrangian free streaming process along such lattice links and a local
collision step specified as a relaxation process. Various elements involved in these two simple steps are constructed based on symmetry
considerations, while obeying certain conservation constraints, in such a way that they recover the dynamics of fluid flow in the near
incompressible limit. The resulting scheme has a number of desirable features. These include the ability to naturally represent complex
fluid physics such as multiphase and multicomponent flows based on kinetic theory, amenability to parallelization due to the locality
of the method and representation of flow through complex geometries. Furthermore, due to the exact conservation in the streaming step
and machine round-off conservation in the collision process, it has considerably low numerical dissipation for a second-order numerical scheme~\cite{ubertini10}. Due to such competitive advantages, the LBM has found applications in the simulation of a wide range of fluid flow problems~\cite{chen98,succi01,luo10}.

Since the LBM is usually developed by means of a bottom-up strategy, there is certain level of flexibility in the construction of its various
elements to recover the macroscopic fluid motion. In particular, the choice of a suitable collision model can have profound influence on
the fidelity as well as the stability of the approach. As such, the construction of the collision step has been the subject of considerable
attention since the inception of the LBM. The simplest among these is the so-called single-relaxation-time (SRT) model~\cite{chen92,qian92},
which is based on the Bhatnagar-Gross-Krook (BGK) approximation~\cite{bhatnagar54}. While it is popular, it has limitations in the
representation of certain flow problems and is generally prone to numerical instability, particularly at high Reynolds numbers. A major
development to address these aspects is the moment approach~\cite{dhumieres92}, which has been constructed based on multiple relaxation
times (MRT) in particular to significantly improve the numerical stability~\cite{lallemand00}. While it is related to its precursor involving
a more general relaxation approximation~\cite{higuera89a,higuera89b}, the characteristic difference being that it performs collision in
an orthogonal moment space leading to an efficient and flexible numerical scheme. This moment approach, which is designated as the standard
MRT formulation in this paper, has recently been studied and compared with some of the other collision models in detail~\cite{luo11}. A
simpler version that is intermediate between the SRT and MRT model is the so-called two-relaxation-time (TRT) model~\cite{ginzburg05},
in which the moments of even and odd orders are relaxed to their equilibrium at different rates. This, along with the MRT model, can be
adjusted such that it results in a minimization of undesirable discrete kinetic effects near walls. Another significant development is the
so-called entropic LBM~\cite{karlin99}. It involves an equilibria, which is based on a constrained minimization of a Lyapunov-type functional.
By modulating the collision process through enforcing entropy involution locally, this approach aims to maintain non-linear stability. This
approach has resulted in a number of simplified variants recently~\cite{asinari09,karlin11}.

An important physical feature of the fluid motion is that their description be independent of any inertial frame of reference (e.g.~\cite{pope00}). This invariance property, which is termed as the Galilean invariance, should be satisfied by any model or numerical scheme for its general
applicability. Furthermore, it has recently been shown that stabilization of classical schemes for compressible flow can be achieved when they are specifically constructed to respect this physical property~\cite{scovazzi07a,scovazzi07b,hughes10}. Keeping these general notions in mind, Galilean invariance can be naturally prescribed in the LBM when its various elements are represented in terms of the \emph{central} moments, i.e. moments obtained by shifting the particle velocity by the local fluid velocity. That is, any dynamical changes due to the collision process and impressed forces can be represented in terms of suitable variations of a set of such central moments. In particular, a collision model based on the relaxation of central moments was constructed recently~\cite{geier06}. The model exhibits a cascaded structure, which was later shown to be equivalent to considering a generalized equilibrium in the lattice or rest frame of reference~\cite{asinari08}. These central moments can be relaxed at different rates during
collision leading to a cascaded MRT or central moment MRT formulation, whereas by contrast the standard MRT formulation considers \emph{raw} moments. A systematic derivation of this approach by including the effect of impressed forces based on central moments was presented in~\cite{premnath09b}. This leads to considering generalized sources, analogous to the generalized equilibrium in the rest frame of reference. They also presented a detailed Chapman-Enskog analysis of the cascaded MRT LBM for its consistency with the macroscopic fluid dynamical equations of motion. This approach was further extended to various lattice models in three-dimensions in~\cite{premnath11a}, in the cylindrical coordinate system for axisymmetric flows in~\cite{premnath12} and for accounting of non-equilibrium effects in~\cite{premnath11b}.

Prior work on the cascaded LBM as discussed above have focused mainly on method developments or their mathematical analysis, with little attention
towards their numerics except for few validation cases. In particular, a detailed numerical study of the properties of the cascaded LBM for established benchmark problems and also their performance against other LBM approaches is lacking. The focus of the present work is intended to fill this gap by presenting a systematic study of the numerical properties of the cascaded LBM, viz., grid convergence, accuracy and stability for various canonical problems of differing complexity in terms of flow features and temporal evolution. Establishing the reliability and merits of the method in quantitative terms could provide confidence in their extension and applications to various complex flow problems of interest. To study the numerics of the cascaded LBM, we consider the Poiseuille flow, decaying Taylor-Green vortex flow, and lid-driven cavity flow, for which either analytical solutions or detailed prior numerical results are available for comparison. Much of the literature on the LBM with other collision models on grid convergence studies have focused only on those for the velocity field. In this work, we present numerical results on the grid convergence of the cascaded LBM for the velocity field as well as its derivatives, i.e. the strain rate tensor. Furthermore, an advantage of the kinetic schemes such as the LBM is that the strain rate tensor can be computed locally in terms of non-equilibrium moments. In this work, we also present a direct comparison of the results obtained using the non-equilibrium moments of the cascaded LBM with those involving the finite differencing of the velocity field at various locations
for the lid-driven cavity flow problem to assess their quantitative accuracy. It may be noted that a detailed comparison study of the SRT and the standard MRT models have recently been performed in~\cite{luo11}. Thus, in this work, we present a quantitative accuracy comparison between the standard MRT LBM and the cascaded or central moment MRT LBM for the lid-driven cavity flow. Finally, we will discuss the numerical stability performance of the various LBM schemes for the above benchmark problem.

The paper is organized as follows. Section~\ref{sec:cascadedLBM} presents the details of the particular version of the cascaded MRT LBM used in this work. In Sec.~\ref{sec:convergencestudy}, the results of the grid convergence study of the cascaded MRT LBM together with the raw moment based standard MRT LBM for the three benchmark problems are discussed. Subsequently, the quantitative accuracy of the cascaded LBM is demonstrated by making detailed comparison with either analytical or other numerical solutions for the above problems in Sec.~\ref{sec:accuracystudy}. In Sec.~\ref{sec:stabilitystudy}, numerical stability test results are presented for the lid-driven cavity flow using the SRT LBM, standard MRT LBM and cascaded MRT LBM. Summary and conclusions of this work are given in Sec.~\ref{sec:conclusion}.

\section{\label{sec:cascadedLBM}Cascaded Lattice Boltzmann Method}
We will now discuss the main features of the cascaded LBM. Similar to the standard MRT LBM, the cascaded MRT LBM also performs collisions in moment space, but these moments are obtained by shifting the particle velocity by the local fluid velocity, i.e. using central moments. As a result, the approach can naturally maintain Galilean invariance. Central moment relaxation process was specified in~\cite{geier06}, which was re-interpreted by
considering generalized equilibrium in~\cite{asinari08}. Its detailed mathematical consistency analysis in a MRT formulation with forcing was carried
out in~\cite{premnath09b}. The computations of the cascaded LBM are actually performed after transforming the central moments into raw moments by
means of a binomial formula. In this work, the specific formulation of the cascaded LBM given in~\cite{premnath09b}, whose details are somewhat
different from that given in~\cite{geier06}, is used. This is briefly discussed in what follows.

In this work, the standard two-dimensional, nine velocity (D2Q9) lattice is employed. We consider the usual bra-ket notations in the description of the method as it provides a convenient representation. That is, we consider the depiction of vectors as $\langle \phi |$ and $| \phi \rangle$, where $\langle \phi |$ represents a row vector of $\phi$ of any state in the corresponding direction $(\phi_0, \phi_1, \phi_2, \cdots, \phi_8)$ and $| \phi \rangle$ represents a column vector $(\phi_0, \phi_1, \phi_2, \cdots, \phi_8)^T$. The inner product $\sum^8_{\alpha=0}\phi_{\alpha}\varphi_{\alpha}$ is then denoted by $\langle \phi | \varphi \rangle$. As the cascaded LBM is a moment approach, we need a set of nine linearly independent moment basis vectors for its specification. The (raw) moments of the distribution function $f_\alpha$ of different orders can be defined as $\sum^8_{\alpha=0}e^m_{\alpha x}e^n_{\alpha y}f_{\alpha}$. Here, $\alpha$ is the discrete particle direction, and $m$ and $n$ are integers. Thus, a set of nine linearly independent nonorthogonal basis vectors obtained using the monomials $e^m_{\alpha x}e^n_{\alpha y}$ in an ascending order can be written as
\begin{equation*}
\begin{split}
& | \rho \rangle  = ||\vec{e}_{\alpha}|^0 \rangle = (1,1,1,1,1,1,1,1,1)^T, \\
& |e_{\alpha x} \rangle  = (0,1,0,-1,0,1,-1,-1,1)^T, \\
& |e_{\alpha y} \rangle  = (0,0,1,0,-1,1,1,-1,-1)^T, \\
& |e^2_{\alpha x}+e^2_{\alpha y} \rangle  = (0,1,1,1,1,2,2,2,2)^T, \\
\end{split}
\end{equation*}
\begin{equation}
\begin{split}
& |e^2_{\alpha x}-e^2_{\alpha y} \rangle  = (0,1,-1,1,-1,0,0,0,0)^T, \\
& |e_{\alpha x}e_{\alpha y} \rangle  = (0,0,0,0,0,1,-1,1,-1)^T, \\
& |e^2_{\alpha x}e_{\alpha y} \rangle  = (0,0,0,0,0,1,1,-1,-1)^T, \\
& |e_{\alpha x}e^2_{\alpha y} \rangle  = (0,0,0,0,0,1,-1,-1,1)^T, \\
& |e^2_{\alpha x}e^2_{\alpha y} \rangle  = (0,0,0,0,0,1,1,1,1)^T.
\end{split}
\end{equation}
This can be transformed by means of the Gram-Schmidt procedure into an equivalent set of \emph{orthogonal} basis vectors, which provides a computationally more efficient and convenient setting for the description of the method. As a result, we have the following orthogonal set~\cite{premnath09b}:
\begin{equation}\label{eq:K value}
\begin{split}
& | K_0 \rangle  = | \rho \rangle, \\
& | K_1 \rangle  = |e_{\alpha x} \rangle, \\
& | K_2 \rangle  = |e_{\alpha y} \rangle, \\
& | K_3 \rangle  = 3|e^2_{\alpha x}+e^2_{\alpha y} \rangle -4| \rho \rangle, \\
& | K_4 \rangle  = |e^2_{\alpha x}-e^2_{\alpha y} \rangle, \\
& | K_5 \rangle  = |e_{\alpha x}e_{\alpha y} \rangle, \\
& | K_6 \rangle  = -3|e^2_{\alpha x}e_{\alpha y} \rangle +2|e_{\alpha y} \rangle, \\
& | K_7 \rangle  = -3|e_{\alpha x}e^2_{\alpha y} \rangle +2|e_{\alpha x} \rangle, \\
& | K_8 \rangle  = 9|e^2_{\alpha x}e^2_{\alpha y} \rangle-6|e^2_{\alpha x}+e^2_{\alpha y} \rangle + 4| \rho \rangle.
\end{split}
\end{equation}
Collecting the above set of vectors as a matrix $\mathcal{K}$, it immediately follows that $\mathcal{K} \mathcal{K}^T$ is a diagonal matrix, owing to
orthogonality. This orthogonal matrix $\mathcal{K}$ can be written in component form as
\begin{equation}
\begin{split}
\mathcal{K} &=\bigl [|K_0\rangle,|K_1\rangle,|K_2\rangle,|K_3\rangle,|K_4\rangle,|K_5\rangle,|K_6\rangle,|K_7\rangle,|K_8\rangle)\bigr ]\\
 & =
\begin{bmatrix}
1 & 0 & 0 &-4 & 0 & 0 & 0 & 0 & 4\\
1 & 1 & 0 &-1 & 1 & 0 & 0 & 2 & -2\\
1 & 0 & 1 &-1 &-1 & 0 & 2 & 0 & -2\\
1 &-1 & 0 &-1 & 1 & 0 & 0 &-2 & -2\\
1 & 0 &-1 &-1 &-1 & 0 &-2 & 0 & -2\\
1 & 1 & 1 & 2 & 0 & 1 &-1 &-1 & 1\\
1 &-1 & 1 & 2 & 0 &-1 &-1 & 1 & 1\\
1 &-1 &-1 & 2 & 0 & 1 & 1 & 1 & 1\\
1 & 1 &-1 & 2 & 0 &-1 & 1 &-1 & 1.
\end{bmatrix}
\end{split}
\end{equation}

To specify the collision step and forcing, we need the central moments of the local equilibrium and sources, which can be obtained as follows.
First, the local Maxwell-Boltzmann distribution function in continuous particle velocity space $(\xi_x, \xi_y)$ is written as
$f^{\mathcal{M}} \equiv f^{\mathcal{M}}(\rho, \vec{u}, \xi_x, \xi_y) = \frac{\rho}{2\pi c_s^2}\exp{\left[-\frac{(\vec{\xi}-\vec{u})^2}{2c_s^2}\right]}$, where $c_s$ is the speed of sound. Typically, $c_s^2=1/3$. Based on this, the continuous central moments of the equilibrium of order $(m+n)$ can be defined as $\widehat{\Pi}^{\mathcal{M}}_{x^m y^n} = \int_{-\infty}^{\infty} \int_{-\infty}^{\infty} f^{\mathcal{M}}(\xi_x-u_x)^m(\xi_y-u_y)^n d\xi_x d\xi_y$, which yields
\begin{equation}
\begin{split}
|\widehat{\Pi}^{\mathcal{M}}_{x^m y^n} \rangle  & = (\widehat{\Pi}^{\mathcal{M}}_0, \widehat{\Pi}^{\mathcal{M}}_x,\widehat{\Pi}^{\mathcal{M}}_y,\widehat{\Pi}^{\mathcal{M}}_{xx},\widehat{\Pi}^{\mathcal{M}}_{yy},\widehat{\Pi}^{\mathcal{M}}_{xy},\widehat{\Pi}^{\mathcal{M}}_{xxy},\widehat{\Pi}^{\mathcal{M}}_{xyy},\widehat{\Pi}^{\mathcal{M}}_{xxyy})^T, \\
& = (\rho,0,0,c_s^2\rho,c_s^2\rho,0,0,0,c_s^4\rho)^T.
\end{split}
\end{equation}
Considering that the impressed forces only influence the fluid momentum, the central moments of the sources of order $(m+n)$ due to a force field $(F_x, F_y)$ defined by
$\widehat{\Gamma}^{\mathcal{F}}_{x^m y^n} = \int_{-\infty}^{\infty} \int_{-\infty}^{\infty} \Delta f^{\mathcal{F}}(\xi_x-u_x)^m(\xi_y-u_y)^n d\xi_x d\xi_y$, where $\Delta f^{\mathcal{F}}$ is the change in the distribution function due to force fields, can be simply written as~\cite{premnath09b}
\begin{equation}
\begin{split}
|\widehat{\Gamma}^{\mathcal{F}}_{x^m y^n} \rangle  & = (\widehat{\Gamma}^{\mathcal{F}}_0, \widehat{\Gamma}^{\mathcal{F}}_x,\widehat{\Gamma}^{\mathcal{F}}_y,\widehat{\Gamma}^{\mathcal{F}}_{xx},\widehat{\Gamma}^{\mathcal{F}}_{yy},\widehat{\Gamma}^{\mathcal{F}}_{xy},\widehat{\Gamma}^{\mathcal{F}}_{xxy},\widehat{\Gamma}^{\mathcal{F}}_{xyy},\widehat{\Gamma}^{\mathcal{F}}_{xxyy})^T, \\
& = (0,F_x,F_y,0,0,0,0,0,0)^T.
\end{split}
\end{equation}

Based on the above continuous central moments, the elements of the cascaded LBE can be formulated. Using the trepezoidal rule representation of the source term, the cascaded LBE can be written as~\cite{premnath09b}
\begin{equation}\label{eq:semi cascaded LBM}
 f_{\alpha}(\vec{x}+\vec{e}_{\alpha}{\delta_t}, t+\delta_t) = f_{\alpha}(\vec{x},t) + \Omega^{\mathcal{C}}_{\alpha (\vec{x},t)} +\frac{1}{2} \bigl [S_{\alpha (\vec{x},t)} + S_{\alpha (\vec{x}+\vec{e}_{\alpha},t+\delta_t)} \bigr].
\end{equation}
Here, the collision term $\Omega^{\mathcal{C}}_{\alpha}$ can be represented as $\Omega^{\mathcal{C}}_{\alpha} \equiv \Omega^{\mathcal{C}}_{\alpha}(\mathbf{f},\bf{\widehat{g}})=(\mathcal{K}\cdot \mathbf{\widehat{g}})_{\alpha}$, where $\mathbf{f}$ $\equiv | f_{\alpha} \rangle = (f_0, f_1, \cdots, f_8)^T$ is the vector of distribution functions and $\mathbf{\widehat{g}}$ $\equiv | \widehat{g}_{\alpha} \rangle = (\widehat{g}_0, \widehat{g}_1, \cdots, \widehat{g}_8)^T$ is the vector of unknown collision kernel to be obtained later. Owing to the cascaded nature of the central moment based approach, it satisfies the following functional relation $\widehat{g}_{\alpha} \equiv \widehat{g}_{\alpha}(\mathbf{f}, \widehat{g}_{\beta}),\ \ \ \ \  \beta=0,1,\cdots, \alpha-1$. The discrete form of the source term $S_{\alpha}$ in the cascaded LBE given above represents the influence of the force field $(F_x,F_y)$ in the velocity space and is defined as $\mathbf{S} \equiv |S_{\alpha}\rangle = (S_0, S_1, S_2, \cdots, S_8)^T$. Noting that Eq.~(\ref{eq:semi cascaded LBM}) is semi-implicit, by using the
standard variable transformation $\overline{f} = f_{\alpha} - \frac{1}{2}S_{\alpha}$, its implicitness can be effectively removed. This yields
\begin{equation}
 \overline{f}_{\alpha}(\vec{x}+\vec{e}_{\alpha}{\delta_t}, t+\delta_t) = \overline{f}_{\alpha}(\vec{x},t) + \Omega^{\mathcal{C}}_{\alpha (\vec{x},t)} +S_{\alpha (\vec{x},t)}.
\end{equation}

The derivation of the collision term, i.e. the collision kernel $\mathbf{\widehat{g}}$ and the source term $\mathbf{S}$ involves matching the \emph{discrete} central moments and the \emph{continuous} central moments of equilibria and sources, which are specified above, of all orders supported by the lattice set. We designate this step as the \emph{Galilean invariance matching principle}. First, the discrete central moments of the distribution functions and sources of order $(m+n)$ can be defined, respectively, as
$\widehat{\kappa}_{x^m y^n} = \langle (e_{\alpha x}-u_x)^m(e_{\alpha y}-u_y)^n| f_\alpha \rangle$ and
$\widehat{\sigma}_{x^m y^n} = \langle (e_{\alpha x}-u_x)^m(e_{\alpha y}-u_y)^n| S_\alpha \rangle$. Also, in terms of the transformed distribution functions
we define $\widehat{\overline{\kappa}}_{x^m y^n}=\langle (e_{\alpha x}-u_x)^m(e_{\alpha y}-u_y)^n| \overline{f}_\alpha \rangle$, which satisfies
$\widehat{\overline{\kappa}}_{x^m y^n} = \widehat{\kappa}_{x^m y^n} - \frac{1}{2}\widehat{\sigma}_{x^m y^n}$, and similarly for the local equilibria
$\widehat{\overline{\kappa}}^{eq}_{x^m y^n}=\langle (e_{\alpha x}-u_x)^m(e_{\alpha y}-u_y)^n| \overline{f}^{eq}_\alpha \rangle$. Then, the Galilean
invariance matching principle reads
\begin{gather}
\widehat{\kappa}^{eq}_{x^m y^n} =  \widehat{\Pi}^{\mathcal{M}}_{x^m y^n}, \\
\widehat{\sigma}_{x^m y^n} =  \widehat{\Gamma}^{\mathcal{F}}_{x^m y^n}.
\end{gather}
This immediately specifies the various discrete central moments. Hence, we get
\begin{equation}
\begin{split}
\hspace{15 mm} |\widehat{\kappa}^{eq}_{x^m y^n} \rangle  & = (\widehat{\kappa}^{eq}_0, \widehat{\kappa}^{eq}_x,\widehat{\kappa}^{eq}_y,\widehat{\kappa}^{eq}_{xx},\widehat{\kappa}^{eq}_{yy},\widehat{\kappa}^{eq}_{xy},\widehat{\kappa}^{eq}_{xxy},\widehat{\kappa}^{eq}_{xyy},\widehat{\kappa}^{eq}_{xxyy})^T \\
& = (\rho,0,0,c_s^2\rho,c_s^2\rho,0,0,0,c_s^4\rho)^T,
\end{split}
\end{equation}
\begin{equation}\label{eq:discrete sigma}
\begin{split}
|\widehat{\sigma}_{x^m y^n} \rangle  & = (\widehat{\sigma}_0, \widehat{\sigma}_x,\widehat{\sigma}_y,\widehat{\sigma}_{xx},\widehat{\sigma}_{yy},\widehat{\sigma}_{xy},\widehat{\sigma}_{xxy},\widehat{\sigma}_{xyy},\widehat{\sigma}_{xxyy})^T \\
& = (0,F_x,F_y,0,0,0,0,0,0)^T,
\end{split}
\end{equation}
and
\begin{equation}\label{eq:transformed kappa}
\begin{split}
| \widehat{\overline{\kappa}}^{eq}_{x^m y^n} \rangle  = & (\widehat{\overline{\kappa}}^{eq}_0, \widehat{\overline{\kappa}}^{eq}_x,\widehat{\overline{\kappa}}^{eq}_y,\widehat{\overline{\kappa}}^{eq}_{xx},\widehat{\overline{\kappa}}^{eq}_{yy},\widehat{\overline{\kappa}}^{eq}_{xy},\widehat{\overline{\kappa}}^{eq}_{xxy},\widehat{\overline{\kappa}}^{eq}_{xyy},\widehat{\overline{\kappa}}^{eq}_{xxyy})^T, \\
                                 = & (\rho, -\frac{1}{2}F_x, -\frac{1}{2}F_y, c_s^2 \rho, c_s^2 \rho, 0, 0, 0, c_s^4\rho)^T.
\end{split}
\end{equation}

The next important step is to transform all the above discrete central moments in terms of raw moments, which can be readily accomplished by
means of the following binomial formula:
$
\langle (e_{\alpha x}-u_x)^m (e_{\alpha y}-u_y)^n | \varphi \rangle
= \langle e_{\alpha x}^m e_{\alpha y}^n | \varphi \rangle + \bigl \langle e_{\alpha x}^m \bigl [ \sum_{j=1}^n C^n_j e^{n-j}_{\alpha y} (-1)^j u^j_y \bigr ] | \varphi \bigr \rangle
+ \bigl \langle e_{\alpha y}^m \bigl [ \sum_{i=1}^m C^m_i e^{m-i}_{\alpha x} (-1)^i u^i_x \bigr ] | \varphi \bigr \rangle
+  \bigl \langle \bigl [ \sum_{i=1}^m C^m_i e^{m-i}_{\alpha x} (-1)^i u^i_x \bigr ] \bigl [ \sum_{j=1}^n C^n_j e^{n-j}_{\alpha y} (-1)^j u^j_y \bigr ] | \varphi \bigr \rangle
$, where $C^p_q = p!/\bigl( q! (p-q)!)$.
Thus, we obtain the following discrete raw moments of sources $\widehat{\sigma}_{x^my^n}^{'}$ as
\begin{equation}\label{eq:S value}
\begin{split}
& \widehat{\sigma}_{0}^{'}=\langle S_{\alpha} | \rho \rangle = 0, \\
& \widehat{\sigma}_{x}^{'}=\langle S_{\alpha} | e_{\alpha x} \rangle  =  F_x, \\
& \widehat{\sigma}_{y}^{'}=\langle S_{\alpha} | e_{\alpha y} \rangle  =  F_y, \\
& \widehat{\sigma}_{xx}^{'}=\langle S_{\alpha} | e_{\alpha x}^2 \rangle = 2F_x u_x, \\
& \widehat{\sigma}_{yy}^{'}=\langle S_{\alpha} | e_{\alpha y}^2 \rangle = 2F_y u_y, \\
& \widehat{\sigma}_{xy}^{'}=\langle S_{\alpha} | e_{\alpha x}e_{\alpha y} \rangle = F_x u_y+F_y u_x, \\
& \widehat{\sigma}_{xxy}^{'}=\langle S_{\alpha} | e_{\alpha x}^2 e_{\alpha y} \rangle = F_y u_x^2 + 2F_x u_x u_y, \\
& \widehat{\sigma}_{xyy}^{'}=\langle S_{\alpha} | e_{\alpha x} e_{\alpha y}^2 \rangle = F_x u_y^2 + 2F_y u_y u_x,
\end{split}
\end{equation}
\begin{equation*}
\begin{split}
& \widehat{\sigma}_{xxyy}^{'}=\langle S_{\alpha} | e_{\alpha x}^2 e_{\alpha y}^2 \rangle = 2F_x u_xu_y^2 + 2F_y u_yu_x^2.
\end{split}
\end{equation*}
Based on the above, we now obtain the source terms projected to the orthogonal moment basis vectors, i.e.
$\braket{K_{\beta}|S_{\alpha}}$, $\beta=0,1,2,\ldots,8$. This intermediate step is needed to obtain the source terms in the velocity space.
It immediately follows that
\begin{eqnarray}
\widehat{m}^{s}_{0}=\braket{K_0|S_{\alpha}}&=&0,\nonumber\\
\widehat{m}^{s}_{1}=\braket{K_1|S_{\alpha}}&=&F_x,\nonumber\\
\widehat{m}^{s}_{2}=\braket{K_2|S_{\alpha}}&=&F_y,\nonumber\\
\widehat{m}^{s}_{3}=\braket{K_3|S_{\alpha}}&=&6(F_xu_x+F_yu_y),\nonumber\\
\widehat{m}^{s}_{4}=\braket{K_4|S_{\alpha}}&=&2(F_xu_x-F_yu_y),\nonumber\\
\widehat{m}^{s}_{5}=\braket{K_5|S_{\alpha}}&=&(F_xu_y+F_yu_x),\nonumber\\
\widehat{m}^{s}_{6}=\braket{K_6|S_{\alpha}}&=&(2-3u_x^2)F_y-6F_xu_xu_y,\nonumber\\
\widehat{m}^{s}_{7}=\braket{K_7|S_{\alpha}}&=&(2-3u_y^2)F_x-6F_yu_yu_x,\nonumber\\
\widehat{m}^{s}_{8}=\braket{K_8|S_{\alpha}}&=&6\left[(3u_y^2-2)F_xu_x+(3u_x^2-2)F_yu_y\right].\nonumber
\end{eqnarray}
Equivalently, this can be written in matrix form as
$\mathcal{K}^T\mathbf{S}=(\mathcal{K}\cdot\mathbf{S})_{\alpha}=(\braket{K_0|S_{\alpha}},\braket{K_1|S_{\alpha}},\braket{K_2|S_{\alpha}},\ldots,\braket{K_8|S_{\alpha}})
=(\widehat{m}^{s}_{0},\widehat{m}^{s}_{1},\widehat{m}^{s}_{2},\ldots,\widehat{m}^{s}_{8})^T\equiv \ket{\widehat{m}^{s}_{\alpha}}$. By exploiting the orthogonal property of $\mathcal{K}$, i.e. $\mathcal{K}^{-1}=\mathcal{K}^T \cdot D^{-1}$, where the diagonal matrix is
$D=\mbox{diag}(\braket{K_0|K_0},\braket{K_1|K_1},\braket{K_2|K_2},\ldots,\braket{K_8|K_8})$, we exactly invert the above to obtain the source terms in velocity space $S_{\alpha}$ as
\begin{equation}
\begin{split}
S_0 = & \frac{1}{9} \bigl( -m_3^s+m_8^s \bigr), \\
S_1 = & \frac{1}{36} \bigl( 6m_1^s-m_3^s+ 9m_4^s+6m_7^s-2m_8^s \bigr), \\
S_2 = & \frac{1}{36} \bigl( 6m_2^s-m_3^s- 9m_4^s+6m_6^s-2m_8^s \bigr), \\
S_3 = & \frac{1}{36} \bigl(-6m_1^s-m_3^s+ 9m_4^s-6m_7^s-2m_8^s \bigr), \\
S_4 = & \frac{1}{36} \bigl(-6m_2^s-m_3^s- 9m_4^s-6m_6^s-2m_8^s \bigr), \\
S_5 = & \frac{1}{36} \bigl( 6m_1^s+6m_2^s+2m_3^s+ 9m_5^s-3m_6^s-3m_7^s+m_8^s \bigr), \\
S_6 = & \frac{1}{36} \bigl(-6m_1^s+6m_2^s+2m_3^s- 9m_5^s-3m_6^s+3m_7^s+m_8^s \bigr), \\
S_7 = & \frac{1}{36} \bigl(-6m_1^s-6m_2^s+2m_3^s+ 9m_5^s+3m_6^s+3m_7^s+m_8^s \bigr), \\
S_8 = & \frac{1}{36} \bigl( 6m_1^s-6m_2^s+2m_3^s- 9m_5^s+3m_6^s-3m_7^s+m_8^s \bigr).
\end{split}
\end{equation}
The discrete raw moments of the transformed distribution functions $\widehat{\overline{\kappa}}_{x^my^n}^{'}$, which will be needed in the evaluation of the collision kernel, can be conveniently written as follows:
\begin{equation}
\begin{split}
\widehat{\overline{\kappa}}_{0}^{'}=\langle \overline{f}_\alpha | \rho \rangle & = \rho, \\
\widehat{\overline{\kappa}}_{x}^{'}=\langle \overline{f}_\alpha | e_{\alpha x} \rangle & = \rho u_x -\frac{1}{2}F_x, \\
\widehat{\overline{\kappa}}_{y}^{'}=\langle \overline{f}_\alpha | e_{\alpha y} \rangle & = \rho u_y -\frac{1}{2}F_y, \\
\widehat{\overline{\kappa}}_{xx}^{'}=\langle \overline{f}_\alpha | e_{\alpha x}^2 \rangle & =  \left(\sum_{\alpha}^{\{1,3,5,6,7,8\}} \right )\otimes \overline{f}_{\alpha}, \\
\widehat{\overline{\kappa}}_{yy}^{'}=\langle \overline{f}_\alpha | e_{\alpha y}^2 \rangle & =  \left(\sum_{\alpha}^{\{2,4,5,6,7,8\}} \right )\otimes \overline{f}_{\alpha}, \\
\end{split}
\end{equation}
\begin{equation*}
\begin{split}
\widehat{\overline{\kappa}}_{xy}^{'}=\langle \overline{f}_\alpha | e_{\alpha x}e_{\alpha y} \rangle & =  \left(\sum_{\alpha}^{\{5,7\}}- \sum_{\alpha}^{\{6,8\}} \right )\otimes \overline{f}_{\alpha}, \\
\widehat{\overline{\kappa}}_{xxy}^{'}=\langle \overline{f}_\alpha | e_{\alpha x}^2e_{\alpha y} \rangle & =  \left(\sum_{\alpha}^{\{5,6\}}- \sum_{\alpha}^{\{7,8\}} \right )\otimes \overline{f}_{\alpha}, \\
\widehat{\overline{\kappa}}_{xyy}^{'}=\langle \overline{f}_\alpha | e_{\alpha x}e_{\alpha y}^2 \rangle & =  \left(\sum_{\alpha}^{\{5,8\}}- \sum_{\alpha}^{\{6,7\}} \right )\otimes \overline{f}_{\alpha}, \\
\widehat{\overline{\kappa}}_{xxyy}^{'}=\langle \overline{f}_\alpha | e_{\alpha x}^2e_{\alpha y}^2 \rangle & =  \left(\sum_{\alpha}^{\{5,6,7,8\}} \right )\otimes \overline{f}_{\alpha}.
\end{split}
\end{equation*}
where we have used
$\left( a\sum_{\alpha}^A +b\sum_{\beta}^B + \cdots \right) \otimes \overline{f}_{\alpha}= a(\overline{f}_{\alpha_1}+\overline{f}_{\alpha_2}+\overline{f}_{\alpha_3}+ \cdots)+ b(\overline{f}_{\beta_1}+\overline{f}_{\beta_2}+\overline{f}_{\beta_3}+ \cdots)+ \cdots$, with $A=\{\alpha_1,\alpha_2,\alpha_3,\cdots\}$, $B=\{\beta_1,\beta_2,\beta_3,\cdots\}, \cdots$, as a compact summation operator for ease of presentation. Furthermore, the raw moments of the collision kernels $\sum_{\alpha} (\mathcal{K} \cdot \mathbf{\widehat{g}})_{\alpha} e_{\alpha x}^m e_{\alpha y}^n = \sum_{\beta} \langle K_{\beta} |  e_{\alpha x}^m e_{\alpha y}^n \rangle \widehat{g}_{\beta}$ are needed in its construction. Collision invariants of conserved moments imply $\widehat{g}_0 = \widehat{g}_1 = \widehat{g}_2=0$. Exploiting the orthogonal property of the matrix $\mathcal{K}$, the non-conserved moments of $\widehat{g}_{\beta}$ at higher orders, i.e. $\beta = 3,4,\cdots, 8$ can be obtained as follows~\cite{premnath09b}:
\begin{eqnarray}
\sum_{\alpha}(\mathcal{K}\cdot \mathbf{\widehat{g}})_{\alpha}=
\sum_{\beta} \braket{K_{\beta}|\rho}\widehat{g}_{\beta}&=&0, \nonumber\\
\sum_{\alpha}(\mathcal{K}\cdot \mathbf{\widehat{g}})_{\alpha}e_{\alpha x}=
\sum_{\beta} \braket{K_{\beta}|e_{\alpha x}}\widehat{g}_{\beta}&=&0, \nonumber\\
\sum_{\alpha}(\mathcal{K}\cdot \mathbf{\widehat{g}})_{\alpha}e_{\alpha y}= \sum_{\beta} \braket{K_{\beta}|e_{\alpha y}}\widehat{g}_{\beta}&=&0, \nonumber\\
\sum_{\alpha}(\mathcal{K}\cdot \mathbf{\widehat{g}})_{\alpha}e_{\alpha x}^2= \sum_{\beta} \braket{K_{\beta}|e_{\alpha x}^2}\widehat{g}_{\beta}&=&6\widehat{g}_{3}+2\widehat{g}_{4}, \nonumber\\
\sum_{\alpha}(\mathcal{K}\cdot \mathbf{\widehat{g}})_{\alpha}e_{\alpha y}^2= \sum_{\beta} \braket{K_{\beta}|e_{\alpha y}^2}\widehat{g}_{\beta}&=&6\widehat{g}_{3}-2\widehat{g}_{4}, \label{eq:collisionkernelmoment}\\
\sum_{\alpha}(\mathcal{K}\cdot \mathbf{\widehat{g}})_{\alpha}e_{\alpha x}e_{\alpha y}= \sum_{\beta} \braket{K_{\beta}|e_{\alpha x}e_{\alpha y}}\widehat{g}_{\beta}&=&4\widehat{g}_{5}, \nonumber\\
\sum_{\alpha}(\mathcal{K}\cdot \mathbf{\widehat{g}})_{\alpha}e_{\alpha x}^2e_{\alpha y}= \sum_{\beta} \braket{K_{\beta}|e_{\alpha x}^2e_{\alpha y}}\widehat{g}_{\beta}&=&-4\widehat{g}_{6}, \nonumber\\
\sum_{\alpha}(\mathcal{K}\cdot \mathbf{\widehat{g}})_{\alpha}e_{\alpha x}e_{\alpha y}^2= \sum_{\beta} \braket{K_{\beta}|e_{\alpha x}e_{\alpha y}^2}\widehat{g}_{\beta}&=&-4\widehat{g}_{7}, \nonumber\\
\sum_{\alpha}(\mathcal{K}\cdot \mathbf{\widehat{g}})_{\alpha}e_{\alpha x}^2e_{\alpha y}^2=
\sum_{\beta} \braket{K_{\beta}|e_{\alpha x}^2e_{\alpha y}^2}\widehat{g}_{\beta}&=&8\widehat{g}_{3}+4\widehat{g}_{8}. \nonumber
\end{eqnarray}

Using the above, the collision kernel $\widehat{g}_{\beta}$ of the cascaded collision operator $\Omega^{\mathcal{C}}_{\alpha} \equiv \Omega^{\mathcal{C}}_{\alpha}(\mathbf{f},\bf{\widehat{g}})=(\mathcal{K}\cdot \mathbf{\widehat{g}})_{\alpha}$ can be obtained as follows.
Starting from the lowest order central moments that are non-collisional invariants (i.e. $\widehat{\overline{\kappa}}_{xx}$ and higher), they are successively set equal to their local attractors based on the transformed equilibria. This step provides tentative expressions for $\widehat{g}_{\alpha}$ based on the equilibrium assumption. This is then modified to allow for relaxation process during collision. That is, they are multiplied with corresponding relaxation parameters~\cite{geier06}. In this step, care needs to be exercised to multiply the relaxation parameters only with those terms that are not yet in post-collision states (i.e. terms not involving $\widehat{g}_{\beta}, \beta=0,1,2,\ldots,\alpha-1$) for a given $\widehat{g}_{\alpha}$. See~\cite{premnath09b} for various details involved in this procedure. Here,
we summarize the final expressions of the non-conserved collision kernels, which are given as follows:
\begin{eqnarray}
\widehat{g}_3&=&\frac{\omega_3}{12}\left\{ \frac{2}{3}\rho+\rho(u_x^2+u_y^2)
-(\widehat{\overline{\kappa}}_{xx}^{'}+\widehat{\overline{\kappa}}_{yy}^{'})
-\frac{1}{2}(\widehat{\sigma}_{xx}^{'}+\widehat{\sigma}_{yy}^{'})
\right\}, \label{eq:collisionkernelg3}\\
\widehat{g}_4&=&\frac{\omega_4}{4}\left\{\rho(u_x^2-u_y^2)
-(\widehat{\overline{\kappa}}_{xx}^{'}-\widehat{\overline{\kappa}}_{yy}^{'})
-\frac{1}{2}(\widehat{\sigma}_{xx}^{'}-\widehat{\sigma}_{yy}^{'})
\right\}, \label{eq:collisionkernelg4}\\
\widehat{g}_5&=&\frac{\omega_5}{4}\left\{\rho u_x u_y
-\widehat{\overline{\kappa}}_{xy}^{'}
-\frac{1}{2}\widehat{\sigma}_{xy}^{'}
\right\}, \label{eq:collisionkernelg5}\\
\widehat{g}_6&=&\frac{\omega_6}{4}\left\{2\rho u_x^2 u_y+\widehat{\overline{\kappa}}_{xxy}^{'}
-2u_x\widehat{\overline{\kappa}}_{xy}^{'}-u_y\widehat{\overline{\kappa}}_{xx}^{'}-\frac{1}{2}\widehat{\sigma}_{xxy}
\right\}-\frac{1}{2}u_y(3\widehat{g}_3+\widehat{g}_4)\nonumber\\&&-2u_x\widehat{g}_5, \label{eq:collisionkernelg6}\\
\widehat{g}_7&=&\frac{\omega_7}{4}\left\{2\rho u_x u_y^2+\widehat{\overline{\kappa}}_{xyy}^{'}
-2u_y\widehat{\overline{\kappa}}_{xy}^{'}-u_x\widehat{\overline{\kappa}}_{yy}^{'}-\frac{1}{2}\widehat{\sigma}_{xyy}
\right\}-\frac{1}{2}u_x(3\widehat{g}_3-\widehat{g}_4)\nonumber\\&&-2u_y\widehat{g}_5, \label{eq:collisionkernelg7}\\
\widehat{g}_8&=&\frac{\omega_8}{4}\left\{\frac{1}{9}\rho+3\rho u_x^2 u_y^2-\left[\widehat{\overline{\kappa}}_{xxyy}^{'}
-2u_x\widehat{\overline{\kappa}}_{xyy}^{'}-2u_y\widehat{\overline{\kappa}}_{xxy}^{'}
+u_x^2\widehat{\overline{\kappa}}_{yy}^{'}+u_y^2\widehat{\overline{\kappa}}_{xx}^{'}\right.\right.
\nonumber \\
&&\left.\left.+4u_xu_y\widehat{\overline{\kappa}}_{xy}^{'}
\right]-\frac{1}{2}\widehat{\sigma}_{xxyy}^{'}
\right\}-2\widehat{g}_3-\frac{1}{2}u_y^2(3\widehat{g}_3+\widehat{g}_4)
-\frac{1}{2}u_x^2(3\widehat{g}_3-\widehat{g}_4)\nonumber\\
&&-4u_xu_y\widehat{g}_5-2u_y\widehat{g}_6
-2u_x\widehat{g}_7.\label{eq:collisionkernelg8}
\end{eqnarray}
In the above, $\omega_{\beta}$, where $\beta=3,4,5,\ldots, 8$, are the relaxation parameters, satisfying the usual bounds $0<\omega_{\beta}<2$.
When a Chapman-Enskog expansion~\cite{chapman64} is applied to the cascaded LBM, it can be shown to recover the Navier-Stokes equations with
the relaxation parameters $\omega_3=\omega^\chi$~and~$\omega_4=\omega_5=\omega^\nu$ controlling the fbulk and shear viscosities, respectively (e.g., $\nu=c_s^2\left(\frac{1}{\omega^\nu}-\frac{1}{2}\right)$)~\cite{premnath09b}. The rest of the parameters can be adjusted independently improve numerical stability. In this work, $\omega_4=\omega_5=\frac{1}{\tau}$ is selected based on the specified kinematic viscosity, while the rest of
the relaxation parameters are set to $1$.

The cascaded LBE can now be re-written in the form of the usual stream-and-collide procedure, leading to the following two steps: \begin{align}\label{eq:transformed f}
& \widetilde{\overline{f}}_{\alpha}(\vec{x},t) =  \overline{f}_{\alpha}(\vec{x},t)+ \Omega^{\mathcal{C}}_{\alpha(\vec{x},t)} + S_{\alpha (\vec{x},t)}, \\
& \overline{f}_{\alpha}(\vec{x}+ \vec{e}_{\alpha},t+\delta_t) = \widetilde{\overline{f}}_{\alpha}(\vec{x},t),
\end{align}
where the symbol ``tilde" ($\sim$) in the above equations refers to the post-collision state of the distribution function. Expanding the collision term
in the first step, the components of the post-collision distribution function can be explicitly written as
\begin{eqnarray}
\widetilde{\overline{f}}_{0}&=&\overline{f}_{0}+\left[\widehat{g}_0-4(\widehat{g}_3-\widehat{g}_8)\right]+S_0, \nonumber\\
\widetilde{\overline{f}}_{1}&=&\overline{f}_{1}+\left[\widehat{g}_0+\widehat{g}_1-\widehat{g}_3+\widehat{g}_4    +2(\widehat{g}_7-\widehat{g}_8)\right]+S_1, \nonumber\\
\widetilde{\overline{f}}_{2}&=&\overline{f}_{2}+\left[\widehat{g}_0+\widehat{g}_2-\widehat{g}_3-\widehat{g}_4
+2(\widehat{g}_6-\widehat{g}_8)\right]+S_2, \nonumber\\
\widetilde{\overline{f}}_{3}&=&\overline{f}_{3}+\left[\widehat{g}_0-\widehat{g}_1-\widehat{g}_3+\widehat{g}_4
-2(\widehat{g}_7+\widehat{g}_8)\right]+S_3, \nonumber\\
\widetilde{\overline{f}}_{4}&=&\overline{f}_{4}+\left[\widehat{g}_0-\widehat{g}_2-\widehat{g}_3-\widehat{g}_4
-2(\widehat{g}_6+\widehat{g}_8)\right]+S_4, \\
\widetilde{\overline{f}}_{5}&=&\overline{f}_{5}+\left[\widehat{g}_0+\widehat{g}_1+\widehat{g}_2+2\widehat{g}_3
+\widehat{g}_5-\widehat{g}_6-\widehat{g}_7+\widehat{g}_8\right]+S_5, \nonumber\\
\widetilde{\overline{f}}_{6}&=&\overline{f}_{6}+\left[\widehat{g}_0-\widehat{g}_1+\widehat{g}_2+2\widehat{g}_3
-\widehat{g}_5-\widehat{g}_6+\widehat{g}_7+\widehat{g}_8\right]+S_6, \nonumber\\
\widetilde{\overline{f}}_{7}&=&\overline{f}_{7}+\left[\widehat{g}_0-\widehat{g}_1-\widehat{g}_2+2\widehat{g}_3
+\widehat{g}_5+\widehat{g}_6+\widehat{g}_7+\widehat{g}_8\right]+S_7, \nonumber\\
\widetilde{\overline{f}}_{8}&=&\overline{f}_{8}+\left[\widehat{g}_0+\widehat{g}_1-\widehat{g}_2+2\widehat{g}_3
-\widehat{g}_5+\widehat{g}_6-\widehat{g}_7+\widehat{g}_8\right]+S_8. \nonumber
\end{eqnarray}
The hydrodynamic fields, i.e. the fluid density and the velocity then follow from taking the zeroth and first moments of the distribution function,
yielding
\begin{align}
\rho = & \sum_{\alpha=0}^8 \overline{f}_{\alpha} = \langle \overline{f}_\alpha | \rho \rangle,\\
\rho u_i = & \sum_{\alpha=0}^8 \overline{f}_{\alpha}e_{\alpha i} +\frac{1}{2}F_i = \langle \overline{f}_\alpha | e_{\alpha i} \rangle +\frac{1}{2} F_i,
i=x,y,
\end{align}                                                                                                                                              and the pressure $p$ satisfies $p=c_s^2\rho$. A particularly useful feature of kinetic schemes such as the cascaded LBM is that the strain-rate tensor
can be computed \emph{locally} from a knowledge of the non-equilibrium moments. In fact, this can be shown by means of the Chapman-Enskog analysis, which was performed on the cascaded LBE in~\cite{premnath09b}. Setting the components of the momentum as $j_x=\rho u_x$ and $j_y=\rho u_y$, such
an analysis shows~\cite{premnath09b}
\begin{align}
\widehat{f_3}^{(neq)}=& -\frac{2}{3\omega_3}\bigl(\partial_xj_x+\partial_yj_y\bigr),\\
\widehat{f_4}^{(neq)}=& -\frac{2}{3\omega_4}\bigl(\partial_xj_x-\partial_yj_y\bigr),\\
\widehat{f_5}^{(neq)}=& -\frac{1}{3\omega_5}\bigl(\partial_xj_y+\partial_yj_x\bigr),
\end{align}
where $\widehat{f}_{\beta}^{(neq)}\approx\widehat{f}_{\beta}-\widehat{f}^{eq}_{\beta}$ are the non-equilibrium raw moments. Specifically,
$\widehat{f}_{3}=\widehat{\kappa}_{xx}^{'}+\widehat{\kappa}_{yy}^{'}$,
$\widehat{f}_{4}=\widehat{\kappa}_{xx}^{'}-\widehat{\kappa}_{yy}^{'}$, and
$\widehat{f}_{5}=\widehat{\kappa}_{xy}^{'}$, whose equilibria are $\widehat{f}_{3}^{eq}=2/3\rho+\rho(u_x^2+u_y^2)$,
$\widehat{f}_{4}^{eq}=\rho(u_x^2-u_y^2)$, and
$\widehat{f}_{5}^{eq}=\rho u_xu_y$, respectively~\cite{premnath09b}. It thus follows that
\begin{align}
\partial_{x}j_x = & -\frac{3\omega_3}{2}\biggl[\displaystyle\sum_{\alpha=0}^{8}f_{\alpha}e^2_{\alpha x}-\left(\frac{1}{3}\rho+\rho u_x^2\right)\bigg],\label{eq:strainrate1}\\
\partial_{y}j_y = & -\frac{3\omega_4}{2}\biggl[\displaystyle\sum_{\alpha=0}^{8}f_{\alpha}e^2_{\alpha y}-\left(\frac{1}{3}\rho+\rho u_y^2\right)\bigg],\label{eq:strainrate2}\\
\partial_{x}j_y+\partial_{y}j_x = & -3\omega_5\biggl[\displaystyle\sum_{\alpha=0}^{8}f_{\alpha}e_{\alpha x}e_{\alpha y}-\rho u_xu_y\bigg].\label{eq:strainrate3}
\end{align}
These specific expressions will be exploited in the numerical study of the cascaded LBM in the remainder of this paper. In the sections
that follow, we will present the results obtained with the cascaded LBM for a set of benchmark problems to assess its numerical
properties in terms of grid convergence, accuracy and stability.

\section{\label{sec:convergencestudy}Grid Convergence Study on the Benchmark Problems}
We first perform a numerical study involving grid convergence for canonical flows including a steady 2D Poiseuille flow, a time-dependent 2D decaying Taylor-Green vortex flow, and a 2D lid-driven cavity flow characterized by various complex features. In the various figures presented
in this section, the symbols represent the computed solution using the cascaded MRT LBM, the thin solid lines are the resulting slopes representing
changes in the relative errors as the grid resolution increases, and the thick solid lines are the ideal slopes corresponding to second-order
accuracy. In this work, a \emph{diffusive scaling} is applied to perform the convergence tests~\cite{junk05}. According to this scaling, the errors
due to compressibility effects decrease at the same rate as the errors due to grid discretization thus prescribing a consistent limit process to represent incompressible flow. That is, the velocity scales in the same proportion as the length scales. Equivalently, this means that the ratio of the Mach number and the grid Knudsen number remains constant for different grid resolutions, \emph{i.e.} $Ma/Kn$ = constant.

\subsection{2D Poiseuille Flow}
The 2D Poiseuille flow is first considered. The flow is between two parallel plates of infinite length in the streamwise direction subjected to a constant body force. A periodic boundary condition is applied at the inlet and the outlet and a no-slip boundary condition at the solid boundaries by
employing the standard half-way bounce back approach. The grid convergence is established by considering the following resolutions consisting of $3\times 24, 3\times 36,\ldots, 3\times 192$ lattice nodes under diffusive scaling. The relaxation time for shear modes is set to $\tau=0.55$ that specifies $\omega_4$ and $\omega_5$. The rest of relaxation parameters are set to unity. The flow is driven by a constant body force with the components $F_x$ specified to yield desired condition (see below) and $F_y=0$. This classical flow problem has the well known parabolic profile as the analytical solution given by $u(y)=u_{max}(1-y^2/L^2)$, where $u_{max} = \frac{F_xL^2}{2\nu}$ is the maximum velocity occurring midway between the plates, $\nu$ is the kinematic viscosity related the to relaxation time $\tau$ as given in the previous section, and $L$ denotes the half-width between the plates. Figure~\ref{fig:convergence_channel} illustrates the relative global errors between the computed solutions obtained using the cascaded MRT LBM and the analytical solutions for such flow at different Reynolds numbers of $100$, $200$ and $400$. The relative global error, which quantifies the difference between the computed and analytical solutions, is defined as
\begin{equation}
\text{Relative Error} = \frac{\sum_i||(u_{c,i}-u_{a,i})||}{\sum_i||u_{a,i}||},
\end{equation}
where $u_{c,i}$ and $u_{a,i}$ are the computed and the analytical solutions, respectively, and a standard Euclidean norm is used in the above measurements. It is seen that the relative errors have slopes of almost equal to $2.00$, which tells that the cascaded MRT LBM is well-posed second-order accurate for this problem. In addition, the relative errors are seen to slightly increases as the Reynolds number increases.
\begin{figure}
  \vspace{10pt}
  \centering
  \subfigure{\label{fig:convergence_channel}\includegraphics[width=140mm,angle=0]{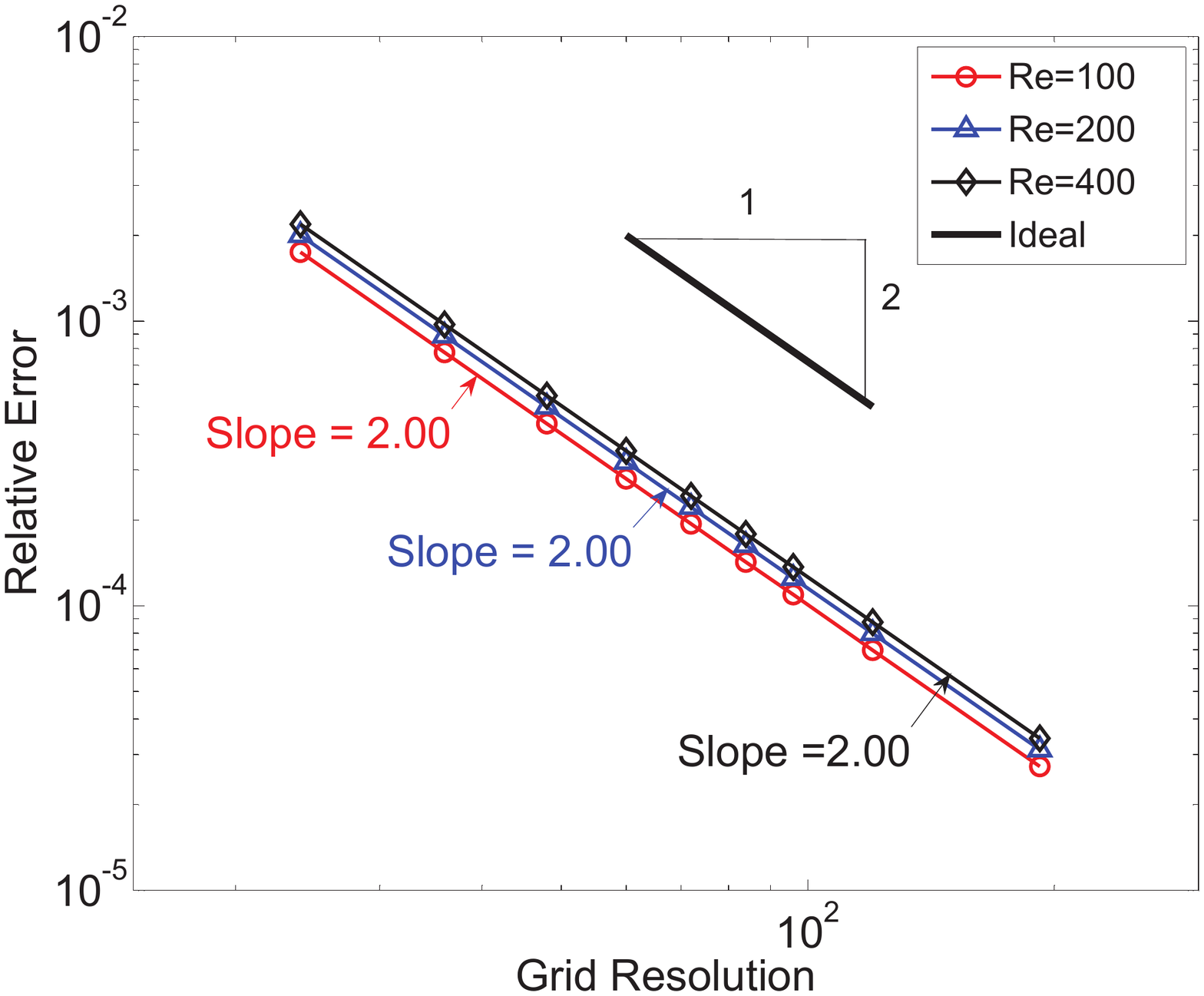}}
  \vspace{-15pt}
  \caption{\label{fig:convergence_channel}Grid convergence of the cascaded MRT LBM for the velocity field in a 2D Poiseuille flow with constant body force under diffusive scaling.}
  \vspace{25pt}
\end{figure}

\subsection{2D Decaying Taylor-Green Vortex Flow}
The second problem considered is the decaying Taylor-Green vortex~\cite{taylor23}, which is a 2D unsteady flow induced by a prescribed initial vortex distribution  and decaying due to fluid viscosity. The fluid domain is a square of side $2\pi$ with no inflow/outflow and wall boundaries. The initial condition is set to be periodic array of vortices in both x and y directions as follows\\
\begin{align}
u(x,y,0)=& -u_0\cos(k x)\sin(k y),\\
v(x,y,0)=& +u_0\sin(k x)\cos(k y),\\
p(x,y,0)=& p_0\biggl[1-\frac{u_0^2}{4c_s^2}\bigl(\cos(2kx)+\cos(2ky)\bigr)\biggr],
\end{align}
where $k=\frac{2\pi}{N}$ is the wavenumber, $u_0$ and $p_0$ are the initial values for velocity and pressure, respectively. Here, $N$ is the number of grid nodes in each direction. The temporal evolution has the characteristic time scale given by $T=\frac{1}{2k^2\nu}$. Since there is no external energy supplied and because of the presence of fluid viscosity, the velocity field will decay with time due to fluid viscous dissipation. There exists an analytical solution for this problem which is a solution of the Navier-Stokes equations in a periodic domain and given by
\begin{align}
u(x,y,t)=& -u_0\cos(kx)\sin(ky)e^{-2k^2\nu t},\label{eq:tayloru}\\
v(x,y,t)=& +u_0\sin(kx)\cos(ky)e^{-2k^2\nu t},\label{eq:taylorv}\\
p(x,y,t)=& p_0-\frac{u_0^2}{4}\biggl[\cos(2kx)+\cos(2ky)\biggr]e^{-4k^2\nu t}.\label{eq:taylorp}
\end{align}
Furthermore, the components of the strain rate tensor also satisfy the following explicit analytical solution:
\begin{align}
S_{xx}=& \frac{\partial u}{\partial x}=k u_0\sin(kx)\sin(ky)e^{-2\nu k^2t}\\
S_{yy}=& \frac{\partial u}{\partial y}=-S_{xx}\\
S_{xy}=& \frac{1}{2}\biggl(\frac{\partial u}{\partial y}+\frac{\partial v}{\partial x}\biggr)=0
\end{align}

\begin{figure}
  \vspace{-20pt}
  \centering
\includegraphics[width=140mm,angle=0]{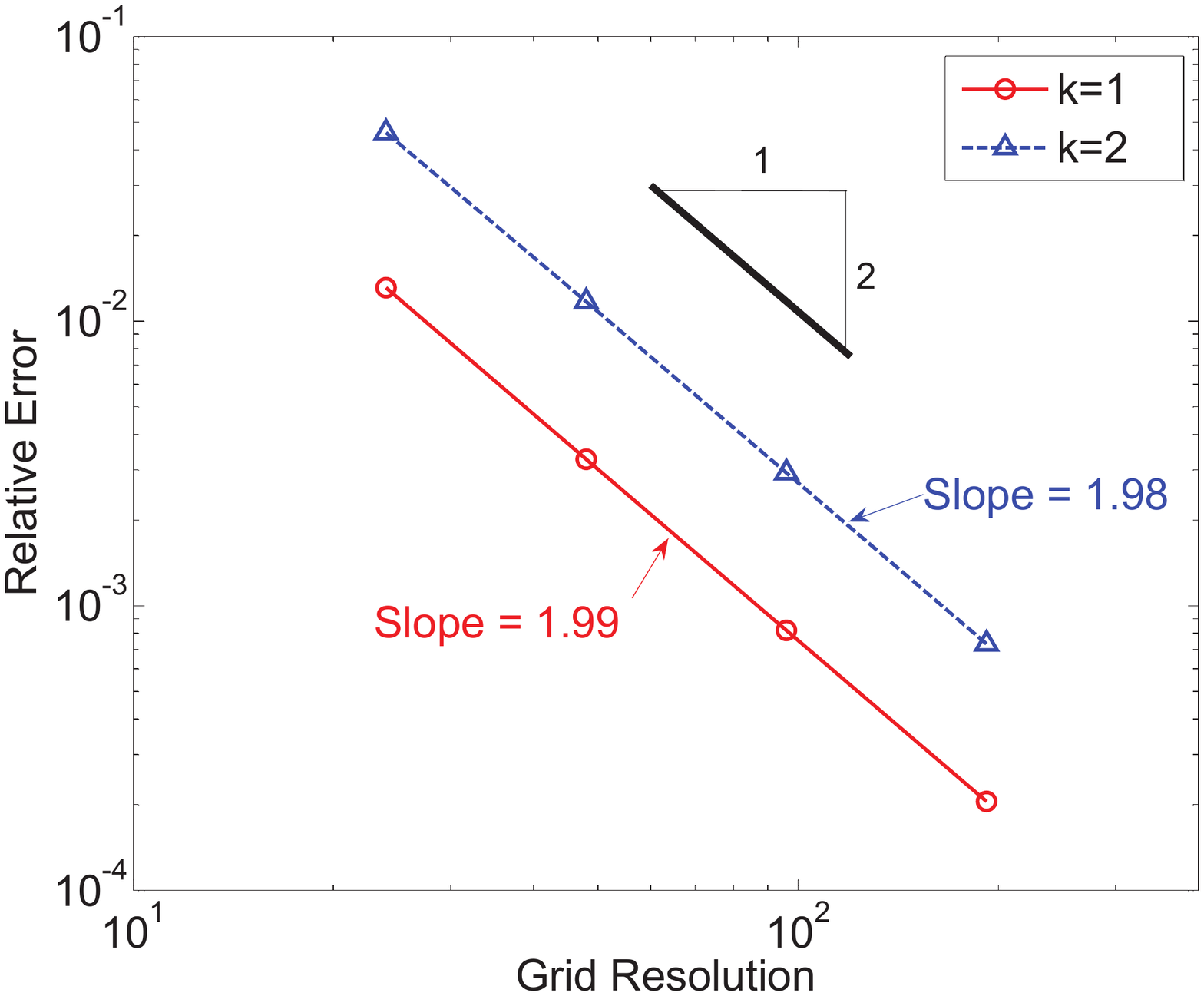}
  \vspace{-15pt}
  \caption{\label{fig:convergence_taylor_velocity}Grid convergence of the cascaded MRT LBM for the velocity field in a 2D Taylor-Green vortex flow with $k=1$ and $k=2$.}
\end{figure}

\begin{figure}
  \vspace{-20pt}
  \centering
\includegraphics[width=140mm,angle=0]{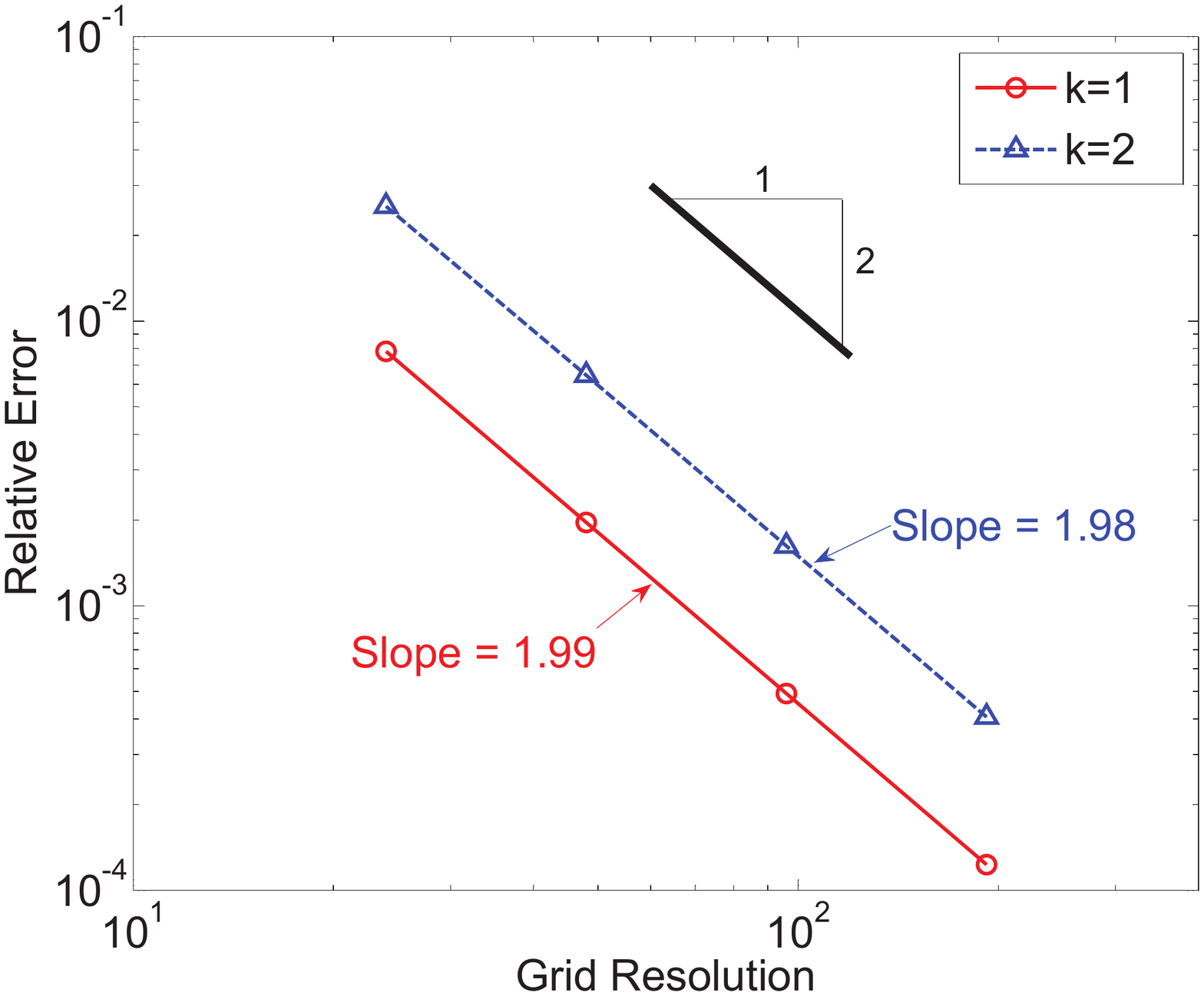}
  \vspace{-15pt}
  \caption{\label{fig:convergence_taylor_strainrate}Grid convergence of the cascaded MRT LBM for the strain rate in a 2D Taylor-Green vortex flow with $k=1$ and $k=2$.}
\end{figure}

In this test, the Reynolds number of the flow is set to $Re=\frac{u_0 l}{\nu}=14.4$, where $l=2\pi$ is the length of the domain. A periodic boundary condition is applied to all the sides of the domain. We consider the following parameters in our grid convergence study: $\tau=0.55$, $k =1, 2$ and $u_0 = 0.01$. Applying the diffusive scaling, we obtain the relative global errors between the computed and the analytical solutions for the grid resolutions of $24\times 24$, $48\times 48$, $96\times 96$, $192\times192$ for a representative time $t=30.1T$. In Fig.~\ref{fig:convergence_taylor_velocity} shown are the relative errors for the u-velocity component, which have the slopes of $1.99$ and $1.98$ for the wavenumbers $k=1$ and $k=2$, respectively. Figure~\ref{fig:convergence_taylor_strainrate} shows the relative errors for the only independent strain rate tensor component $S_{xx}$ with the slopes of $1.99$ and $1.98$ as well for the above two wavenumbers.
Thus, it is evident that the cascaded MRT LBM is second-order accurate not only for the velocity field, but also for the components of the strain rates as well. This finding is consistent with a recent study with the SRT LBM for this problem~\cite{kruger10}.

\subsection{2D Lid-driven Cavity Flow}
\begin{figure}
\centering
\vspace{-50pt}
\includegraphics[width = 150mm, angle = 0]{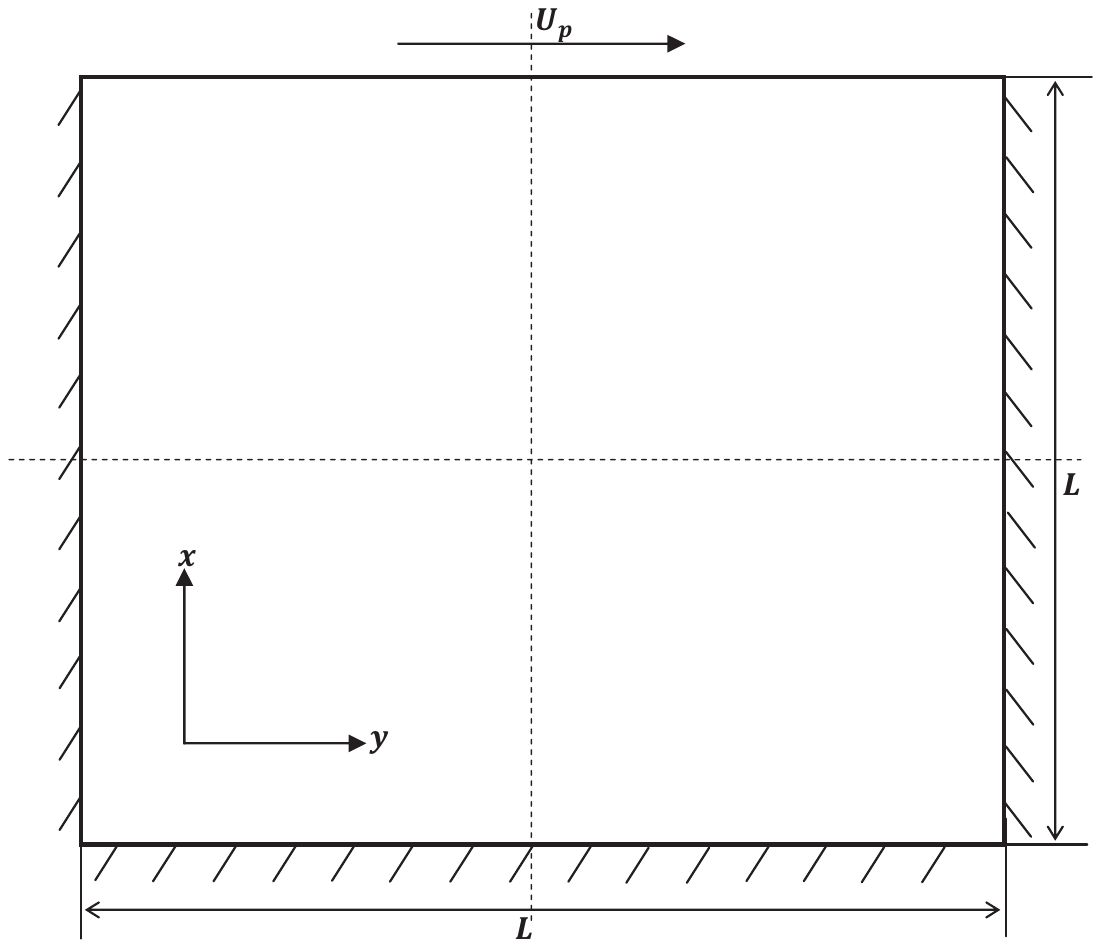}
\vspace{-315pt}
\caption{\label{fig:cavity}Illustration of the geometry of a lid-driven cavity flow.}
\end{figure}

Finally, the 2D lid-driven cavity flow is considered, whose geometric simplicity is contrasted by various complex flow features. It is generally considered a standard benchmark test for CFD methods and has been a subject of many investigations using a variety of methods (see e.g.~\cite{ghia82,schreiber83,vanka86,erturk05,bruneau06}). Grid convergence for this problem has been studied using different collision models (SRT and standard MRT) for the LBM by various researchers (e.g.~\cite{luo11}). In this section, the aim is to analyze the grid convergence and an estimation of the order of accuracy of the cascaded MRT LBM for this flow problem. More detailed accuracy investigation of the various flow features will be
carried out in the next section. While the geometry is simple from the boundary condition implementation point of view, the flow contains singular points and becomes very complicated in terms of flow structures, particularly as the Reynolds number increases (see e.g.~\cite{erturk05} for a review). A schematic of the arrangement of the boundaries in a 2D lid-cavity flow is shown in Fig.~\ref{fig:cavity}.
Fluid is enclosed inside a square cavity of length, $L$, and is set into motion by the moving upper wall that has a constant velocity $U_o$. The side and the bottom walls are considered to be stationary, which allows to implement a simple half-way bounce-back boundary condition on them. However, because the upper wall is in constant motion, a momentum correction needs to be added~\cite{lallemand03} into the regular bounce-back scheme for the upper boundary. This is implemented as $f_{\alpha}(i,N_y-1)=\widetilde{f}_{\overline{\alpha}}(i,N_y-1) + 6\rho w_{\alpha}e_{\alpha y}U_p$, where $\widetilde{f}_{\overline{\alpha}}(i,N_y-1)$ is the post-collision distribution function, for $\alpha=4,7,8$, with $\overline{\alpha}=2,5,6$ as the opposite directions of $\alpha$, and $w_{\alpha}$ is the weighting factor~\cite{lallemand03}. Ghia~\emph{et al.}~\cite{ghia82} have systematically studied this problem in much detail by employing a vorticity-stream function formulation of the 2D incompressible Navier-Stokes equations, which is solved by a multigrid method. Some of their numerical results have been used for making accuracy comparisons in this work which will be discussed in a later section. Because of the lack of analytical solutions, the computed solutions obtained by a relatively very fine grid resolution, i.e. with \emph{i.e.} $801 \times 801$, are treated as the approximate benchmark or reference (``analytical'') solutions. Not only is the convergence of velocity fields tested, but also the grid convergence of the components of the strain rate tensor is considered. It may be noted that the study involving the latter quantity has not so far received enough attention for this problem using the LBM.

\begin{figure}
  \vspace{-10pt}
  \centering
\includegraphics[width=140mm,angle=0]{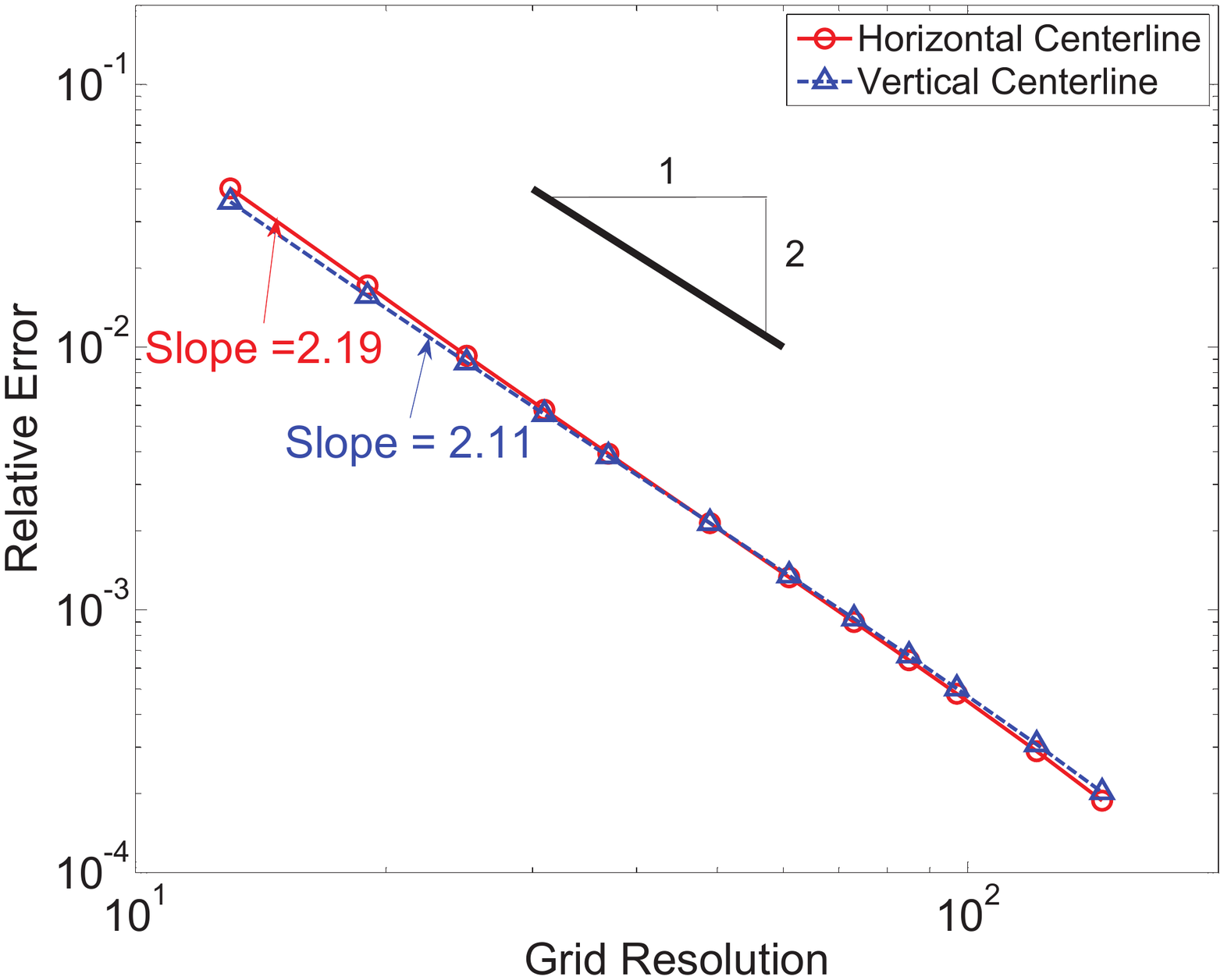}
  \vspace{-15pt}
  \caption{\label{fig:convergence_cavity_centerlinesU}Grid convergence of the cascaded MRT LBM for the U-velocity component in a 2D lid-driven cavity flow for $Re=100$.}
\end{figure}

\begin{figure}
  \vspace{-30pt}
  \centering
\includegraphics[width=140mm,angle=0]{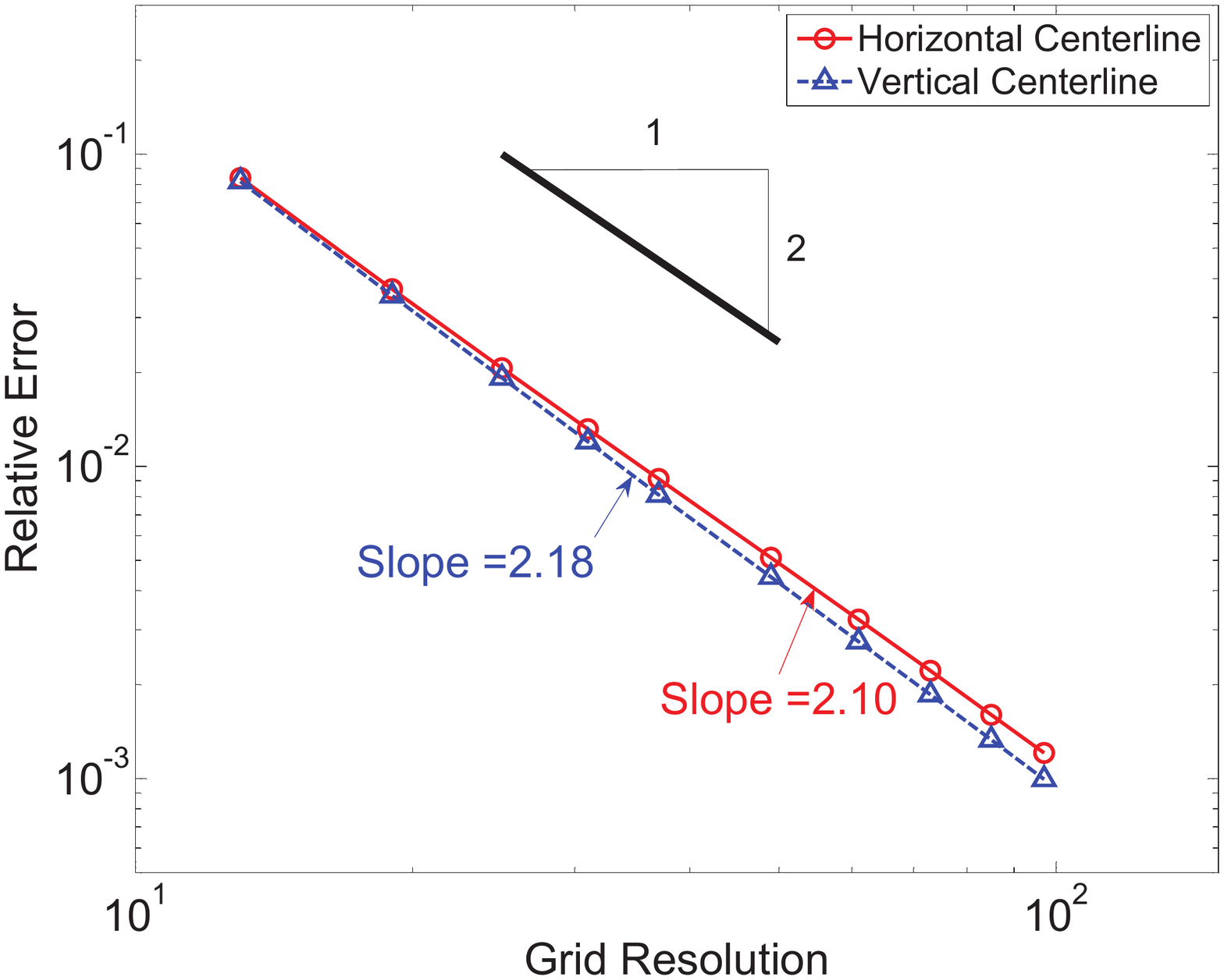}
  \vspace{-15pt}
  \caption{\label{fig:convergence_cavity_centerlinesV}Grid convergence of the cascaded MRT LBM for the V-velocity component in a 2D lid-driven cavity flow for $Re=100$.}
\end{figure}

\begin{figure}[h]
\vspace{-10pt}
\centering
\includegraphics[width = 140mm, angle = 0]{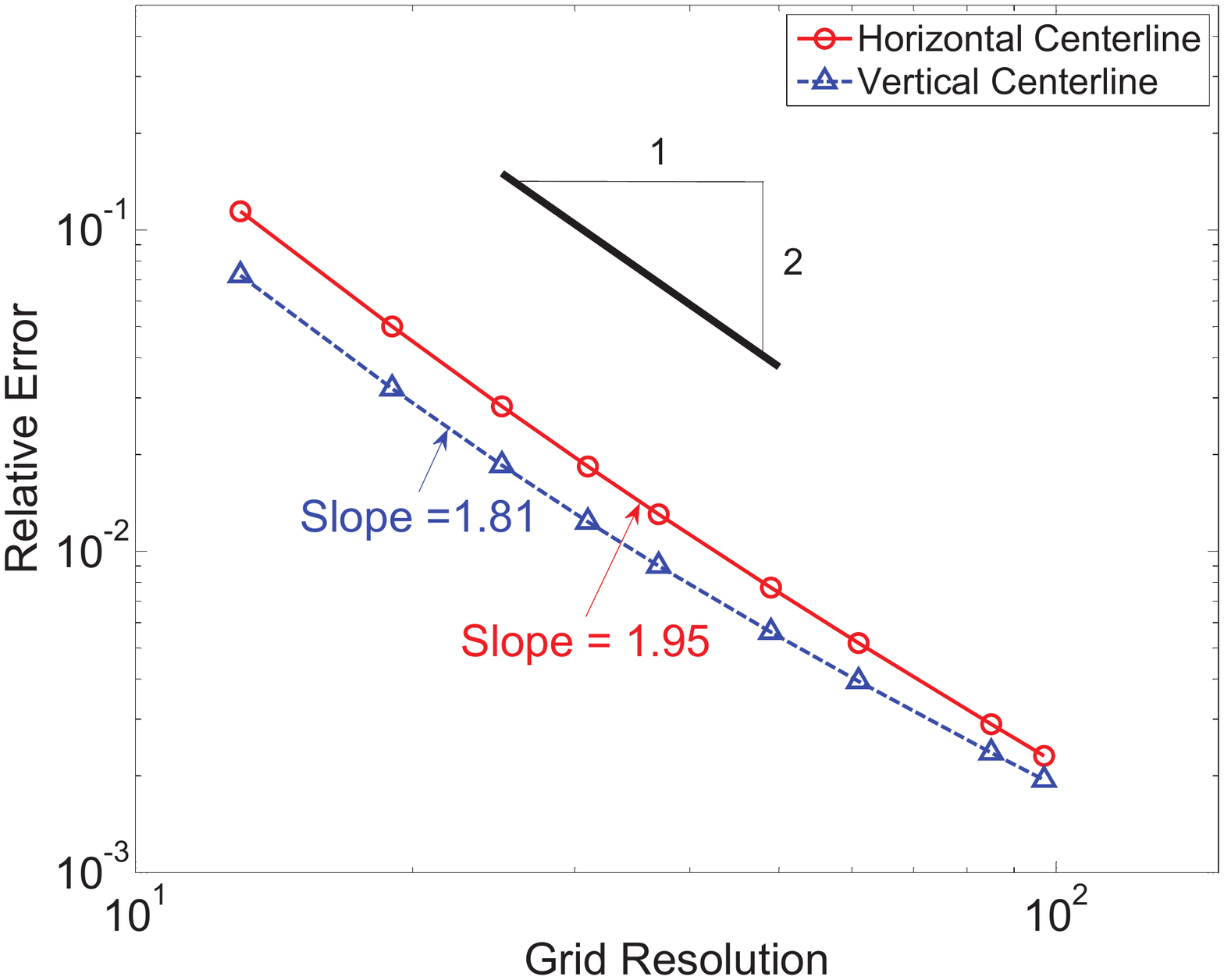}
\vspace{-15pt}
\caption{\label{fig:convergence_cavity_strainrateVY}Grid convergence of the cascaded MRT LBM for the strain rate tensor component $\frac{\partial v}{\partial x}$ in a 2D lid-driven cavity flow for $Re=100$.}
\end{figure}

\begin{figure}
\vspace{-20pt}
\centering
\includegraphics[width = 140mm, angle = 0]{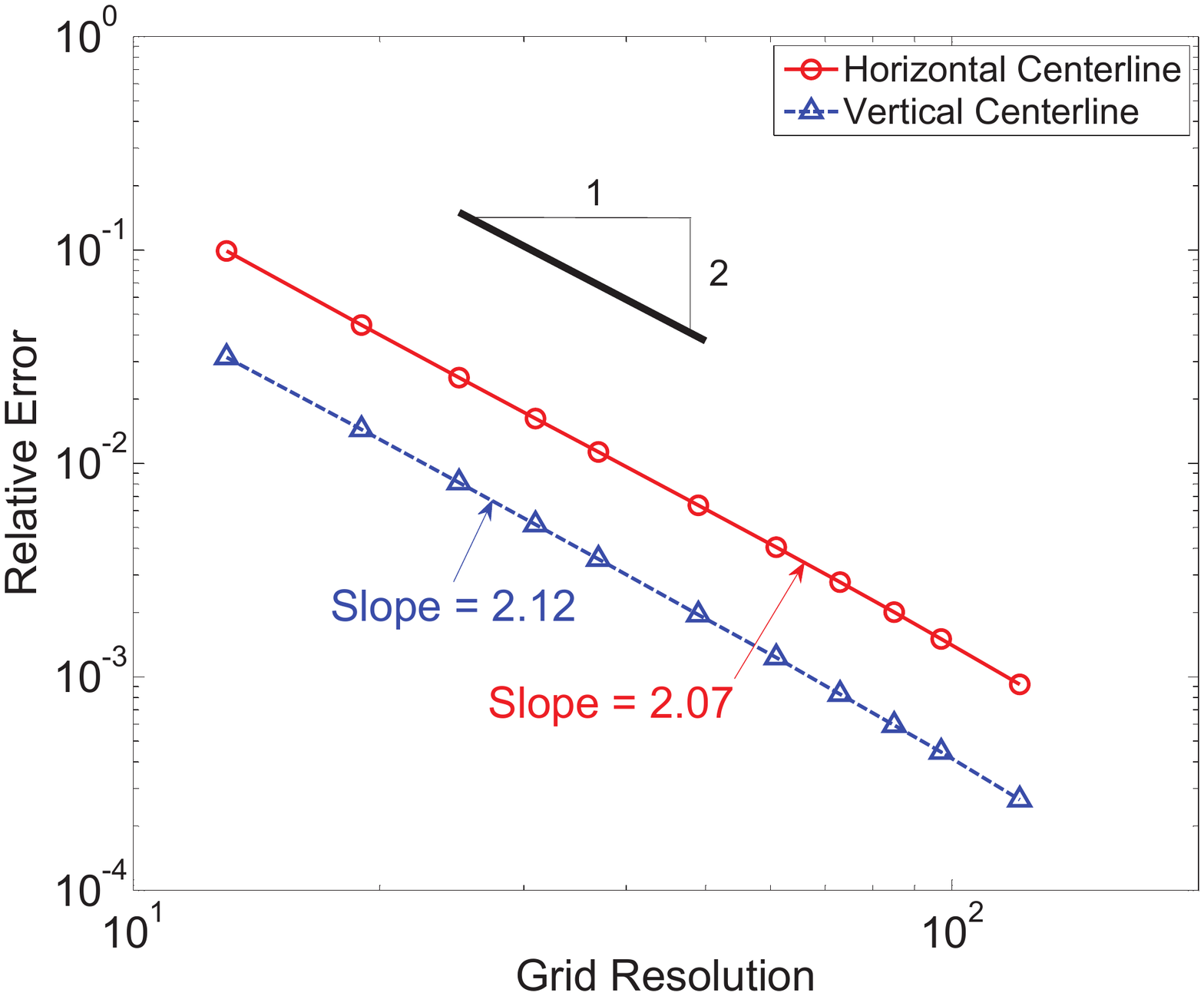}
\vspace{-15pt}
\caption{\label{fig:convergence_cavity_strainrateSUM}Grid convergence of the cascaded MRT LBM for the strain rate tensor component $\frac{\partial u}{\partial y}+\frac{\partial v}{\partial x}$ in a 2D lid-driven cavity flow for $Re=100$.}
\end{figure}

The components of the velocity field and the strain rate tensor at the centerlines of the cavity in both vertical and horizontal directions are computed for a given Reynolds number once the solutions converge to steady state. The solutions are considered to reach steady state convergence
when the relative global errors is small than $10^{-15}$. Again, diffusive scaling is employed to set the parameters for
different grid resolutions consisting of $13\times 13$, $19\times 19$, $25\times 25$, $31\times 31$, $37\times 37$, $49\times 49$, $61\times 61$, $85\times 85$, $97\times 97$ and $121\times 121$ nodes. Figure~\ref{fig:convergence_cavity_centerlinesU} shows the grid convergence of the U-component
of the velocity field at a Reynolds number of $100$. It is found that the best fit slopes are $2.11$ and $2.19$ along the vertical and the
horizontal centerlines, respectively, for the U-velocity. Likewise, the slopes are $2.18$ and $2.11$ respectively along the vertical and the
horizontal centerlines for the V-velocity as shown in Fig.~\ref{fig:convergence_cavity_centerlinesV}.
For the normal strain rate tensor component $\frac{\partial v}{\partial y}$ , the slopes are found to be $1.81$ and $1.95$ respectively for the vertical and the horizontal centerlines, which is shown in Fig.~\ref{fig:convergence_cavity_strainrateVY}. Furthermore, it is seen that along the vertical and horiztonal centerlines, the strain rate tensor component $\frac{\partial u}{\partial y}+\frac{\partial v}{\partial x}$ has the
slopes of $2.12$ and $2.07$, respectively, for grid convergence (see Fig.~\ref{fig:convergence_cavity_strainrateSUM}). One reason why the slopes are either somewhat higher or lower than $2$, rather than very close to the ideal value as seen with the other two problems discussed before, is that the reference solution for obtaining the relative error is taken to be that of the numerical solution with the very fine grid. This is often the practice as the ``analytical'' solution does not exist for this problem.

Overall, it is seen that the cascaded MRT LBM gives a very respectable second order accuracy for a variety of flows, including the relatively simple Poiseuille flow and decaying Taylor-Green vortex flow, and for relatively complex flows such as the lid-driven cavity flow. The method is found
to be second order accurate not only for the velocity field, but also for their derivatives for the above problems.

\section{\label{sec:accuracystudy}Accuracy Studies on the Benchmark Problems}
Let us now make a more detailed comparison of the accuracy of the solutions computed using the cascaded LBM with prior results involving either
analytical or other numerical solution for the flow fields of the three benchmark problems considered in the previous section.

\subsection{2D Poiseuille Flow}
\begin{figure}
\vspace{-20pt}
\centering
\includegraphics[width = 140mm, angle = 0]{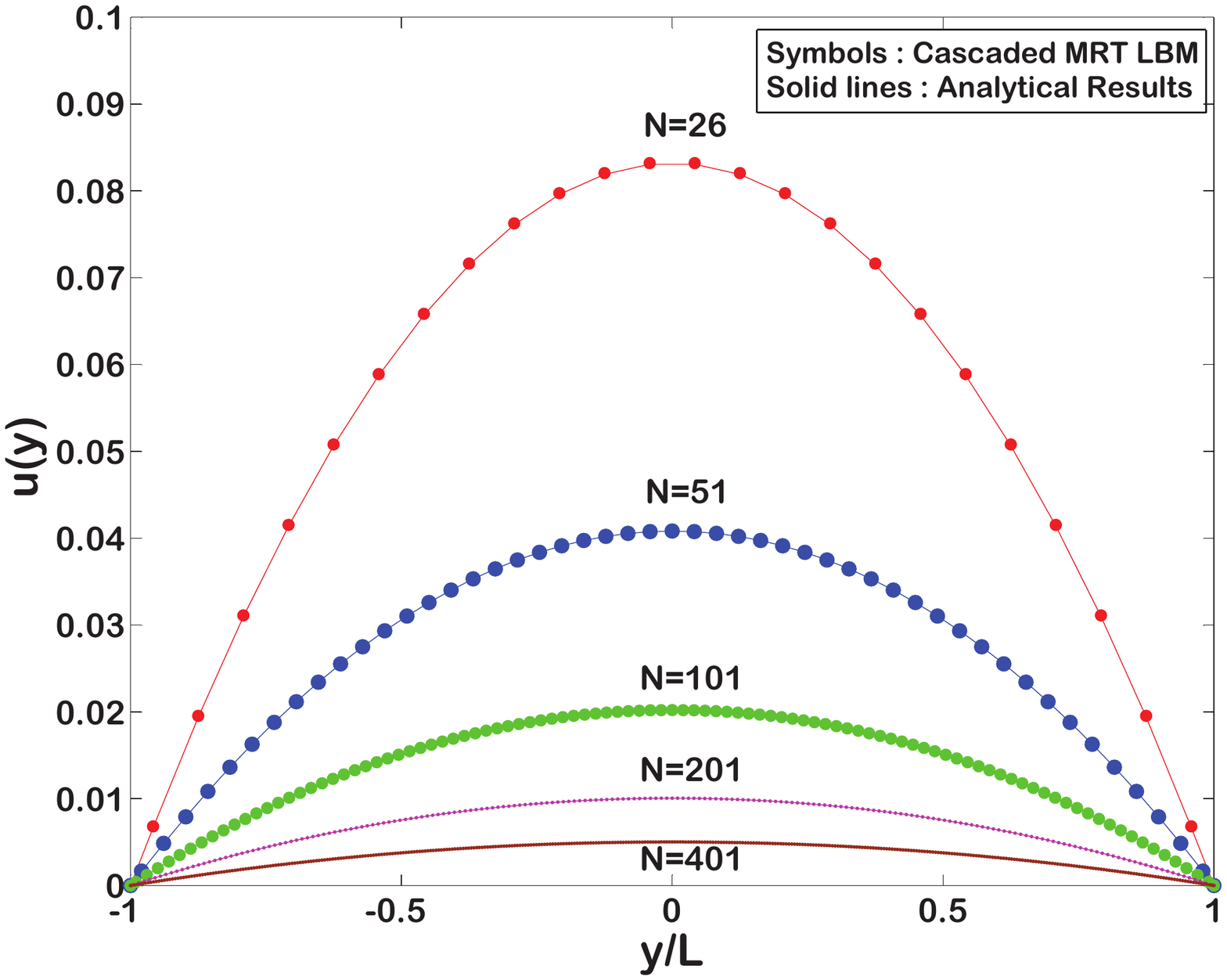}
\vspace{-20pt}
\caption{\label{fig:accuracy_channel_constanttau}Comparison of the velocity profiles in a 2D Poiseuille flow for $Re=200$ at different grid resolutions $N$ in the wall normal direction with a constant relaxation time $\tau=0.55$.}

\end{figure}

Figure~\ref{fig:accuracy_channel_constanttau} shows a comparison of the velocity profiles of the 2D Poiseuille flow between the results obtained using the cascaded LBM and the parabolic analytical solution at a constant Reynolds number of $200$ with constant relaxation time $\tau=0.515$ for different grid resolutions in the wall normal direction starting from $26$ to $401$. Here, diffusive scaling is employed in the selection of parameters. That is, as the resolution is doubled, the maximum flow velocity or the Mach number is decreased by a factor of $2$. The results are in excellent agreement with the analytical solution, in which the maximum relative error is less than $0.22$ percent.

\subsection{2D Decaying Taylor-Green Vortex Flow}
Using the same set of parameters specified for this time-dependent problem in the previous section, we now compare the computed $U-$ and $V-$ velocity components along the vertical and horizontal centerlines, respectively, with the corresponding analytical solutions (Eq.~(\ref{eq:tayloru})-(\ref{eq:taylorv})) at three different representative instants. Figures~\ref{fig:accuracy_taylorU} and~\ref{fig:accuracy_taylorV} show such a comparison of the velocity components at times $t=6.55T$, $13.10T$ and $25.20T$, where the characteristic time $T$ is defined in the previous section, reflecting the decaying of the initial vortex distribution. It is evident that the cascaded MRT LBM is in excellent agreement with the analytical solution at all times shown.
\begin{figure}
\centering
\subfigure{\label{fig:accuracy_taylorU}\includegraphics[width = 140mm, angle = 0]{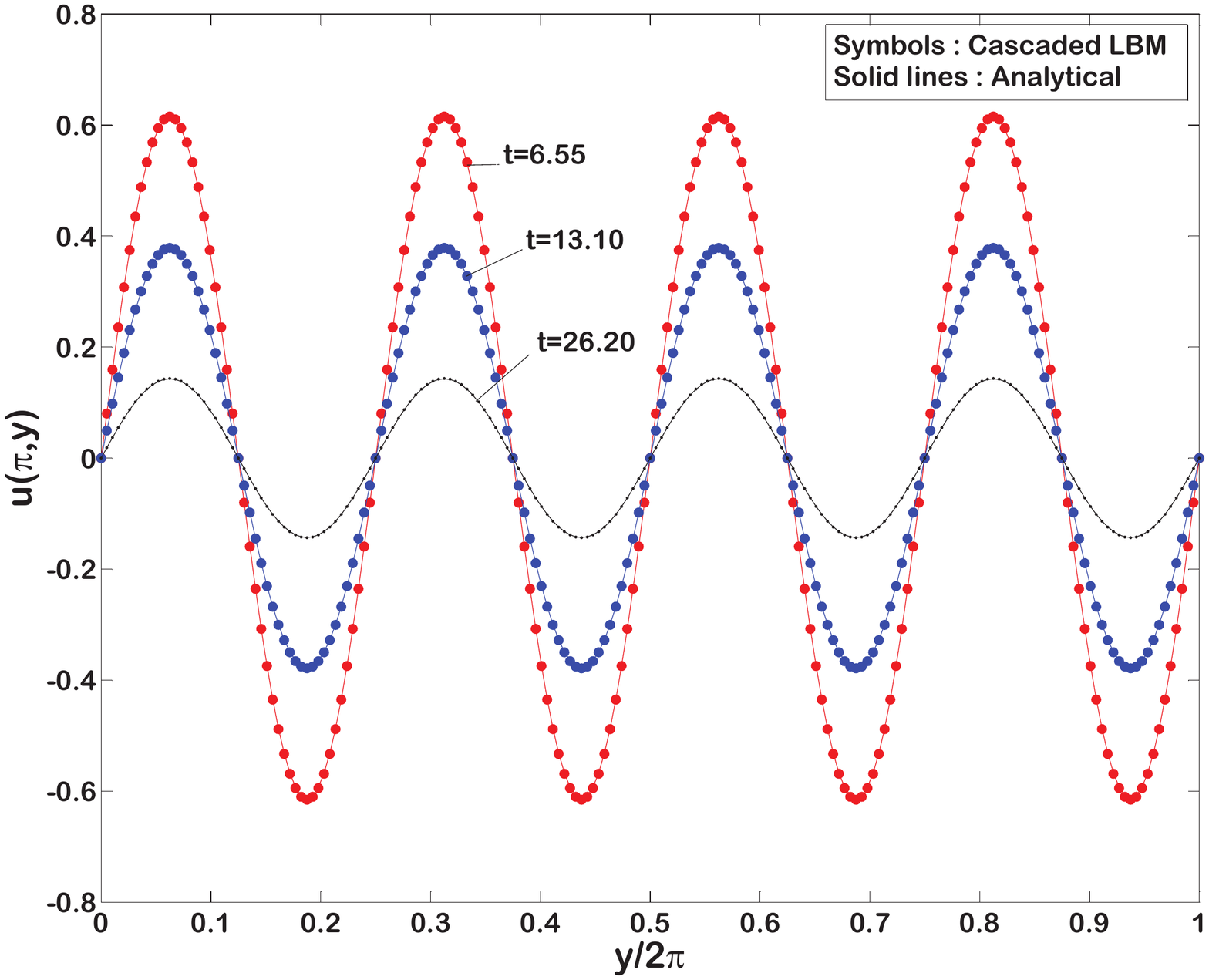}}
\vspace{-5pt}
\caption{\label{fig:accuracy_taylorU}Comparison of the U-velocity component in a decaying Taylor-Green vortex flow for $Re=14.4$ at three different non-dimensional times $T$: $t = 6.55T, 13.10T$ and $26.20T$.}
\subfigure{\label{fig:accuracy_taylorV}\includegraphics[width = 140mm, angle = 0]{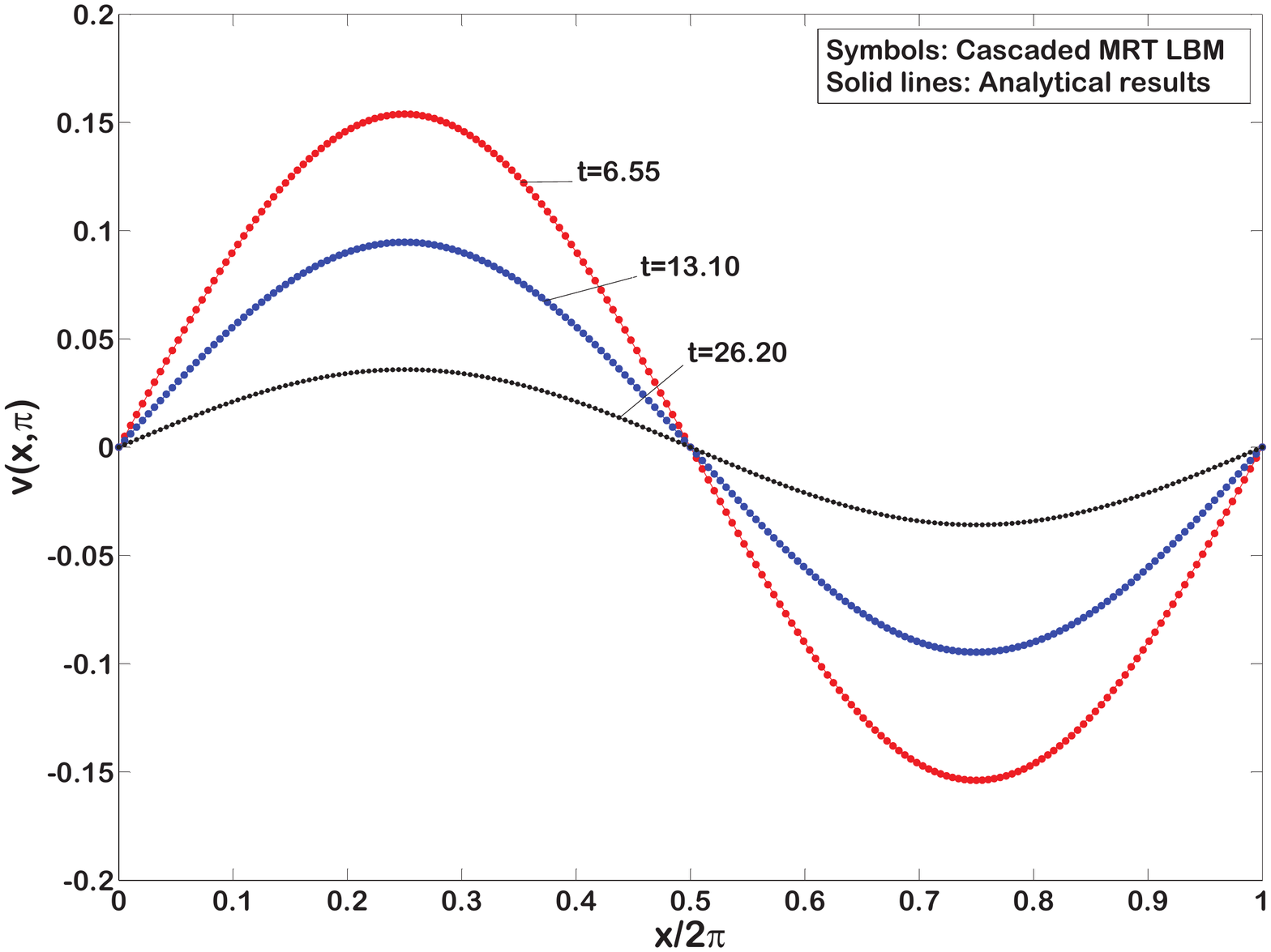}}
\vspace{-12pt}
\caption{\label{fig:accuracy_taylorV}Comparison of the V-velocity component in a decaying Taylor-Green vortex flow for $Re=14.4$ at three different non-dimensional times $T$: $t = 6.55T, 13.10T$ and $26.20T$.}
\end{figure}

\subsection{2D Lid-driven Cavity Flow}
Let us now consider more detailed features of the lid-driven cavity flow problem discussed in the last section at various Reynolds numbers in order
to make quantitative comparisons. Figures~\ref{fig:accuracy_cavity_centerlinesU} and~\ref{fig:accuracy_cavity_centerlinesV} show the U- and V- components of the velocity, respectively, along the centerlines of the square cavity at Reynolds numbers of $100$, $400$, $1000$, $3200$, $5000$, and $7500$ obtained using the cascaded MRT LBM along with the previous numerical data presented by Ghia \emph{et al}~\cite{ghia82}. The cascaded MRT LBM results corresponding to the finest grid considered earlier, i.e. for the $401 \times 401$ grid resolution are chosen to make comparison. In these figures, the solid lines represent the computed results obtained by the cascaded MRT LBM, and the symbols are the prior data provided by Ghia \emph{et al}~\cite{ghia82}. The velocities are normalized by the lid velocity $U_0$. Very good agreement is seen for all the Reynolds numbers considered.
\begin{figure}
\vspace{-20pt}
\centering
\subfigure{\label{fig:accuracy_cavity_centerlinesU}\includegraphics[width = 140mm, angle = 0]{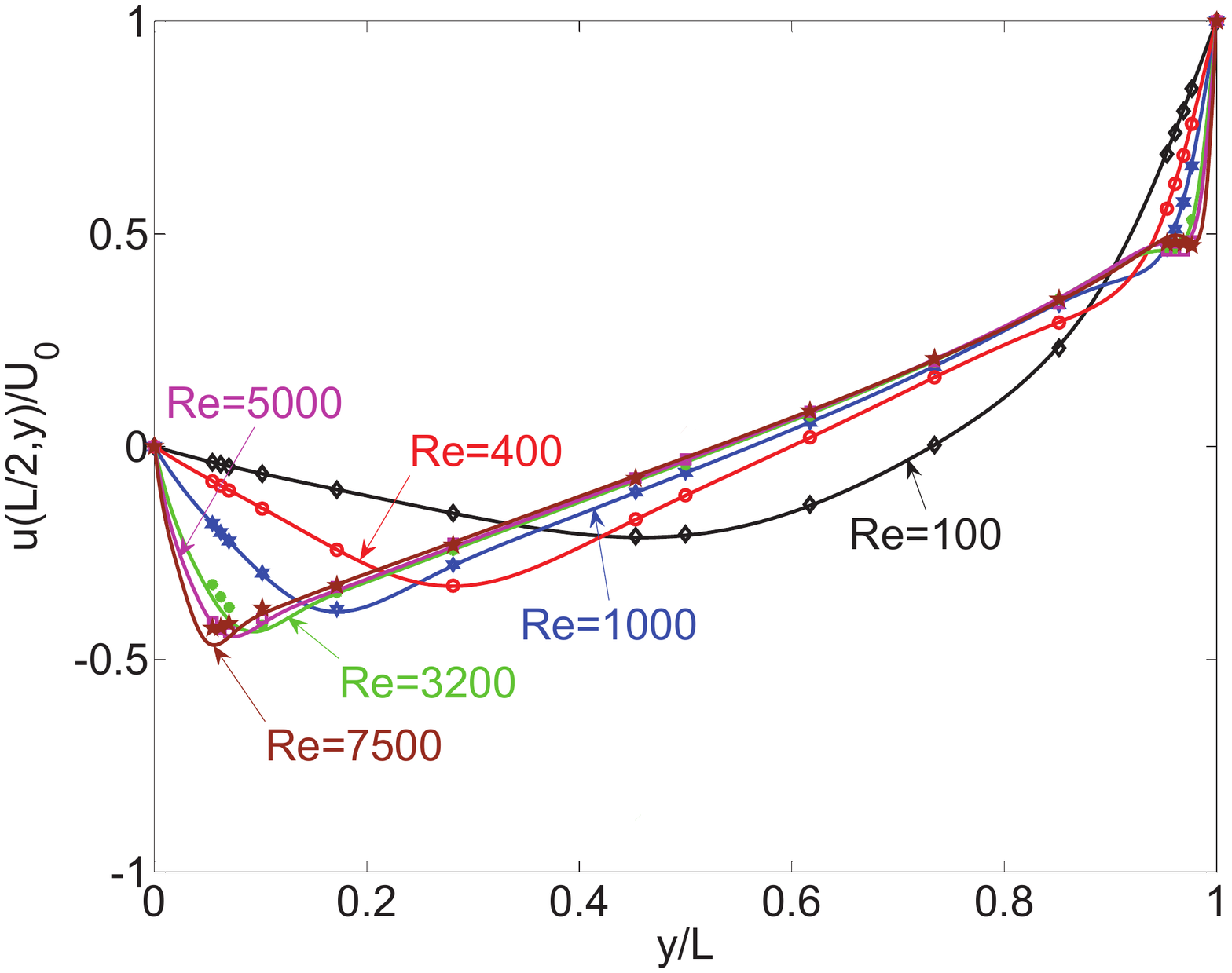}}
\vspace{-20pt}
\caption{\label{fig:accuracy_cavity_centerlinesU}Comparison of the $U$- component of the velocity field along the vertical centerline of the cavity flow at various Reynolds numbers: $Re=100,400,1000,3200,5000$ and $7500$. Lines -- cascaded MRT LBM and symbols -- data by Ghia \emph{et al}~\cite{ghia82}.}
\subfigure{\label{fig:accuracy_cavity_centerlinesV}\includegraphics[width = 140mm, angle = 0]{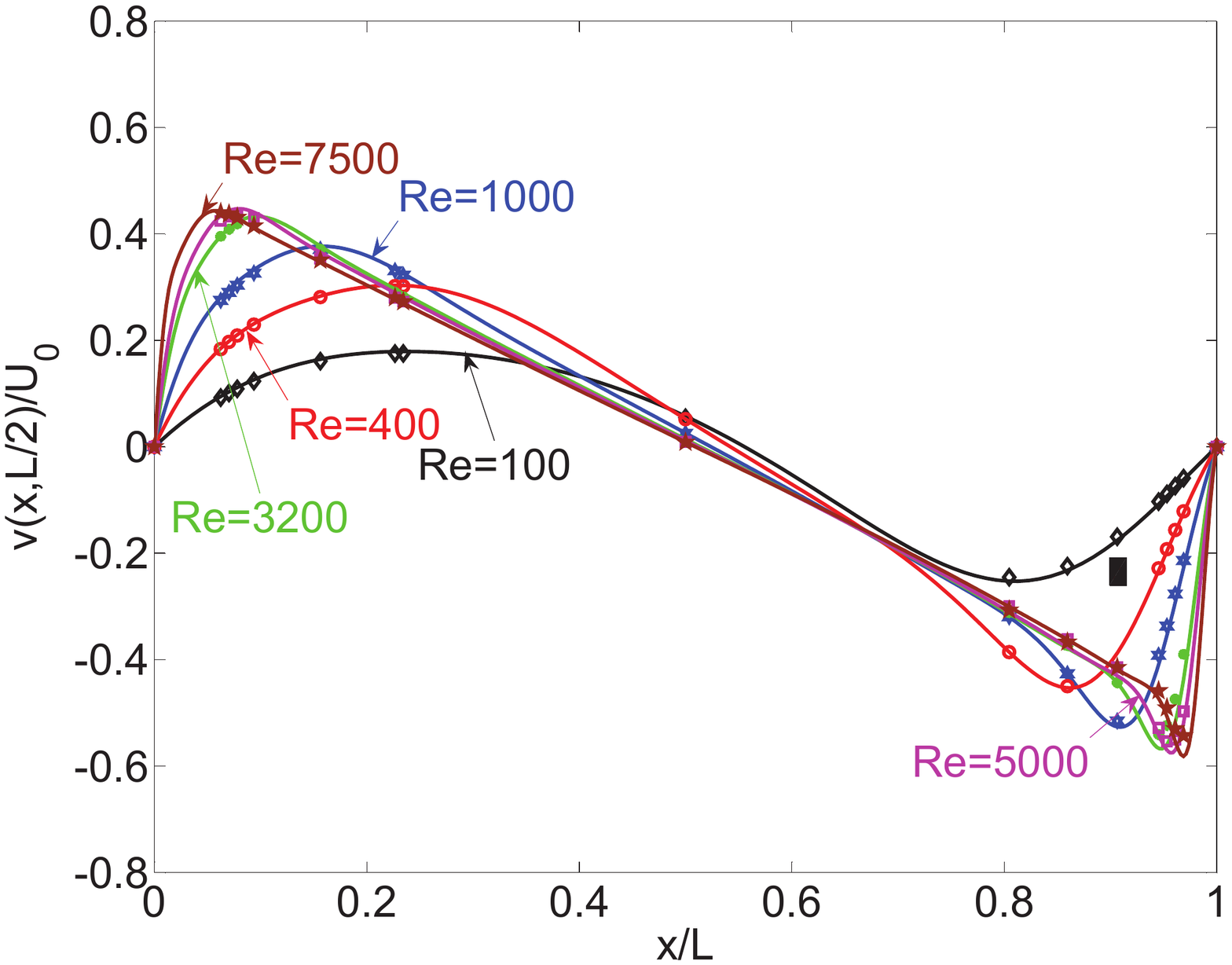}}
\vspace{-20pt}
\caption{\label{fig:accuracy_cavity_centerlinesV}Comparison of the $V$- component of the velocity field along the horizontal centerline of the cavity flow at various Reynolds numbers: $Re=100,400,1000,3200,5000$ and $7500$. Lines -- cascaded MRT LBM and symbols -- data by Ghia \emph{et al}~\cite{ghia82}.}
\end{figure}

\begin{figure}[h!]
  \centering
  \subfigure[Cascaded MRT $Re=3200$]{
   \vspace{20pt}
  \label{fig:subfig:g}
  \includegraphics[width = 2.0in, angle = 90]{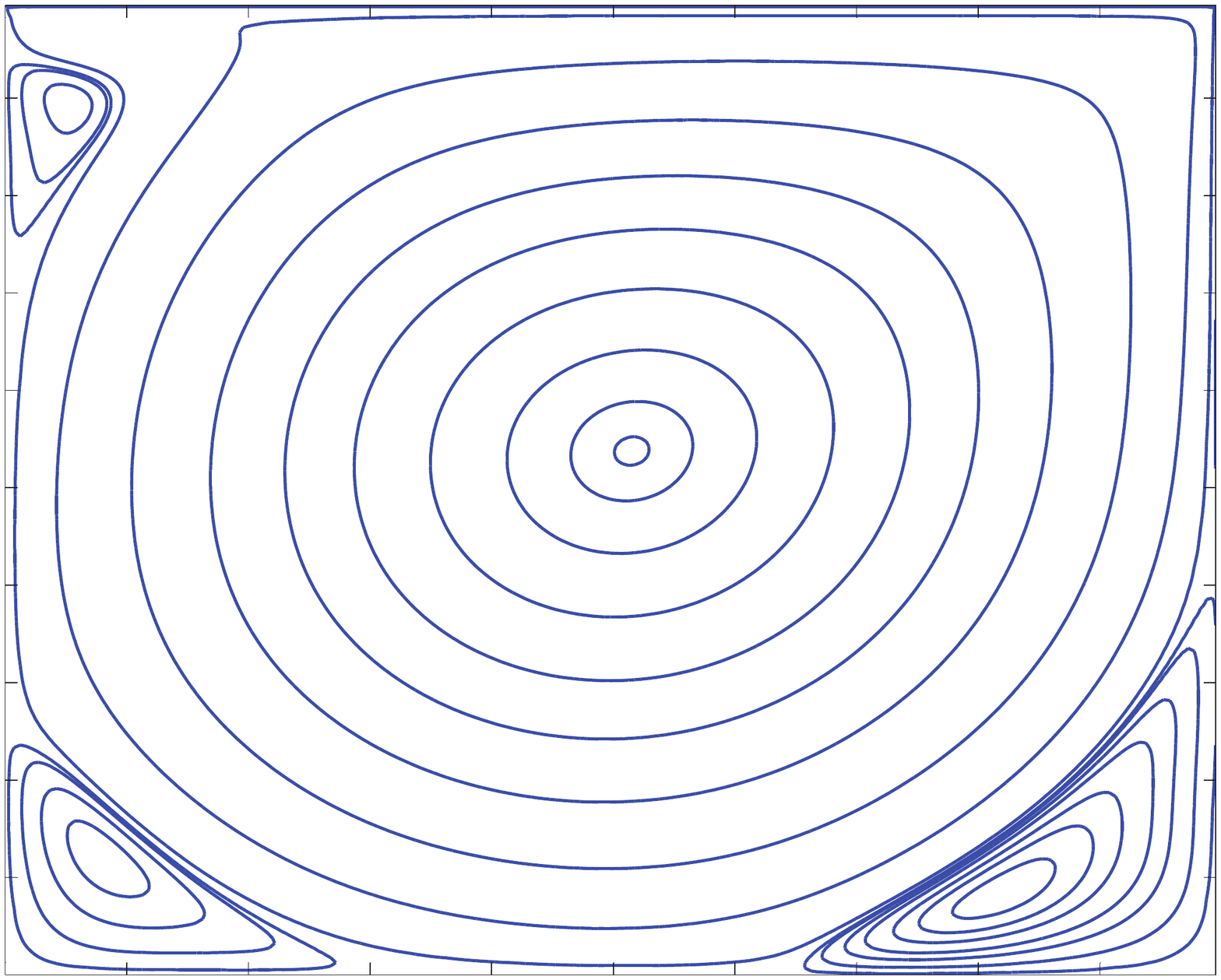}}
  \subfigure[Standard MRT $Re=3200$]{
   \vspace{20pt}
  \label{fig:subfig:h}
  \includegraphics[width = 2.0in, angle = 90]{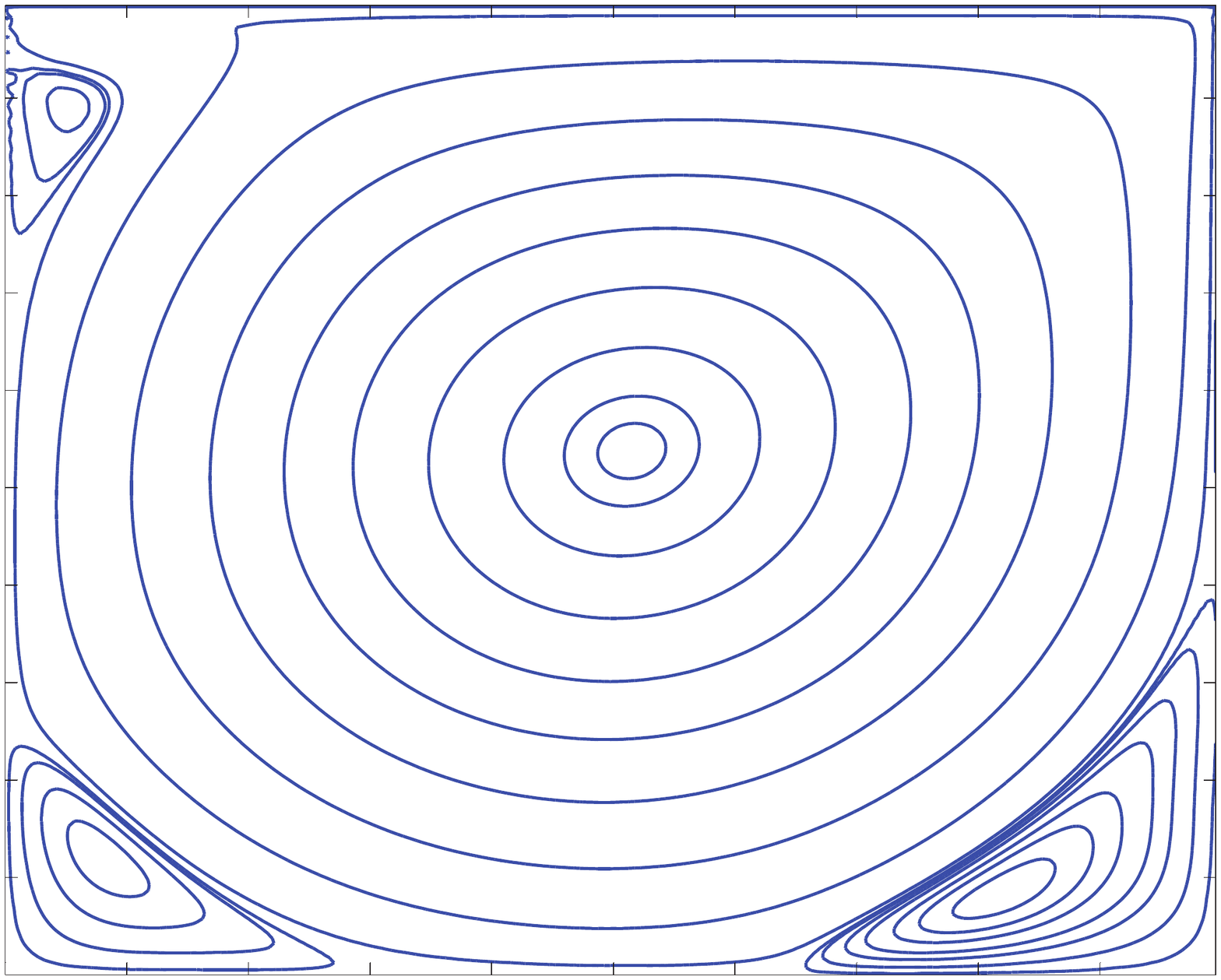}}
  \centering
  \subfigure[Cascaded MRT $Re=5000$]{
   \vspace{20pt}
  \label{fig:subfig:i}
  \includegraphics[width = 2.0in, angle = 90]{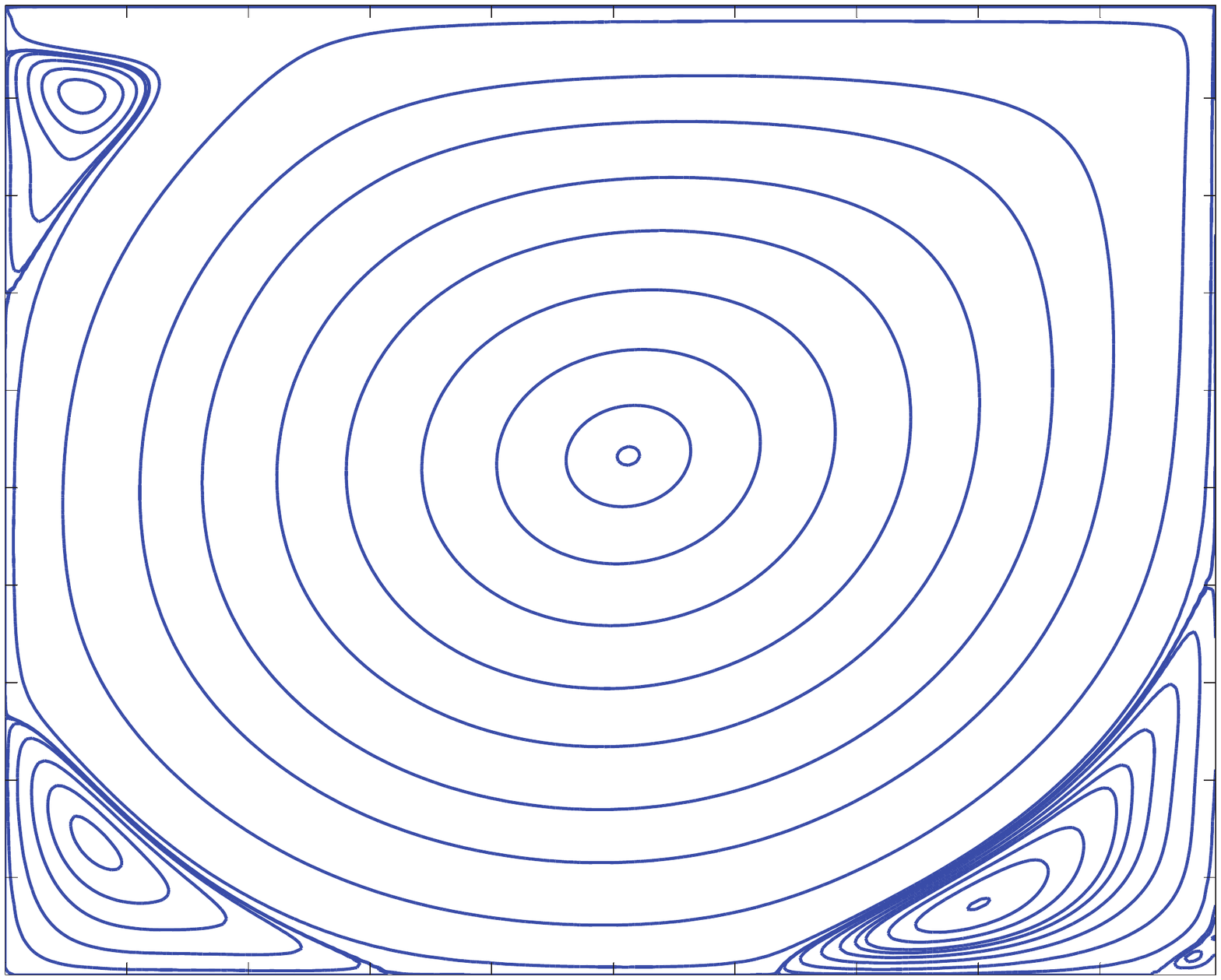}}
  \subfigure[Standard MRT $Re=5000$]{
   \vspace{20pt}
  \label{fig:subfig:j}
  \includegraphics[width = 2.0in, angle = 90]{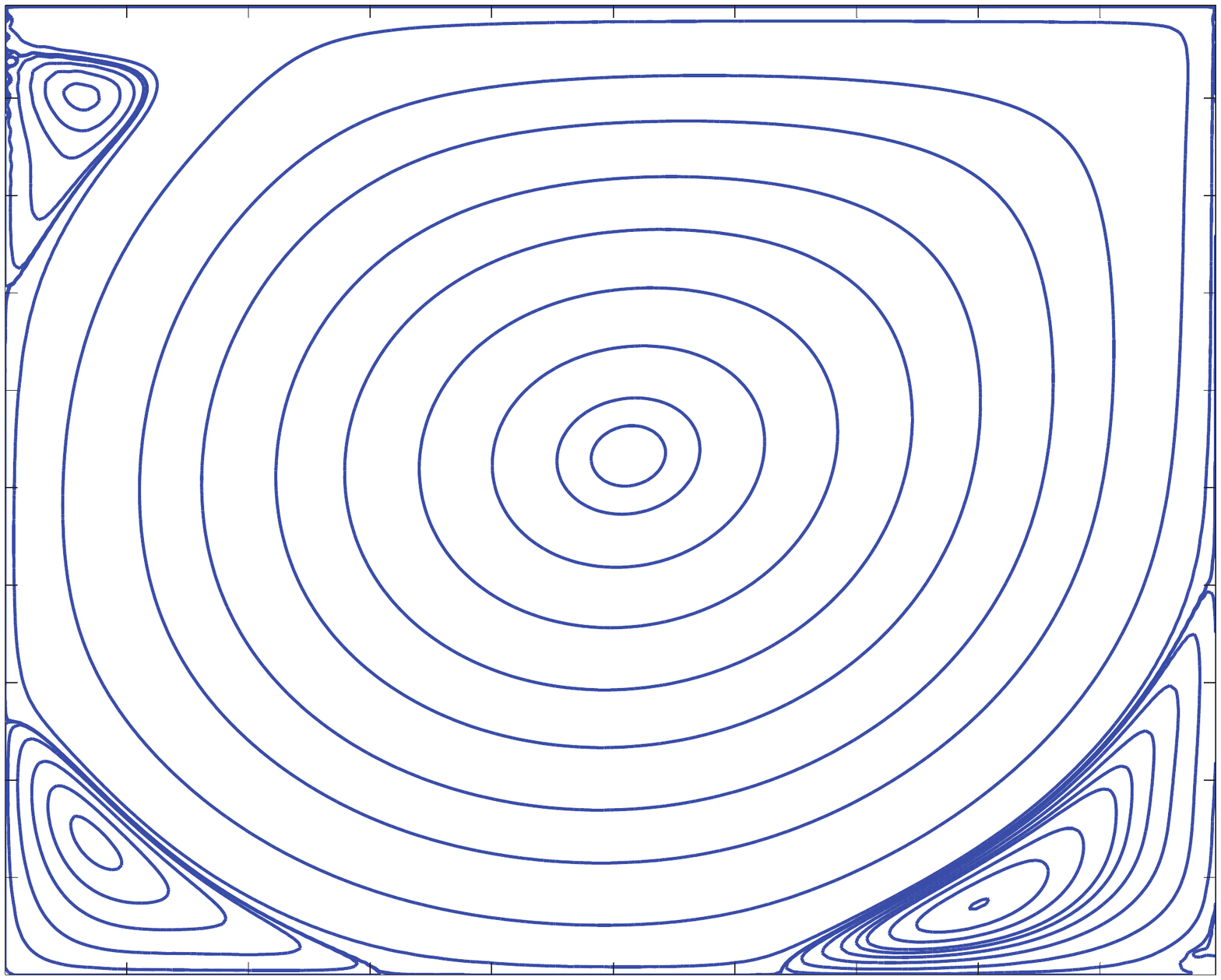}}
  \centering
  \subfigure[Cascaded MRT $Re=7500$]{
   \vspace{20pt}
  \label{fig:subfig:k}
  \includegraphics[width = 2.0in, angle = 90]{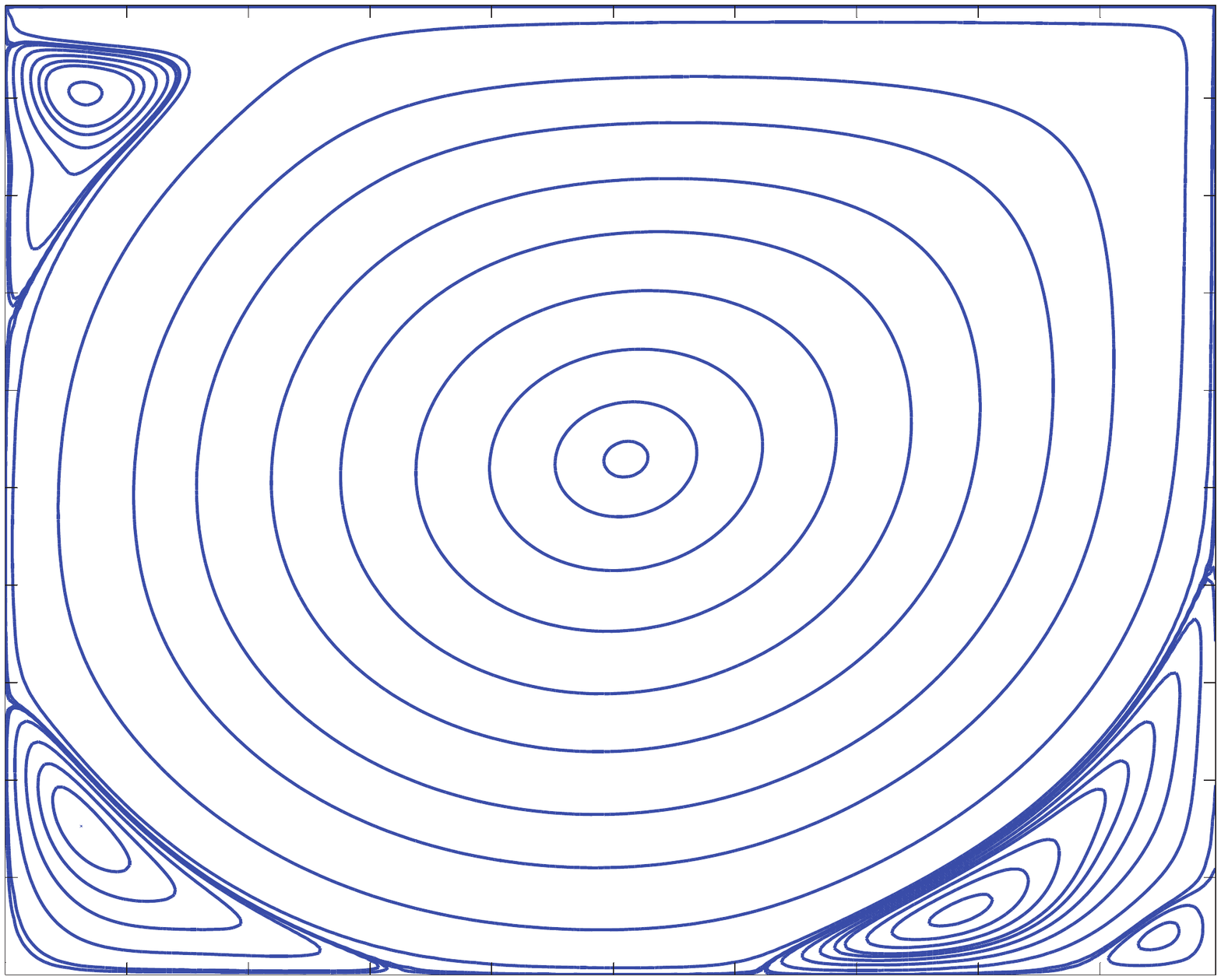}}
  \subfigure[Standard MRT $Re=7500$]{
   \vspace{20pt}
  \label{fig:subfig:l}
  \includegraphics[width = 2.0in, angle = 90]{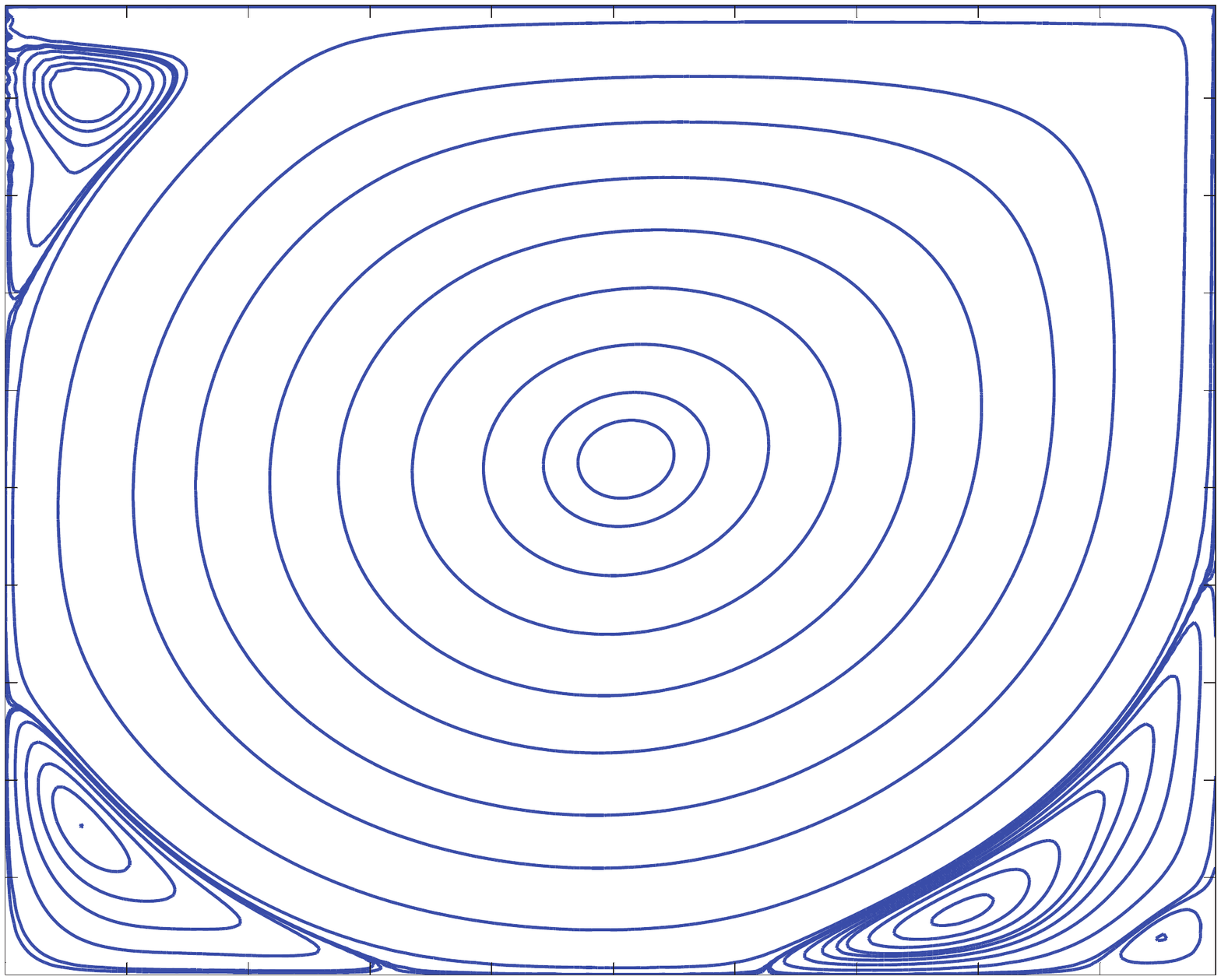}}
  \caption{Comparison of the streamlines in a 2D lid-driven cavity flow at different Reynolds numbers computed with cascaded (central moment) MRT LBM and standard (raw moment) MRT LBM: $Re=100,400,1000,3200,5000$ and $7500$. Solutions obtained using $201^2$ grids with both methods.}
  \label{fig:streamlines}
\end{figure}

\begin{figure}
\centering
  \subfigure[Cascaded MRT Top]{
  \label{fig:subfig:a}
  \includegraphics[width = 3.0in, angle = 0]{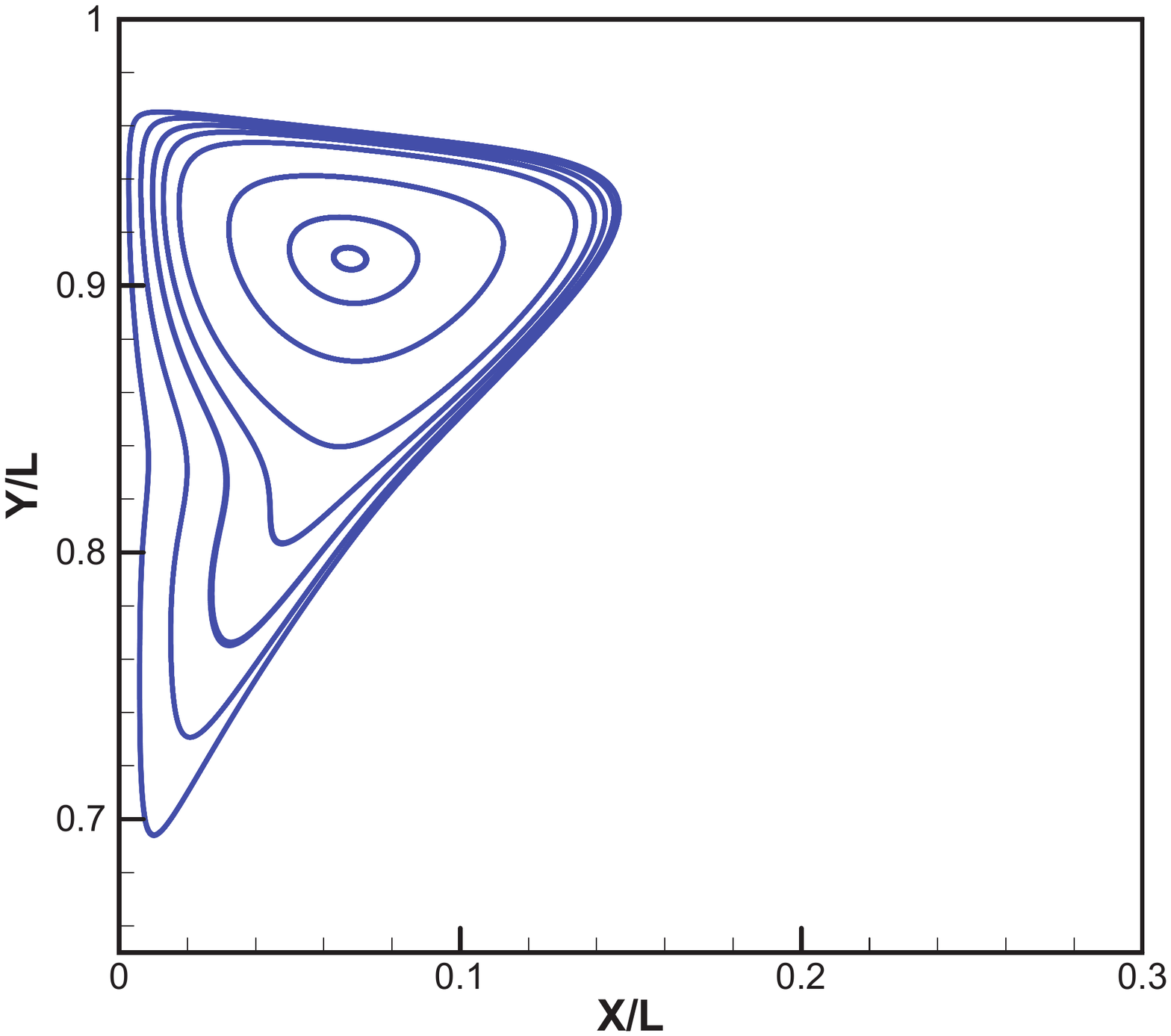}}
  \subfigure[Standard MRT Top]{
  \label{fig:subfig:b}
  \includegraphics[width = 3.0in, angle = 0]{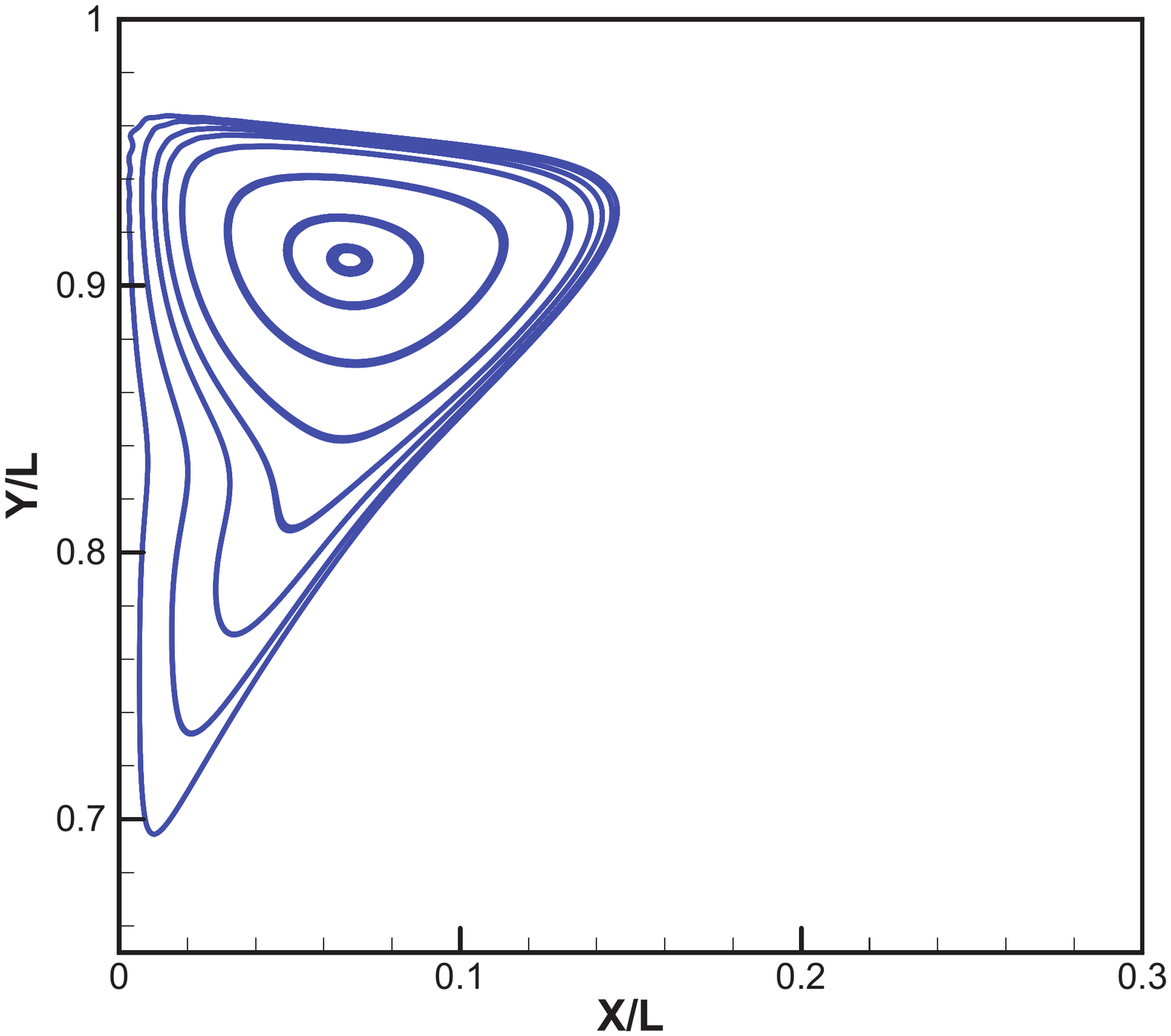}}\\
  \centering
  \subfigure[Cascaded MRT Bottomleft]{
  \label{fig:subfig:c}
  \includegraphics[width = 3.0in, angle = 0]{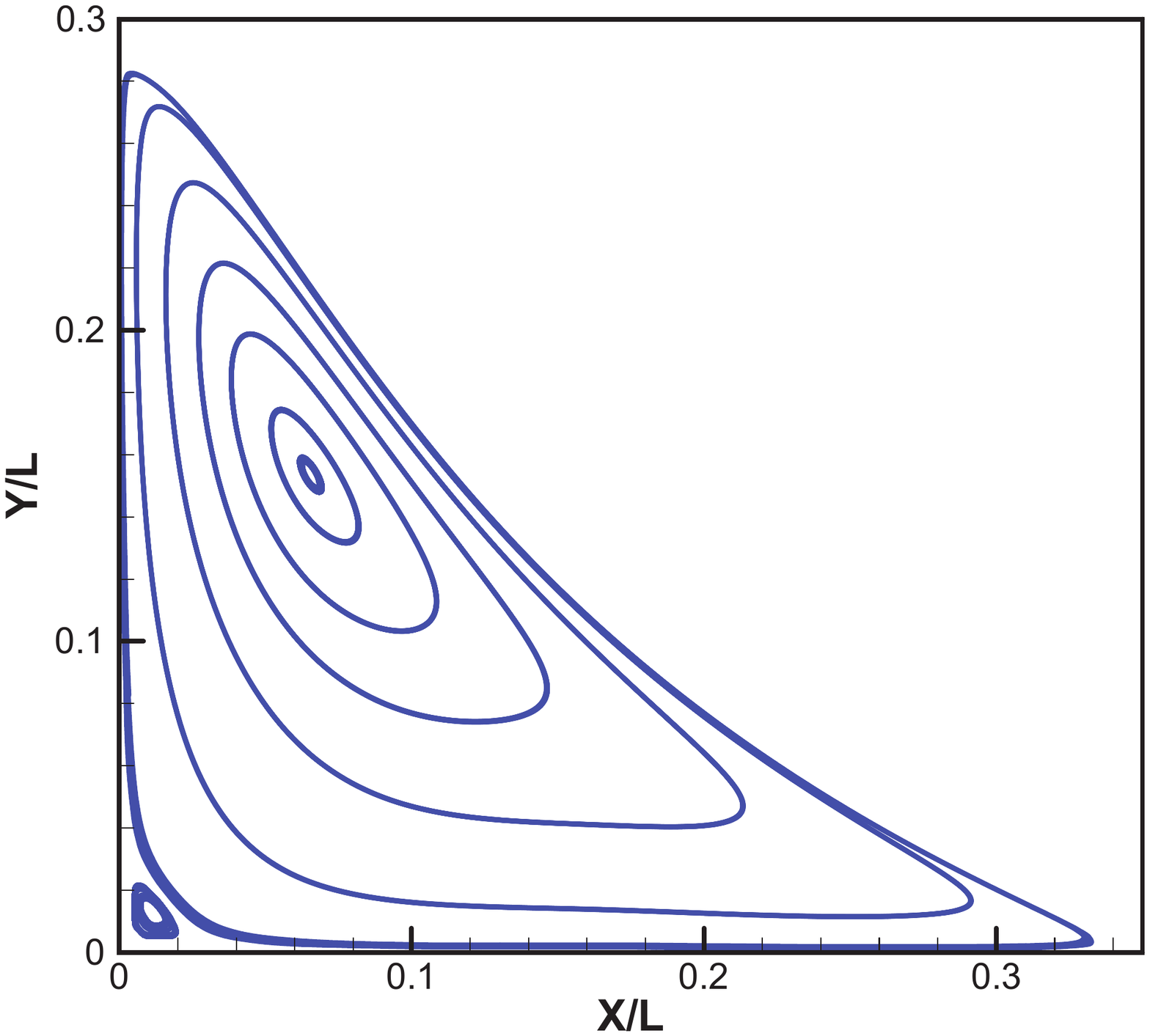}}
  \subfigure[Standard MRT Bottomleft]{
  \label{fig:subfig:d}
  \includegraphics[width = 3.0in, angle = 0]{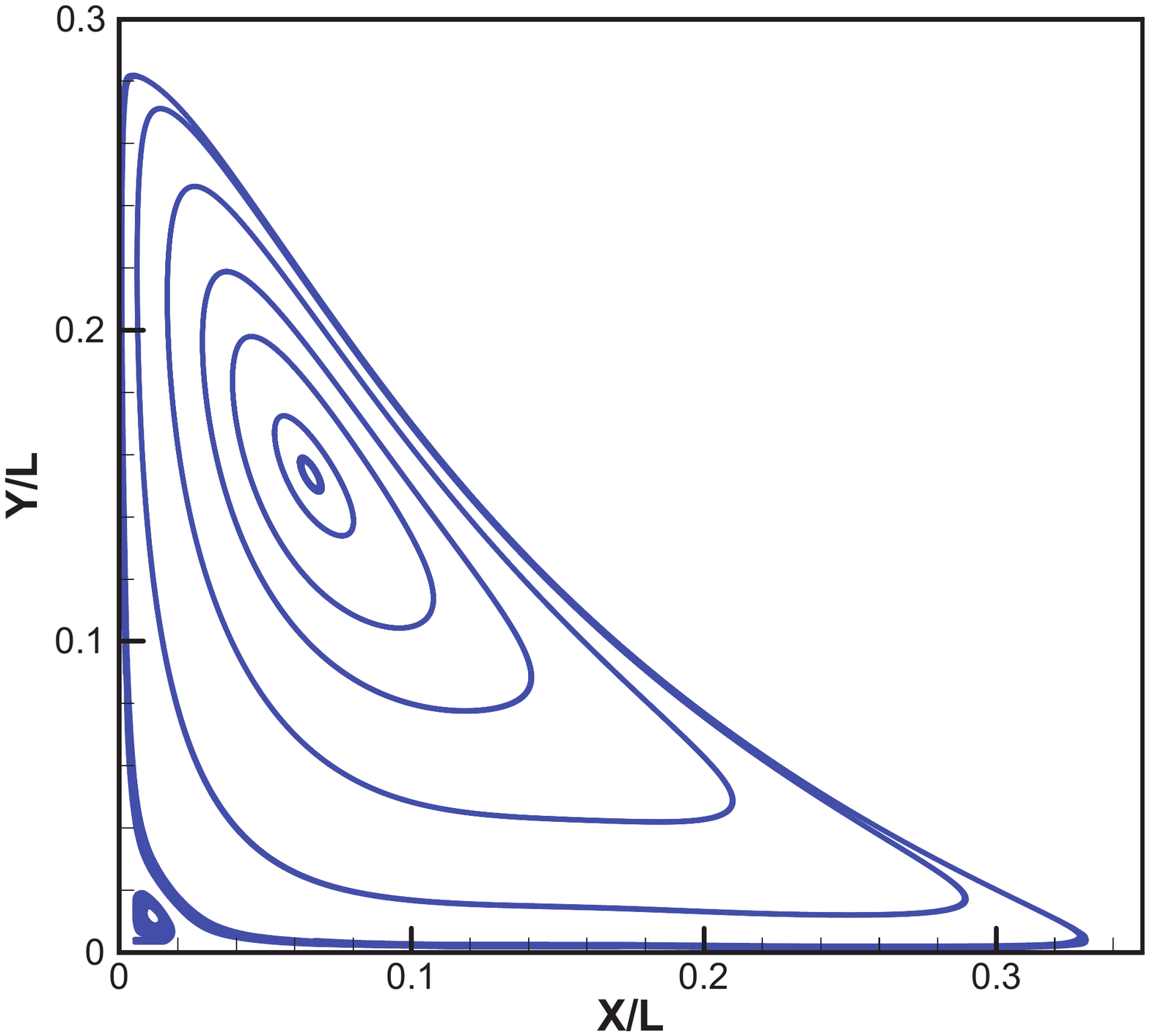}}\\
  \centering
  \subfigure[Cascaded MRT Bottomright]{
  \label{fig:subfig:e}
  \includegraphics[width = 3.0in, angle = 0]{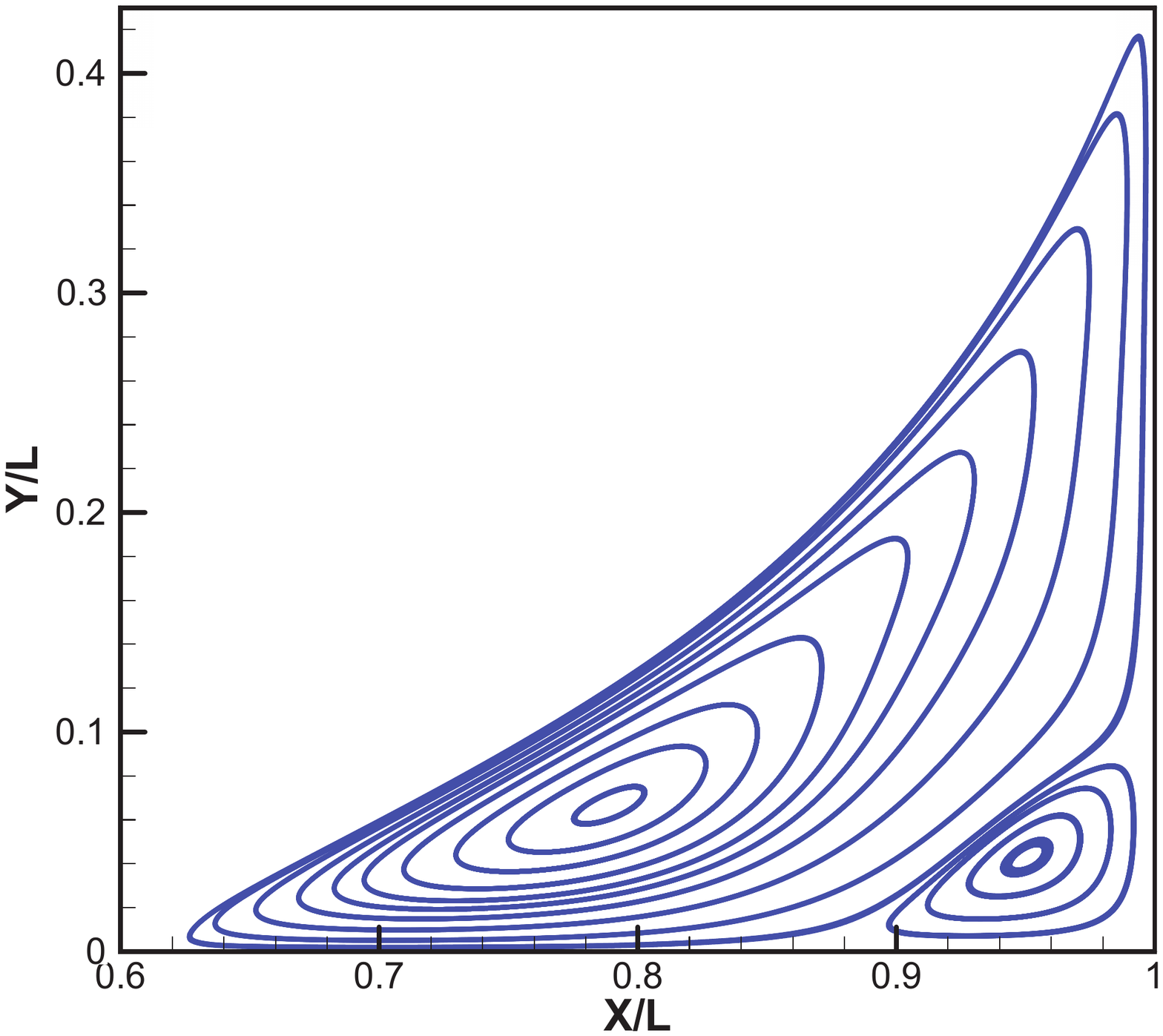}}
  \subfigure[Standard MRT Top]{
  \label{fig:subfig:f}
  \includegraphics[width = 3.0in, angle = 0]{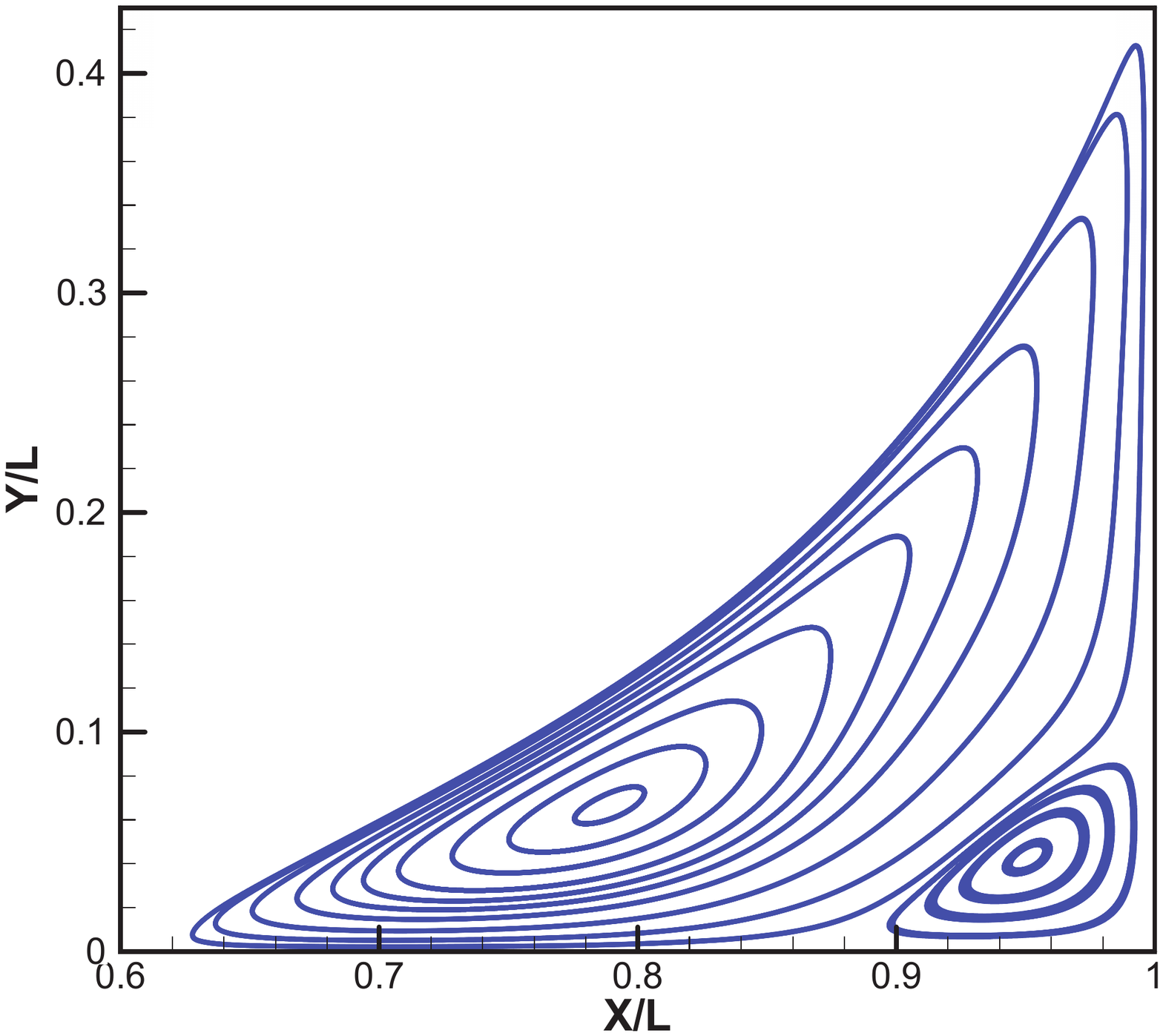}}
  \caption{Comparison of the streamlines of the secondary vortices in a 2D lid-driven cavity flow at $Re=7500$ computed with cascaded (central moment) MRT LBM and standard (raw moment) MRT LBM.}
  \label{fig:secondary}
\end{figure}

In a previous work, it was established that the standard MRT LBM based on raw moments is superior when compared with the SRT LBM for the computation of lid-driven cavity flow~\cite{luo11}. Hence, it would be sufficient to make a direct comparison between the cascaded MRT LBM based on central moments and the standard MRT LBM for various flow characteristics of this problem. First, in order to provide a global characteristics of the flow field, it would be interesting to compare the streamlines in the cavity at various Reynolds numbers. It is known that at a certain Reynolds number above $7500$, the flow field becomes unsteady and we restrict such comparisons for stationary state solutions only. Hence, Fig.~\ref{fig:streamlines} shows the computed streamlines at Reynolds numbers of $100$, $400$, $1000$, $5000$ and $7500$ using both the above methods. The streamlines computed by both these approaches are plotted side-by-side for comparison. It is found that the streamlines appear to be remarkably very similar with both the raw moment and central moment based approaches. At Reynolds numbers of $100$, $400$ and $1000$, a major vortex appears around the geometric center of the cavity with two minor vortices around the lower corners. Since the lid is driven from left to right, the major vortex circulates in a clockwise direction and the two minor vortices circulate in a counter-clockwise direction. At Reynolds numbers of $3200$ and $5000$, in addition to the vortices that exist with the lower Reynolds number cases, there appears another minor vortex on the left upper corner, which circulates in a counter-clockwise direction. When the Reynolds number increases further to $7500$, a fourth minor vortex is found on the right lower corner, which circulates in a clockwise direction. All the above flow features correspond to steady states. Furthermore, in order to provide a more detailed comparison, we present various secondary vortices that appear in the cavity at $Re=7500$ in Fig.~\ref{fig:secondary}. Again, remarkable similarity between the cascaded MRT LBM and the standard MRT LBM is found for these more detailed secondary flow structures.
\begin{figure}[h!]
  \centering
  \subfigure[Cascaded MRT $Re=100$]{
  \vspace{20pt}
  \label{fig:subfig:a}
  \includegraphics[width = 2.0in, angle = 90]{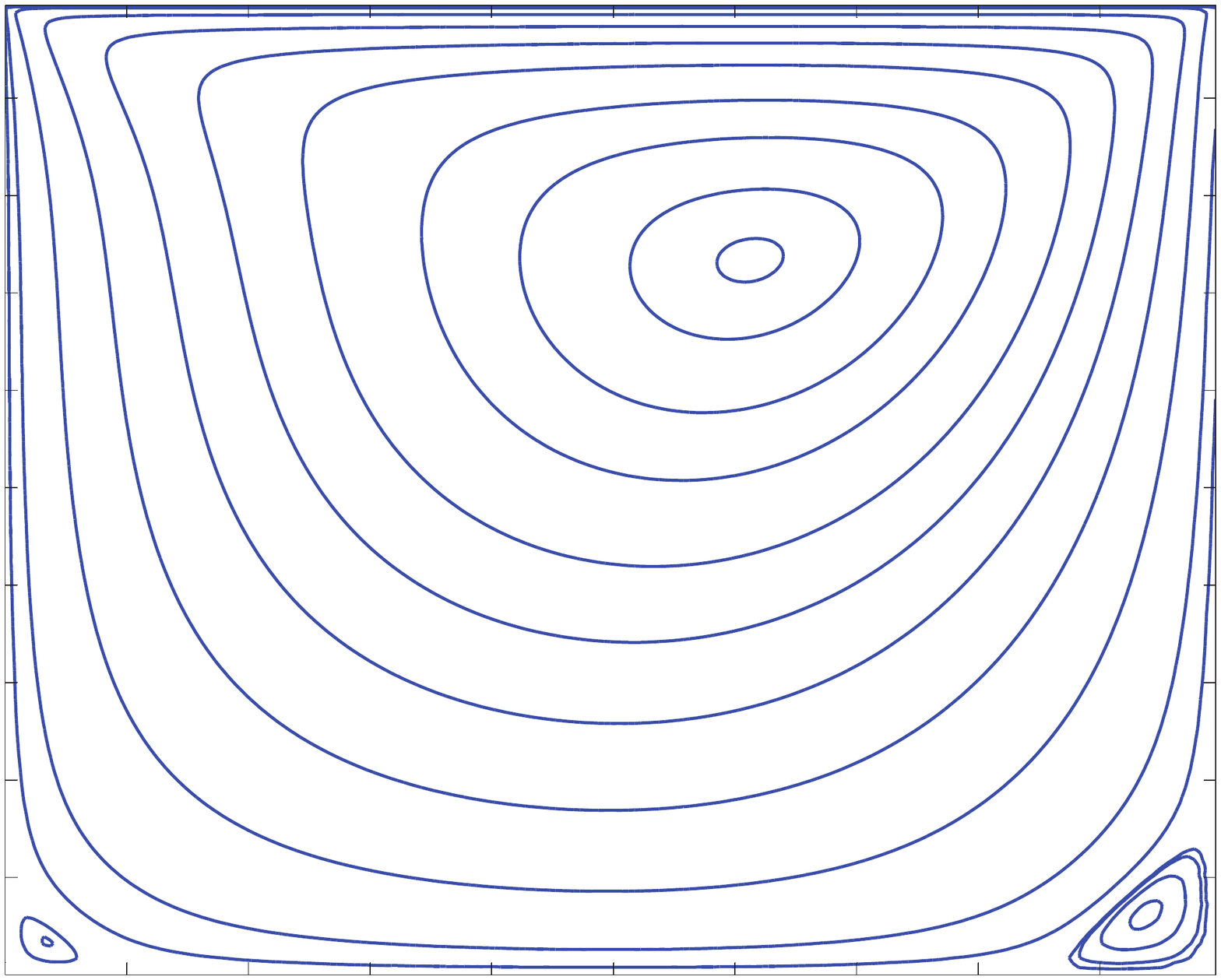}}
  \subfigure[Standard MRT $Re=100$]{
  \vspace{20pt}
  \label{fig:subfig:b}
  \includegraphics[width = 2.0in, angle = 90]{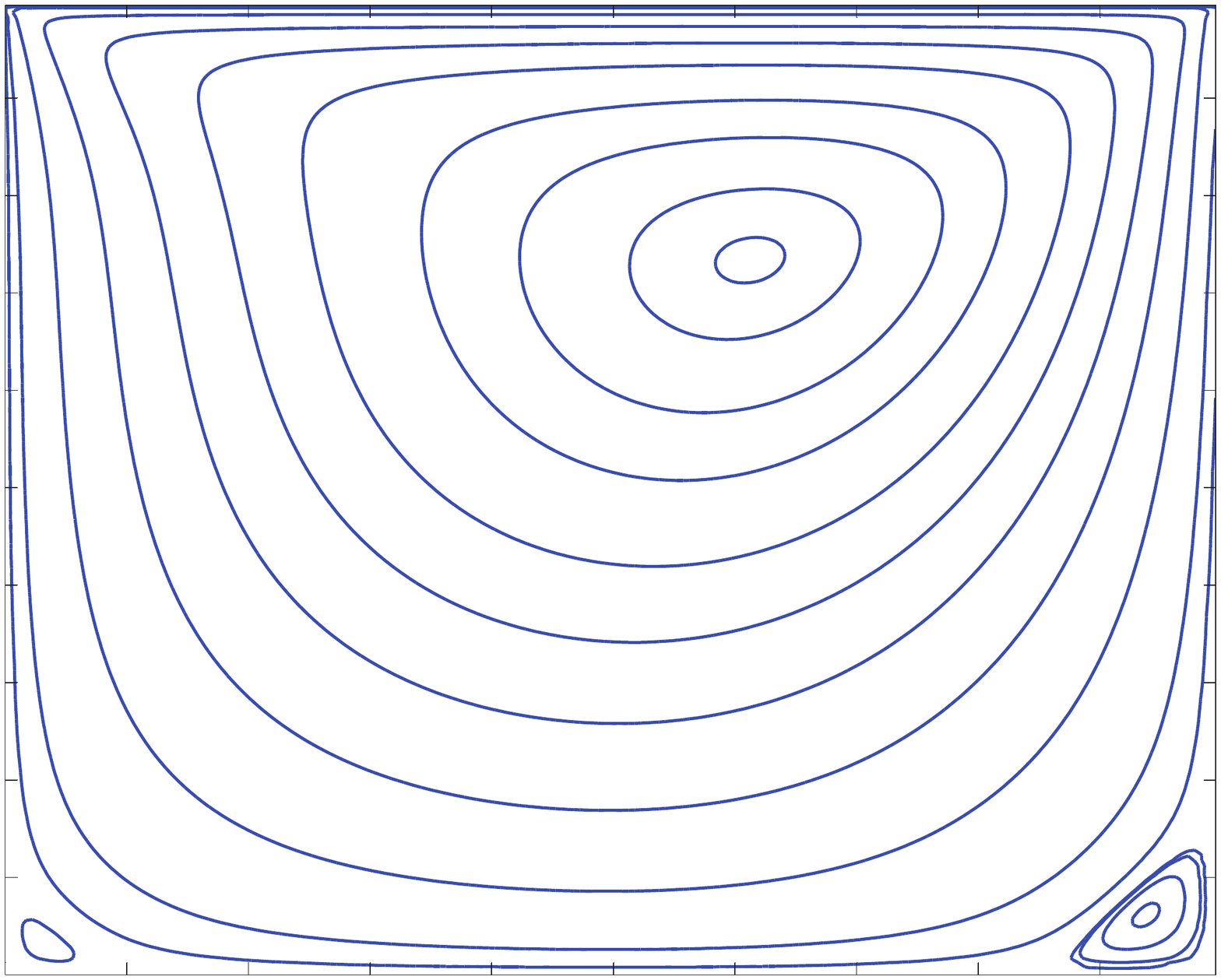}}
  \centering
  \subfigure[Cascaded MRT $Re=400$]{
  \vspace{20pt}
  \label{fig:subfig:c}
  \includegraphics[width = 2.0in, angle = 90]{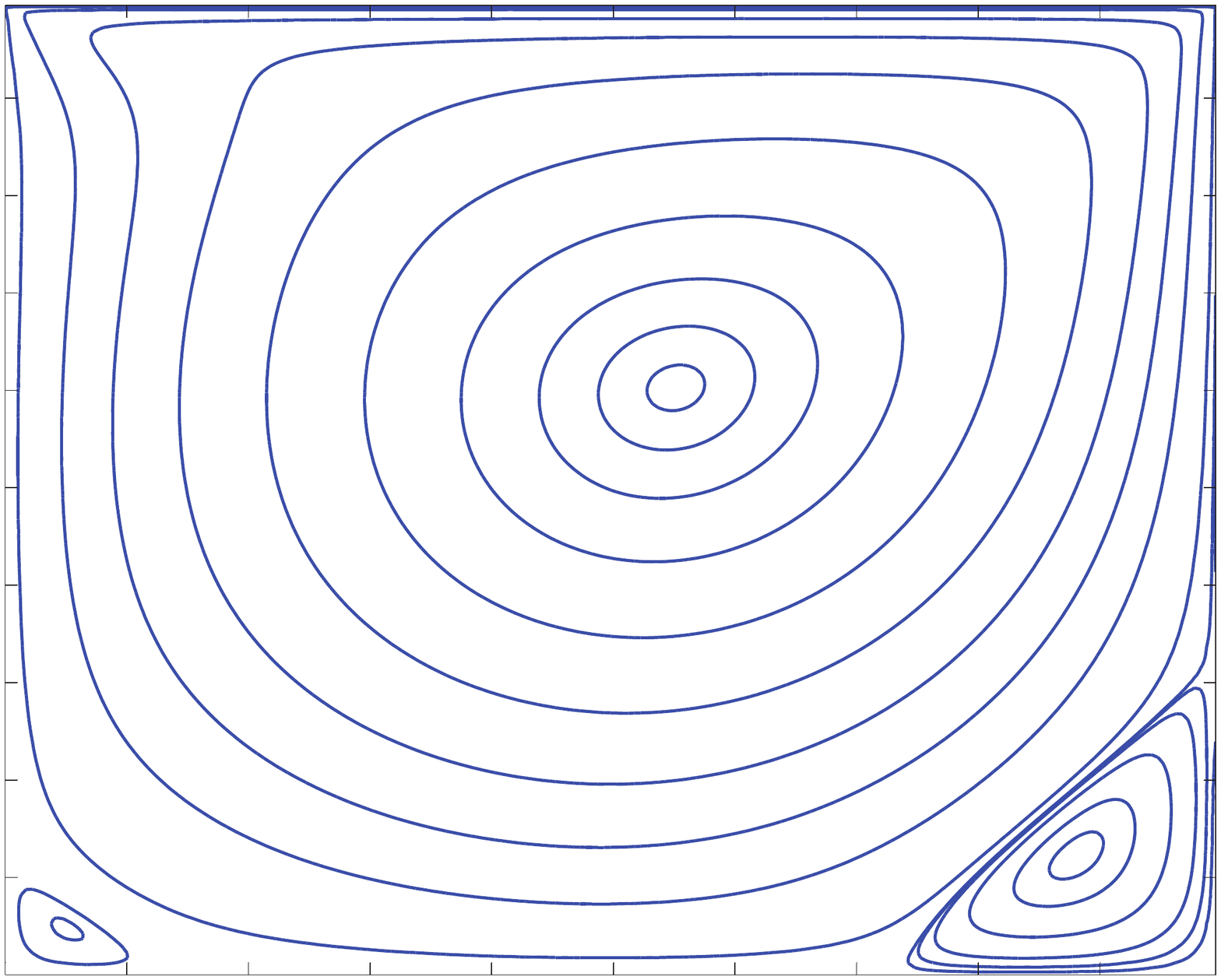}}
  \subfigure[Standard MRT $Re=400$]{
  \vspace{20pt}
  \label{fig:subfig:d}
  \includegraphics[width = 2.0in, angle = 90]{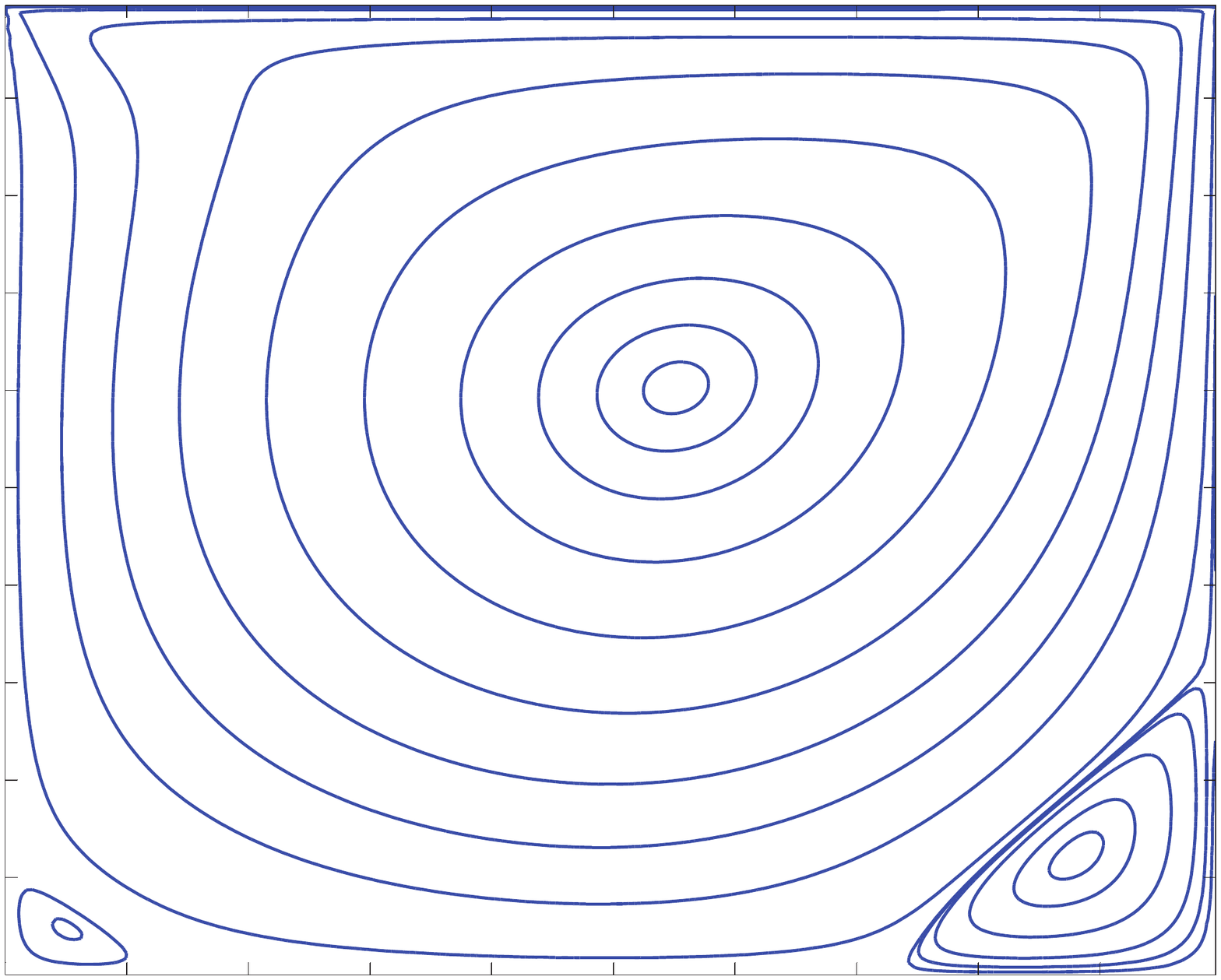}}
  \centering
  \subfigure[Cascaded MRT $Re=1000$]{
  \vspace{20pt}
  \label{fig:subfig:e}
  \includegraphics[width = 2.0in, angle = 90]{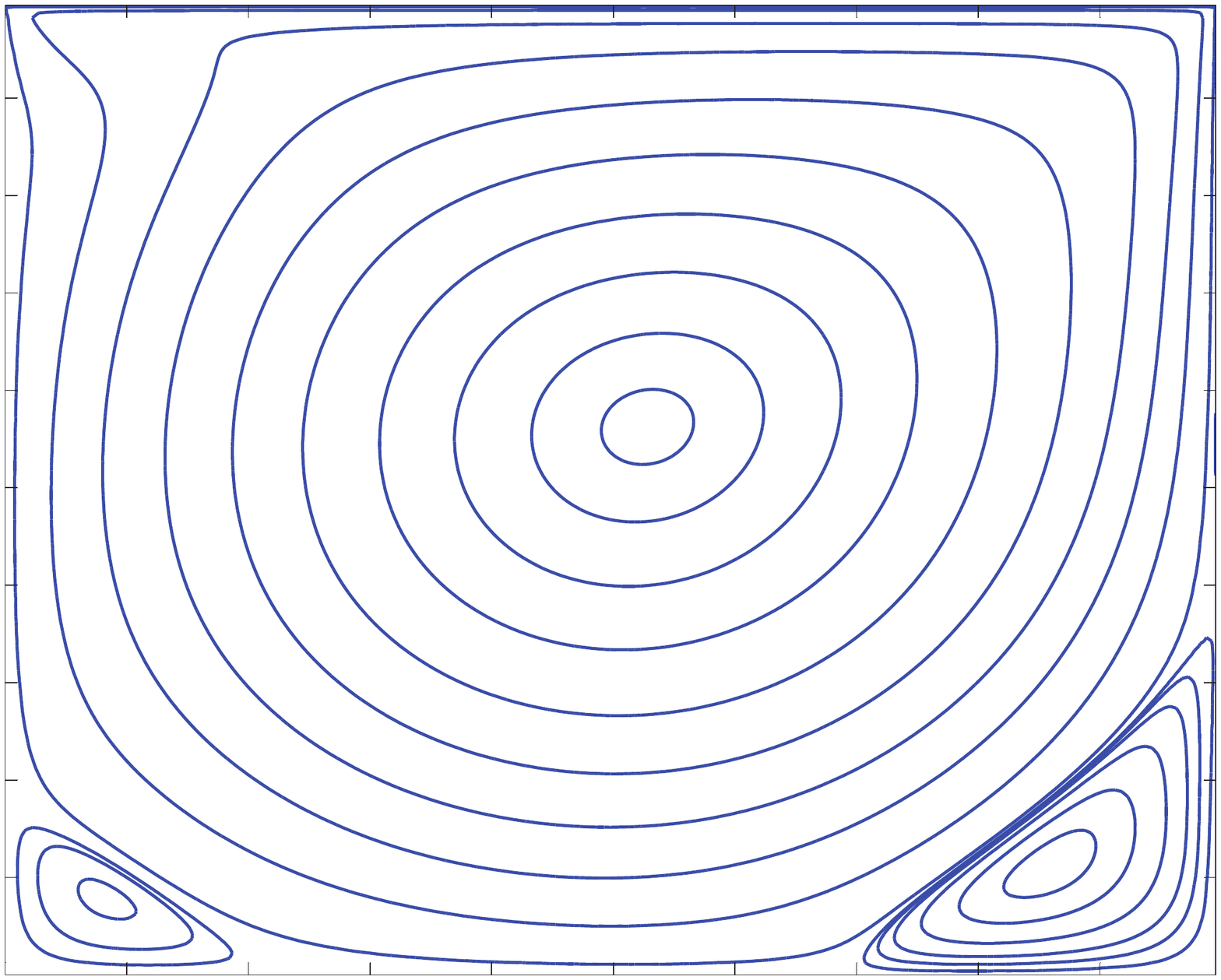}}
  \subfigure[Standard MRT $Re=1000$]{
  \vspace{20pt}
  \label{fig:subfig:f}
  \includegraphics[width = 2.0in, angle = 90]{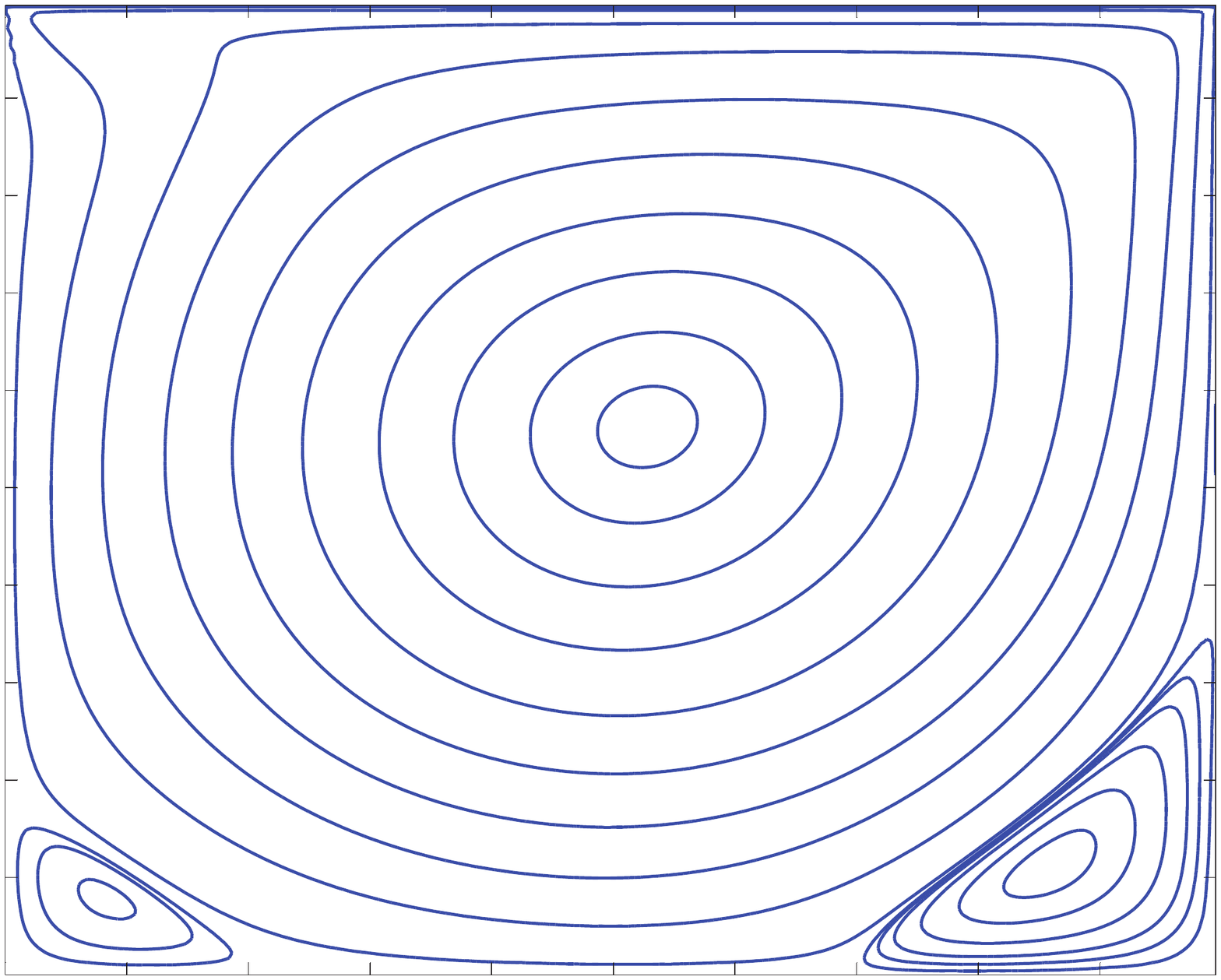}}
\end{figure}

\begin{figure}
\includegraphics[width = 140mm, angle = 0]{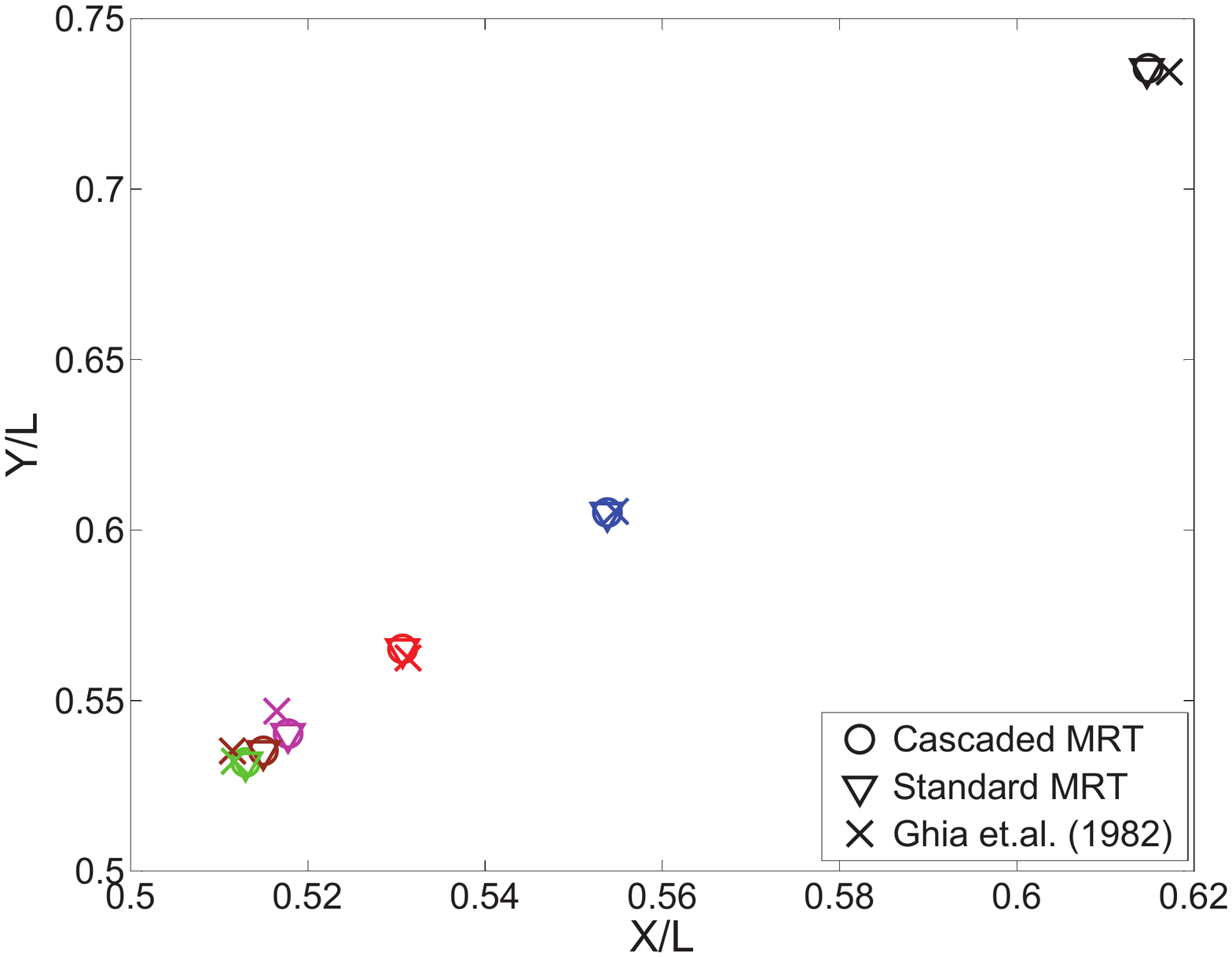}
\caption{\label{fig:location_primary}Comparison of the Cartesian coordinates of the location of the center of the primary vortex in a lid-driven cavity flow at different Reynolds numbers.}
\end{figure}

In order to provide a more quantitative perspective, Fig.~\ref{fig:location_primary} illustrates a comparison of the center of the primary vortex location in the cavity flow at different Reynolds numbers ($Re = 100, 400, 1000, 3200, 5000$, and $7500$) between the cascaded and standard MRT LBM
as well as the data by Ghia \emph{et al}~\cite{ghia82}.
From the earlier streamline plots, it can be observed that the location of the primary vortex moves towards the geometric center of the cavity as the Reynolds number increases. The computed results using the cascaded MRT LBM and the standard MRT LBM are in excellent agreement (within $0.014$ percent) with each other for all Reynolds numbers. In addition, they are both in very good agreement with the data by Ghia \emph{et al}~\cite{ghia82} to within $0.50$ percent for all Reynolds numbers. These quantitative results for the primary vortex
locations are enumerated in Table~\ref{tab:location_primary}.
\begin{table}
\centering
\renewcommand{\arraystretch}{1.1}
{\small
\label{tab:primaryvortex}
\caption{\label{tab:location_primary}Comparison of the location of the primary vortex in a lid-driven cavity flow at different Reynolds numbers.}
\begin{tabularx}{0.7\textwidth}{l|c|c|c}
\toprule[1px]
$Re$            &   Cascaded MRT LBM   &   Standard MRT LBM   &   Ghia \emph{et al}(1982)~\cite{ghia82} \\ \toprule[1px]
100             & $(0.61482, 0.73543)$ & $(0.61467, 0.73524)$ & $(0.61720, 0.73440)$ \\
400	            & $(0.55380, 0.60514)$ & $(0.55380, 0.60514)$ & $(0.55470, 0.60550)$ \\
1000            & $(0.53070, 0.56512)$ & $(0.53070, 0.56512)$ & $(0.53130, 0.56250)$ \\
3200            & $(0.51778, 0.54027)$ & $(0.51777, 0.54028)$ & $(0.51650, 0.54690)$ \\
5000            & $(0.51499, 0.53522)$ & $(0.51497, 0.53524)$ & $(0.51150, 0.53520)$ \\
7500            & $(0.51299, 0.53186)$ & $(0.51298, 0.53188)$ & $(0.51170, 0.53220)$ \\ \bottomrule[1px]
\end{tabularx}}
\end{table}
In addition, Table~\ref{tab:location_secondary} presents a comparison between the above two methods and the prior numerical data for the
location of secondary vortices at different Reynolds numbers. Again, both the cascaded MRT LBM and the standard MRT LBM are in
excellent quantitative agreement for the location of these detailed secondary vortical structures with the data by Ghia \emph{et al}~\cite{ghia82}.
\begin{table}
\centering
\renewcommand{\arraystretch}{1.1}
{\scriptsize
\label{tab:secondaryvortices}
\caption{\label{tab:location_secondary}Comparison of the location of various secondary vortices in a lid-driven cavity flow at differnt Reynolds numbers.}
\begin{tabularx}{0.67\textwidth}{c|c|c|c|c}
\toprule[1px]
\multicolumn{5}{c}{First Secondary Vortex} \\
\toprule[1px]
                     &       $Re$       &   Cascaded MRT LBM   &   Standard MRT LBM   & Ghia \emph{et al}(1982)~\cite{ghia82}\\ \hline
\multirow{6}{*}{Top} &  100             &        NA            &         NA           &       NA             \\
                     &  400	            &        NA            &         NA           &       NA             \\
                     &  1000            &        NA            &         NA           &       NA             \\
                     &  3200            & $(0.0547,0.8976)$    & $(0.0546, 0.8973)$   & $(0.0547, 0.8984)$   \\
                     &  5000            & $(0.0644, 0.9081)$   & $(0.0641, 0.9076)$   & $(0.0625, 0.9102)$   \\
                     &  7500            & $(0.0676, 0.9102)$   & $(0.0677, 0.9099)$   & $(0.0664, 0.9141)$   \\ \hline

\multirow{6}{*}{Bottom Left}&  100      & $(0.0387, 0.0387)$   & $(0.0373, 0.0373)$   & $(0.0313, 0.0391)  $ \\
                     &  400	            & $(0.0533, 0.0493)$   & $(0.0530, 0.0494)$   & $(0.0508, 0.0469)  $ \\
                     &  1000            & $(0.0842, 0.0791)  $ & $(0.0842, 0.0791)  $ & $(0.0859, 0.0781)  $ \\
                     &  3200            & $(0.0821, 0.1207)  $ & $(0.0821, 0.1207)  $ & $(0.0859, 0.1094)  $ \\
                     &  5000            & $(0.0740, 0.1378)  $ & $(0.0740, 0.1378)  $ & $(0.0703, 0.1367)  $ \\
                     &  7500            & $(0.0654, 0.1536)  $ & $(0.0654, 0.1536)  $ & $(0.0645, 0.1504)  $ \\ \hline

\multirow{6}{*}{Bottom Right}&  100     & $(0.9383, 0.0658)  $ & $(0.9386, 0.0654)  $ & $(0.9453, 0.0625)  $ \\
                     &  400	            & $(0.8833, 0.1243)  $ & $(0.883, 0.1243)   $ & $(0.8906, 0.1250)  $ \\
                     &  1000            & $(0.8631, 0.1128)  $ & $(0.8631, 0.1128)  $ & $(0.8594, 0.1094)  $ \\
                     &  3200            & $(0.8229, 0.0853)  $ & $(0.8229, 0.0852)  $ & $(0.8125, 0.0859)  $ \\
                     &  5000            & $(0.8037, 0.0739)  $ & $(0.8037, 0.0739)  $ & $(0.8086, 0.0742)  $ \\
                     &  7500            & $(0.7892, 0.0663)  $ & $(0.7893, 0.0663)  $ & $(0.7813, 0.0625)  $ \\ \toprule[1px]
    \multicolumn{5}{c}{Second Secondary Vortex} \\
\toprule[1px]
\multirow{6}{*}{Bottom Left}&  100      &        NA            &         NA           &       NA             \\
                     &  400	            &        NA            &         NA           &       NA             \\
                     &  1000            &        NA            &         NA           &       NA             \\
                     &  3200            & $(0.0075, 0.0075)  $ & $(0.0073, 0.0073)  $ & $(0.0078, 0.0078)  $ \\
                     &  5000            & $(0.0075, 0.0075)  $ & $(0.0074, 0.0074)  $ & $(0.0117, 0.0078)  $ \\
                     &  7500            & $(0.0125, 0.0125)  $ & $(0.0115, 0.0115)  $ & $(0.0117, 0.0117)  $ \\ \hline
\multirow{6}{*}{Bottom Right}&  100     &        NA            &         NA           &       NA             \\
                     &  400	            & $(0.9926, 0.0075)  $ &         NA           & $(0.9922, 0.0078)  $ \\
                     &  1000            & $(0.9923, 0.0076)  $ & $(0.9928, 0.0073)  $ & $(0.9922, 0.0078)  $ \\
                     &  3200            & $(0.9875, 0.0113)  $ & $(0.9885, 0.0115)  $ & $(0.9844, 0.0078)  $ \\
                     &  5000            & $(0.9775, 0.0200)  $ & $(0.9771, 0.0193)  $ & $(0.9805, 0.0195)  $ \\
                     &  7500            & $(0.9508, 0.0429)  $ & $(0.9509, 0.0429)  $ & $(0.9492, 0.0430)  $ \\ \toprule[1px]
\multicolumn{5}{c}{Third Secondary Vortex} \\
\toprule[1px]
\multirow{6}{*}{Bottom Right}&  100     &        NA            &         NA           &       NA             \\
                     &  400	            &        NA            &         NA           &       NA             \\
                     &  1000            &        NA            &         NA           &       NA             \\
                     &  3200            &        NA            &         NA           &       NA             \\
                     &  5000            &        NA            &         NA           &       NA             \\
                     &  7500            & $(0.9964, 0.0037)  $ &         NA           & $(0.9961, 0.0039)  $ \\ \bottomrule[1px]
\end{tabularx}}
\end{table}

\begin{figure}
  \centering
  \subfigure[Cascaded MRT $Re=100$]{
  \label{fig:subfig:a}
  \includegraphics[width = 3.1in, angle = 0]{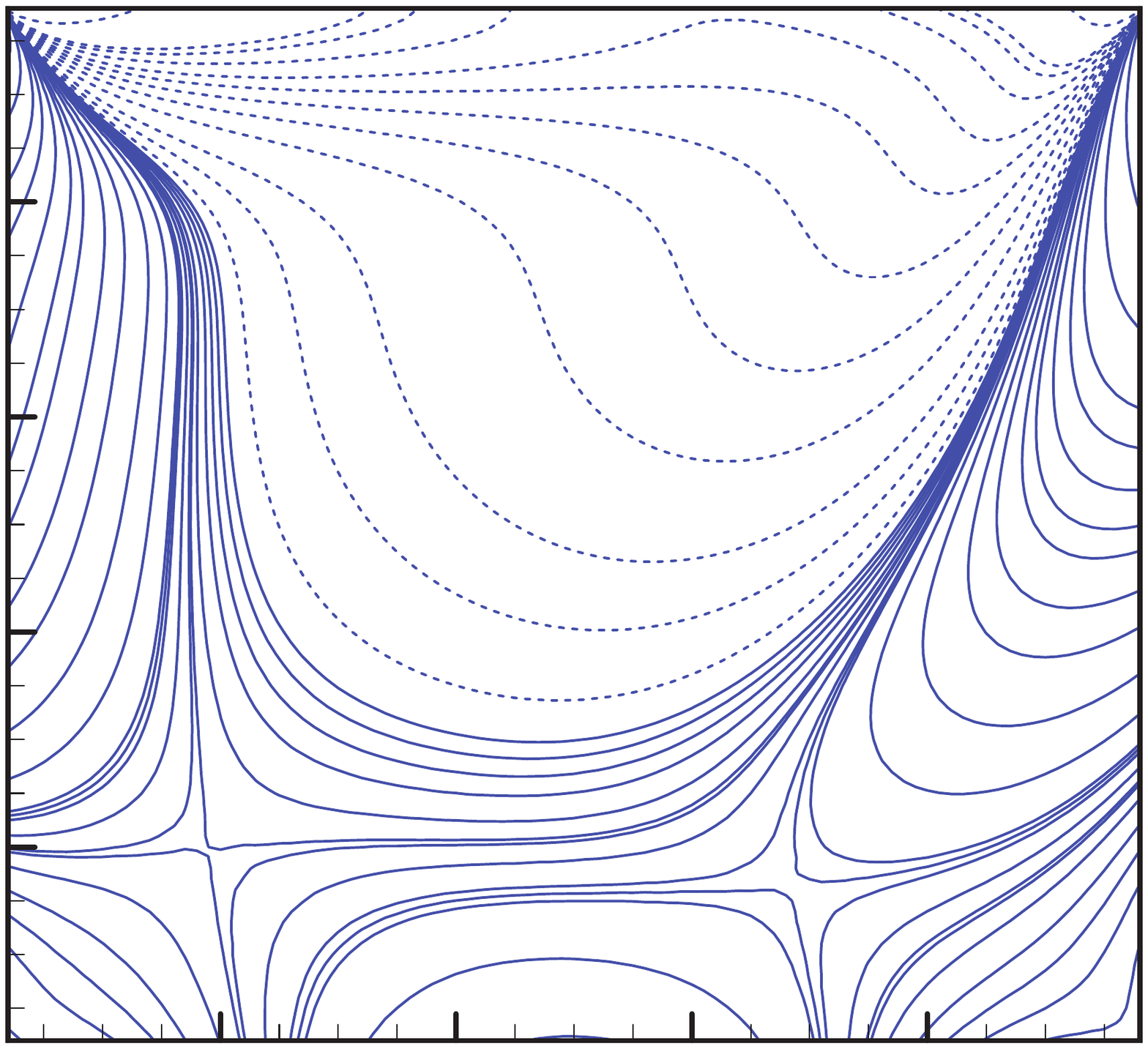}}
  \subfigure[Standard MRT $Re=100$]{
  \label{fig:subfig:b}
  \includegraphics[width = 3.1in, angle = 0]{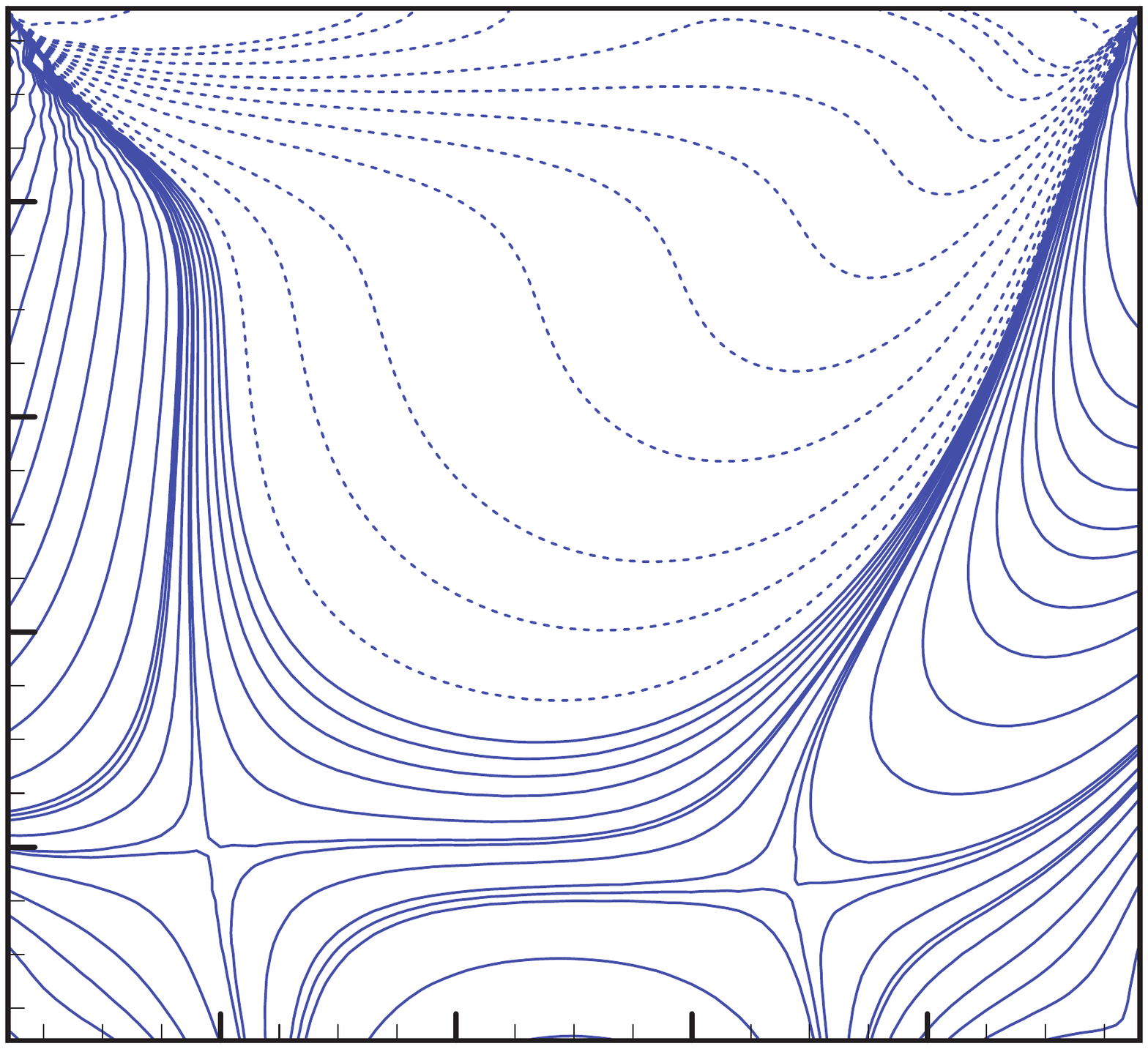}}
  \centering
  \subfigure[Cascaded MRT $Re=400$]{
  \label{fig:subfig:c}
  \includegraphics[width = 3.1in, angle = 0]{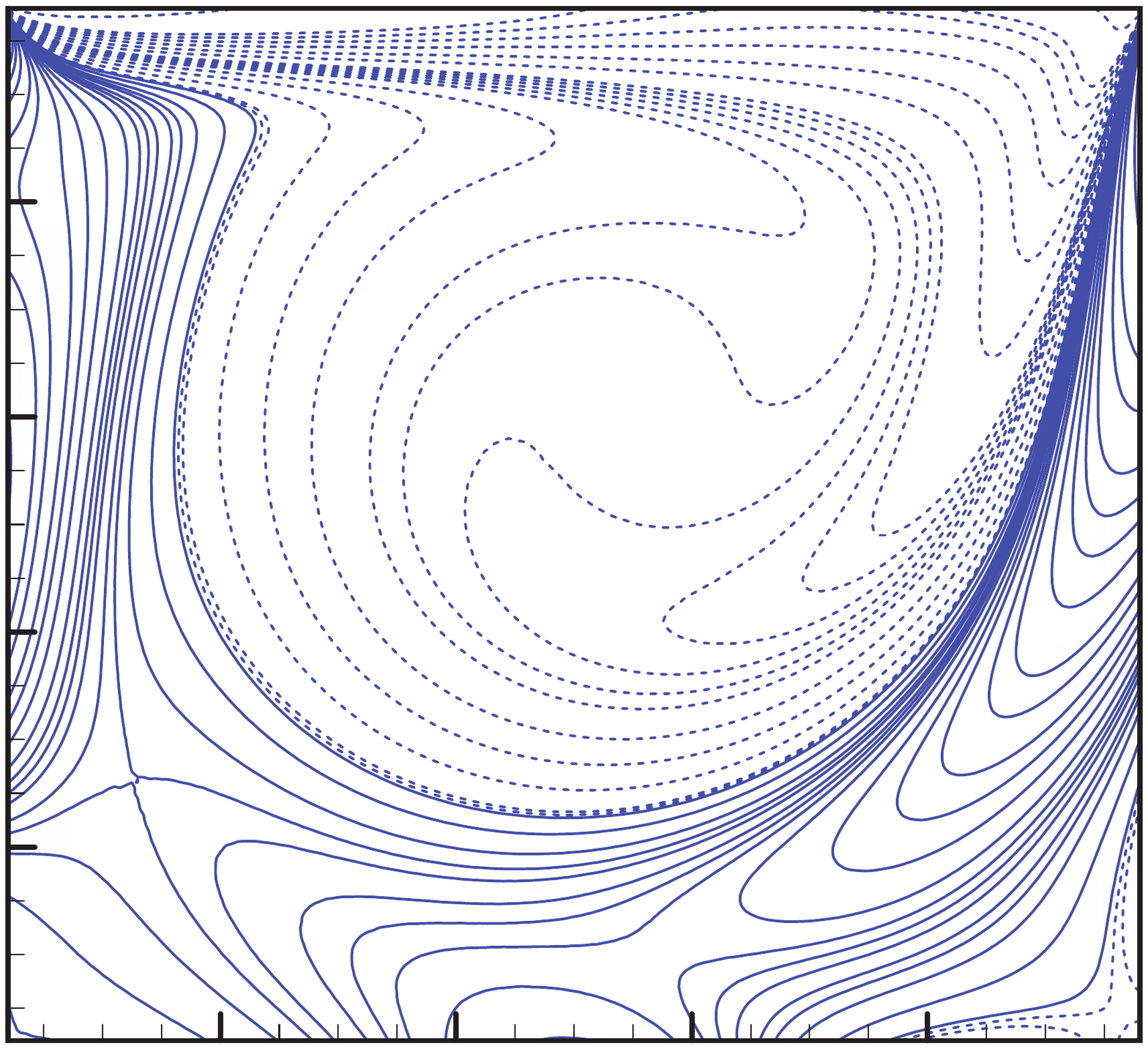}}
  \subfigure[Standard MRT $Re=400$]{
  \label{fig:subfig:d}
  \includegraphics[width = 3.1in, angle = 0]{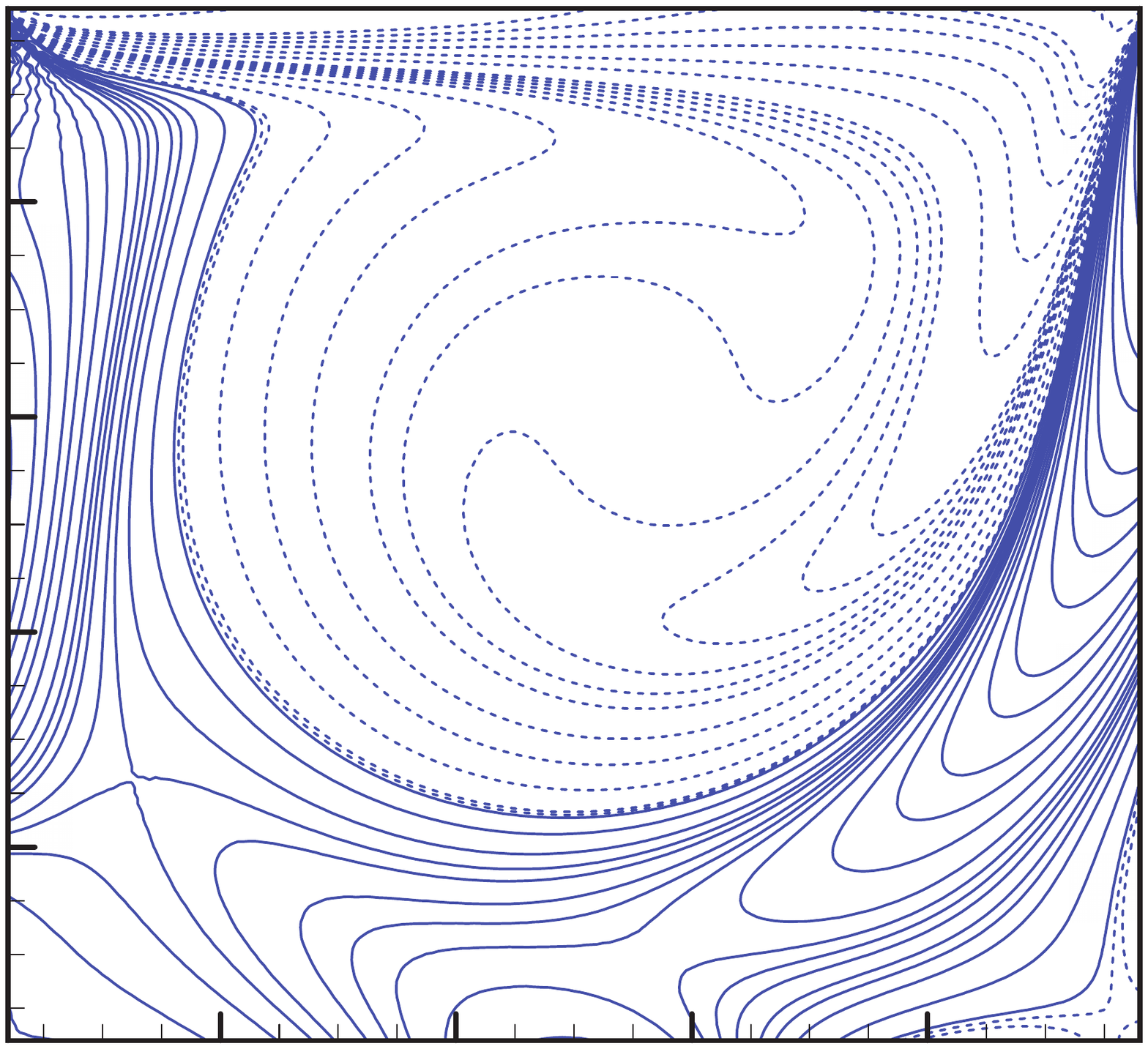}}
  \centering
  \subfigure[Cascaded MRT $Re=1000$]{
  \label{fig:subfig:e}
  \includegraphics[width = 3.1in, angle = 0]{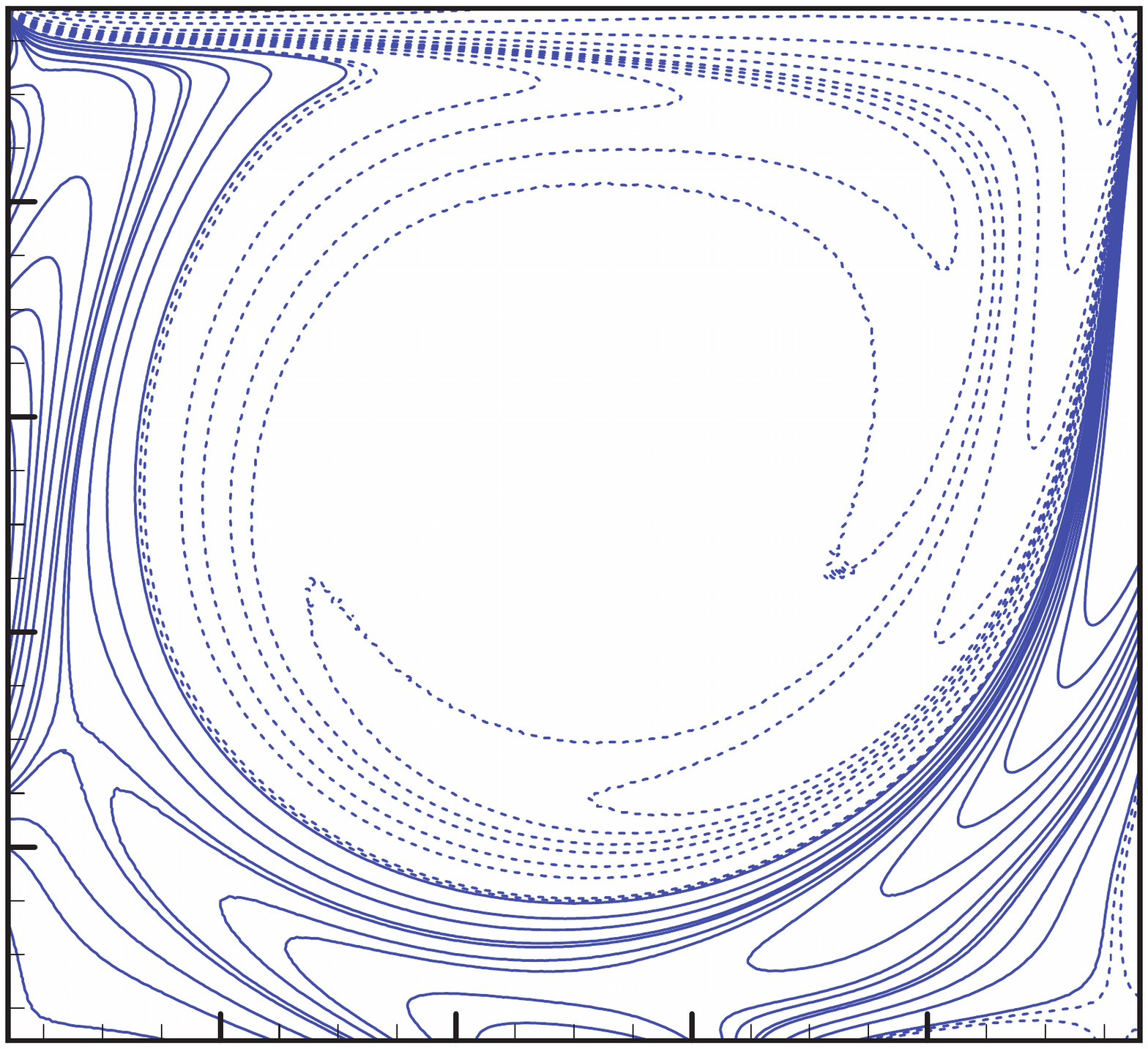}}
  \subfigure[Standard MRT $Re=1000$]{
  \label{fig:subfig:f}
  \includegraphics[width = 3.1in, angle = 0]{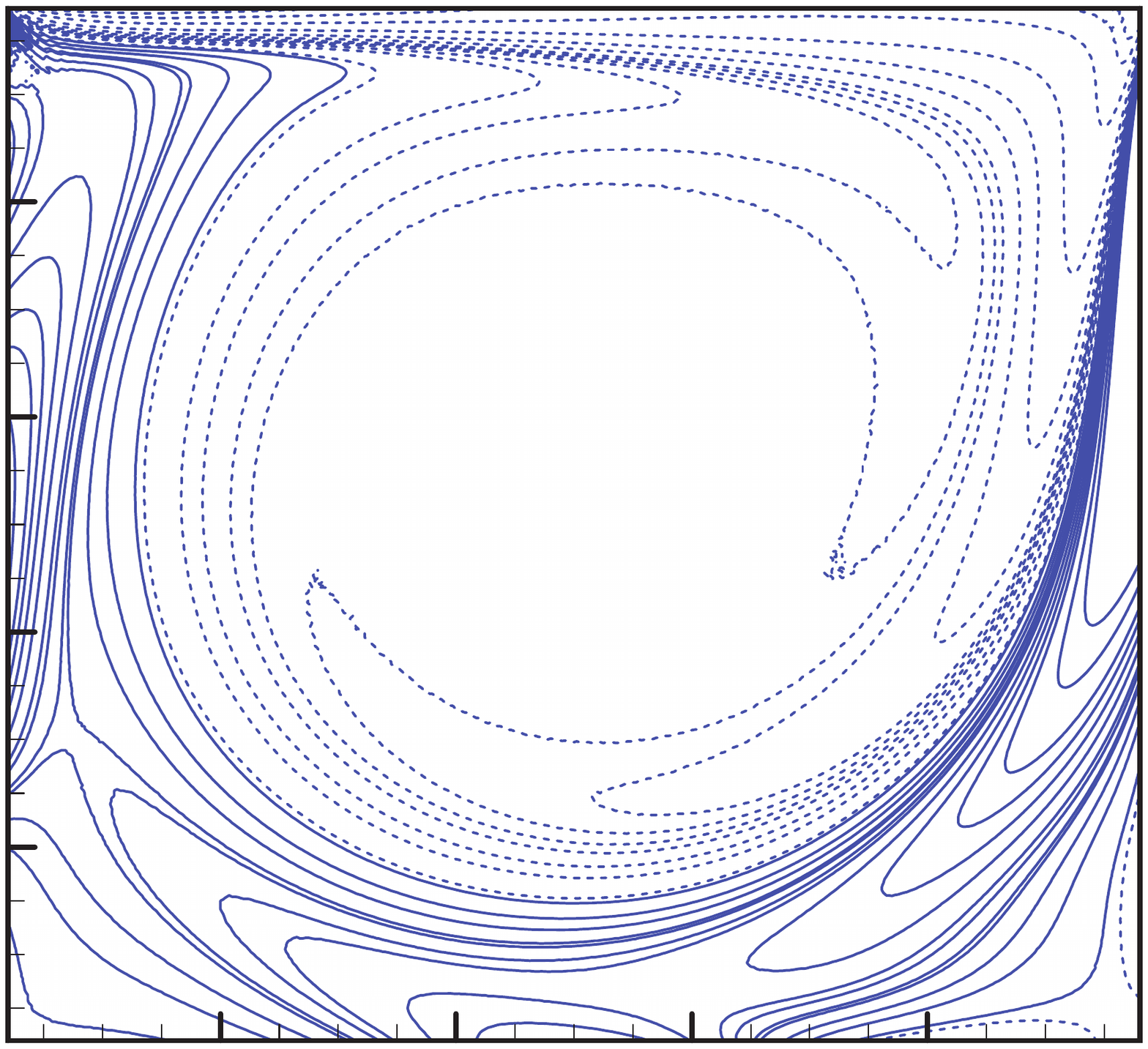}}
  \caption{Comparison of the vorticity contours in a 2D lid-driven cavity flow at different Reynolds numbers computed with cascaded (central moment) MRT LBM and standard (raw moment) MRT LBM: $Re=100,400$ and $1000$.}\label{fig:vorticity}
\end{figure}

Another useful global characteristic for comparison is the vorticity contours in the cavity at different Reynolds numbers. Figure~\ref{fig:vorticity} shows the vorticity contours computed using both the standard MRT LBM and the cascaded MRT LBM at three different Reynolds numbers ($Re=100, 400$, and $1000$). As Reynolds number increases, the vorticity contours become denser and denser approaching the boundary walls. Overall, the vorticity distribution is found to be very similar using both the methods for all the Reynolds numbers considered thus corraborating the earlier results.

As discussed earlier, one of the useful features of kinetic schemes such as the cascaded MRT LBM is that the components of the strain rate
tensor can be obtained locally from the components of the non-equilibrium moments of the distribution function (see Eqs.~(\ref{eq:strainrate1})-(\ref{eq:strainrate3})). The cavity flow being a shear driven problem generally has all the components
of the strain rate tensor non-zero, and whose magnitudes can dramatically change with the Reynolds number. Hence, this problem provides
a good test for the evalution of the accuracy of the computation of strain rate tensor by kinetic theory considerations, i.e. using
non-equilibrium moments (Eqs.~(\ref{eq:strainrate1})-(\ref{eq:strainrate3})). For the sake of comparison, we will make use of the standard second-order central differencing of the velocity field to obtain the usual direct estimation of the strain rate tensor components. In this regard, flow at two different Reynolds numbers are considered ($Re=100$ and $1000$) and the components of the strain rate tensor are obtained at five different locations within the cavity using the above two methods, which are enumerated in Table~\ref{tab:strainrate}. As the Reynolds number is increased from $100$ to $1000$, the magnitudes of the strain rate tensor change significantly, which are quite well captured by the kinetic approach. Indeed, remarkably the local computation using the non-equilibrium moments are in very good agreement with the finite-difference estimation at various locations in the cavity for both the Reynolds numbers, with the maximum difference within 2 percent. This further demonstrates the numerical fidelity of the approach. In particular, such non-equilibrium moments based approach for the strain rate components can be used in the subgrid scale models for large eddy simulation of turbulent flows using the cascaded MRT LBM.
\begin{table}
\centering
\renewcommand{\arraystretch}{1.1}
{\scriptsize
\label{tab:strainrate}
\caption{\label{tab:strainrate}Comparison of the components of the strain rate tensor computed using the local non-equilibrium moments (Eqs.~(\ref{eq:strainrate1})-(\ref{eq:strainrate3})) and the finite-differencing (second-order central) of the velocity field
with the cascaded MRT LBM at five different locations within the cavity for two different Reynolds numbers ($Re=100$ and $1000$).}
\begin{tabularx}{0.62\textwidth}{c|c|c|c|c|c}
\toprule[1px]
\multicolumn{6}{c}{$Re=100$} \\
\toprule[1px]
                     &   &    Location                  &    Non-eqm. Moments   &   Finite Difference  &  Difference \\ \hline
\multirow{5}{*}{\large $\frac{\partial v}{\partial y} $}
                     & A & $(\frac{L}{4},\frac{L}{2}) $ & $2.416\times 10^{-4}$ & $2.415\times 10^{-4}$ &  0.044$\%$    \\
                     & B & $(\frac{L}{2},\frac{L}{4}) $ & $1.711\times 10^{-5}$ & $1.721\times 10^{-5}$ &  0.613$\%$    \\
                     & C & $(\frac{L}{2},\frac{L}{2}) $ & $1.850\times 10^{-4}$ & $1.848\times 10^{-4}$ &  0.102$\%$     \\
                     & D & $(\frac{L}{2},\frac{3L}{4})$ &$-9.020\times 10^{-5}$ &$-9.025\times 10^{-5}$ &  0.057$\%$    \\
                     & E & $(\frac{3L}{4},\frac{L}{2})$ &$-3.541\times 10^{-4}$ &$-3.536\times 10^{-4}$ &  0.125$\%$    \\ \bottomrule[1px]
\multicolumn{6}{c}{$Re=100$} \\
\toprule[1px]
\multirow{5}{*}{\large $\frac{\partial u}{\partial y}+\frac{\partial v}{\partial x} $}
                     & A & $(\frac{L}{4},\frac{L}{2}) $ & $3.516\times 10^{-5}$ & $3.526\times 10^{-5}$ &  0.300$\%$    \\
                     & B & $(\frac{L}{2},\frac{L}{4}) $ & $-4.344\times 10^{-4}$& $-4.342\times 10^{-4}$&  0.045$\%$    \\
                     & C & $(\frac{L}{2},\frac{L}{2}) $ & $-3.368\times 10^{-4}$& $-3.363\times 10^{-4}$&  0.135$\%$    \\
                     & D & $(\frac{L}{2},\frac{3L}{4})$ & $4.599\times 10^{-4}$ & $4.603\times 10^{-4}$ &  0.093$\%$    \\
                     & E & $(\frac{3L}{4},\frac{L}{2})$ &$-5.290\times 10^{-4}$ &$-5.280\times 10^{-4}$ &  0.198$\%$    \\ \bottomrule[1px]
\multicolumn{6}{c}{$Re=1000$} \\
\toprule[1px]
\multirow{5}{*}{\large $\frac{\partial v}{\partial y} $}
                     & A & $(\frac{L}{4},\frac{L}{2}) $ & $4.220\times 10^{-5}$ & $4.217\times 10^{-5}$ &  0.050$\%$    \\
                     & B & $(\frac{L}{2},\frac{L}{4}) $ & $2.196\times 10^{-5}$ & $2.209\times 10^{-5}$ &  0.596$\%$    \\
                     & C & $(\frac{L}{2},\frac{L}{2}) $ & $5.017\times 10^{-5}$ & $5.017\times 10^{-5}$ &  0.008$\%$    \\
                     & D & $(\frac{L}{2},\frac{3L}{4})$ & $2.370\times 10^{-5}$ & $2.372\times 10^{-5}$ &  0.073$\%$    \\
                     & E & $(\frac{3L}{4},\frac{L}{2})$ & $6.397\times 10^{-5}$ & $6.446\times 10^{-5}$ &  0.750$\%$    \\ \bottomrule[1px]
\multicolumn{6}{c}{$Re=1000$} \\
\toprule[1px]
\multirow{5}{*}{\large $\frac{\partial u}{\partial y}+\frac{\partial v}{\partial x} $}
                     & A & $(\frac{L}{4},\frac{L}{2}) $ &$-1.344\times 10^{-4}$ &$-1.334\times 10^{-4}$ &  0.782$\%$    \\
                     & B & $(\frac{L}{2},\frac{L}{4}) $ & $8.137\times 10^{-5}$ & $7.984\times 10^{-5}$ &  1.914$\%$    \\
                     & C & $(\frac{L}{2},\frac{L}{2}) $ &$-3.980\times 10^{-5}$ &$-3.979\times 10^{-5}$ &  0.021$\%$    \\
                     & D & $(\frac{L}{2},\frac{3L}{4})$ & $2.291\times 10^{-4}$ & $2.289\times 10^{-4}$ &  0.049$\%$    \\
                     & E & $(\frac{3L}{4},\frac{L}{2})$ &$-1.688\times 10^{-4}$ &$-1.699\times 10^{-4}$ &  0.626$\%$    \\ \bottomrule[1px]
\end{tabularx}}
\end{table}

\section{\label{sec:stabilitystudy}Numerical Stability Studies on the Benchmark Problems}
We will now discuss the results of numerical stability studies. Among the three benchmark problems discussed earlier, the lid-driven cavity flow
presents the most stringent test since it is a fully 2D problem with boundaries containing singularity and the flow is shear driven. In fact, such
a cavity flow problem was considered in detail to determine stability regimes of the SRT and the standard MRT collision models in a recent work~\cite{luo11}. Earlier, its three-dimensional counterpart was also considered from this viewpoint~\cite{premnath09a}. These studies have demonstrated the superiority of the use of multiple relaxation times in providing controlled additional numerical dissipation to enhance numerical stability on either coarser grids or at high Reynolds numbers when compared with the single relaxation time models. Hence, it is appropriate to
consider the 2D lid-driven cavity flow to establish the stability regime of the cascaded MRT LBM in the context of other collision models. We now make a direct comparison of the maximum threshold Reynolds number for numerical stability of the SRT LBM, the standard MRT LBM and the cascaded MRT LBM for this problem. With the cascaded MRT LBM, the relaxation parameters $\omega_4=\omega_5=1/\tau$ are selected based on the specified kinematic
viscosity, while the rest of relaxation parameters are set to unity for simplicity. For each approach, for a given grid resolution, the lid velocity was fixed and the relaxation time $\tau$ was decreased gradually until the computation became unstable.

\begin{figure}
\includegraphics[width = 140mm, angle = 0]{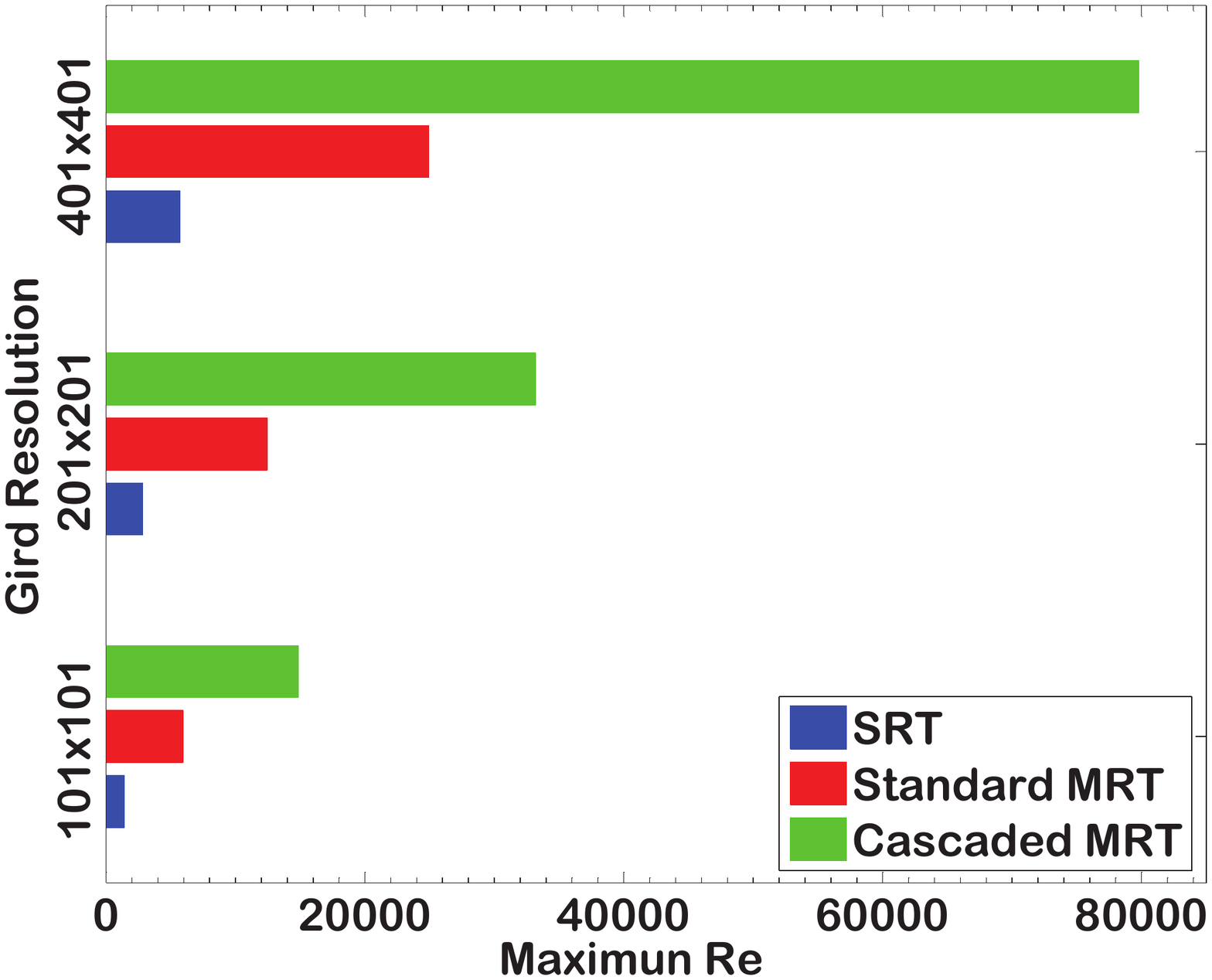}
\vspace{-5pt}
\caption{\label{fig:stability/standard_standard_stability}Comparison of the maximum Reynolds number for numerical stability of different methods for simulation of the lid-driven cavity flow.}
\end{figure}

\begin{figure}
\includegraphics[width = 140mm, angle = 0]{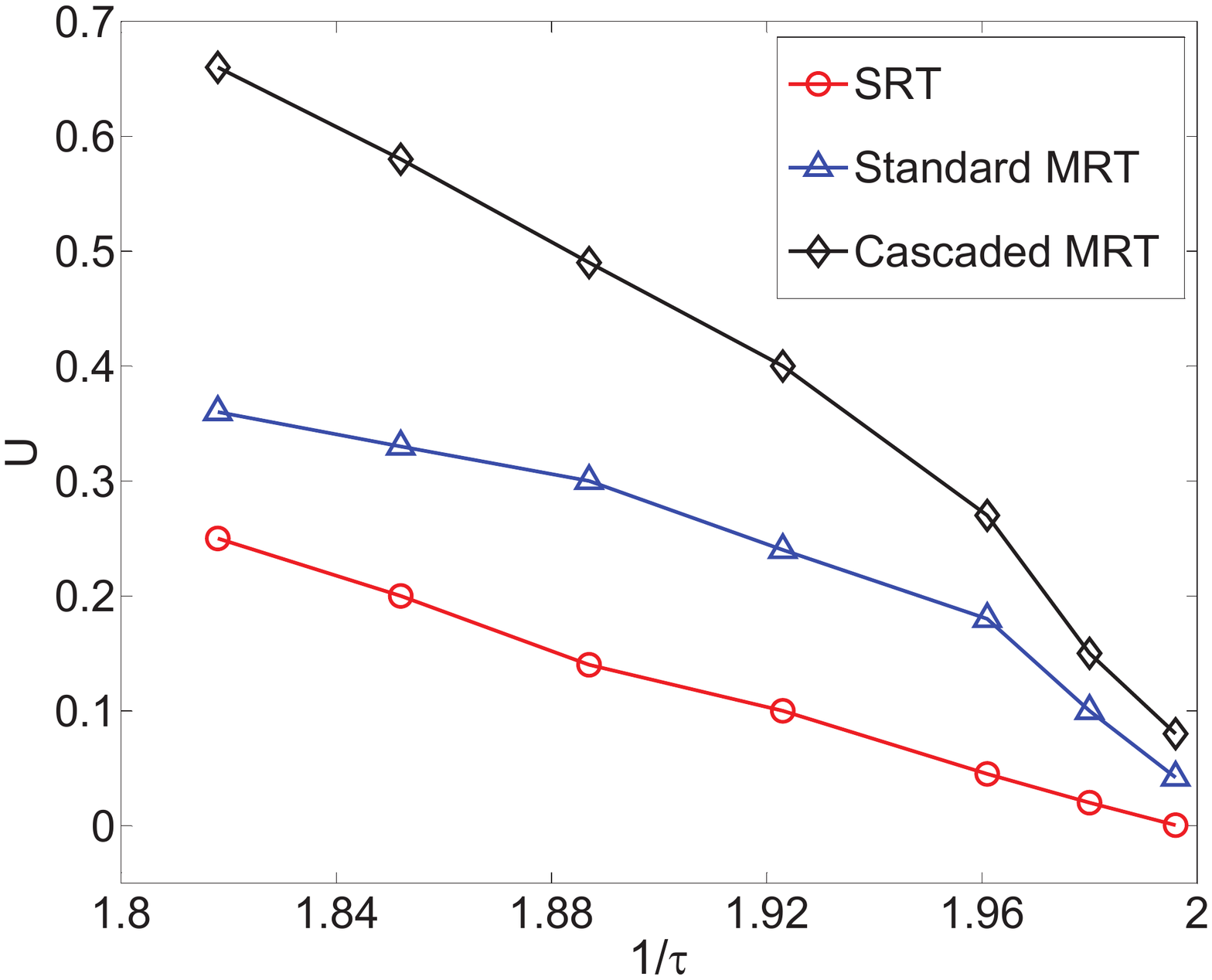}
\vspace{-5pt}
\caption{\label{fig:stability/alternative_standard_stability}Alternative stability test to determine the maximum threshold lid velocity for different methods for a chosen coarse resolution ($26\times 26$).}
\end{figure}

Figure~\ref{fig:stability/standard_standard_stability} shows the maximum Reynolds number ($Re=U_0L/\nu$) that could be attained for each method before the computations became unstable, i.e. when the relative global error increases rapidly or becomes exponentially large as the simulation progresses. Results are provided for different grid resolutions for these three approaches. It is clear that the cascaded MRT computations can reach Reynolds numbers that are about $2$ or $3$ times higher than that of the standard MRT approach and the standard MRT computations can reach Reynolds numbers that are $3$ or $4$ times higher than that of the SRT approach. The latter results are consistent with prior findings~\cite{luo11,premnath09a}.  Relaxation of different \emph{central} moments at different rates provides a controlled additional numerical dissipation to maintain numerical
stability. That is, maintaining frame invariance in conjunction with the use of multiple relaxation times further promotes the stability of the method. It may be noted that stabilization of certain classical methods have been achieved by constructing discretization operators that enforce Galilean
invariance~\cite{scovazzi07a,scovazzi07b,hughes10}. Hence, it may be expected that explicitly incorporating an invariance property could aid with
other standard mechanisms of stabilization of the LBM. As carried out in~\cite{luo11}, we also perform an alternate stability test with the three approaches on a chosen coarse grid for this problem. In this test, the grid resolution is fixed at a relatively coarse resolution of $26\times 26$, and then viscosity $\nu$ (or equivalently $\tau$) is also set for all the three approaches. We then intend to find the maximum lid velocity which can maintain the stability of computations for $50,000$ time steps~\cite{luo11}. Figure~\ref{fig:stability/alternative_standard_stability} shows how the three methods behave for this test. It is seen that the parameter regime or the maximum lid velocity for stability is considerably higher with the cascaded MRT LBM when compared with the other approaches. This further establishes the merits of the use of multiple relaxation times for central moment relaxation. Often, the stability of the CFD methods are characterized in terms of the grid or cell Reynolds number given by $Re_c=U_0\Delta x/\nu$ (e.g.~\cite{wesseling00}). Thus, we also present the maximum cell Reynolds number for stability of the three approaches for this problem in Table~\ref{tab:cellReynoldsnumber}, which demonstrates the advantages of the cascaded MRT LBM.
\begin{table}
\vspace{0pt}
\centering
\renewcommand{\arraystretch}{1.1}
{\small
\label{tab:cellReynoldsnumber}
\caption{\label{tab:cellReynoldsnumber}Comparison of the maximum cell Reynolds number ($Re_c=U_0\Delta x/\nu$) for numerical stability of different methods for simulation of the lid-driven cavity flow problem.}
\begin{tabularx}{0.72\textwidth}{l|c|c|c}
\toprule[1px]
Grid Resolution   &    SRT LBM         &  Standard MRT LBM    &  Cascaded MRT LBM \\ \toprule[1px]
$101\times101$    & $     14.14      $ & $       59.40      $ & $     148.50    $ \\
$201\times201$    & $     14.21      $ & $       62.18      $ & $     165.83    $ \\
$401\times401$    & $     14.25      $ & $       62.34      $ & $     199.50    $ \\ \bottomrule[1px]
\end{tabularx}}
\end{table}

Another important aspect is the computational cost. As shown previously, the cascaded MRT approach can be more stable with similar accuracy compared with the standard MRT for the lid-driven cavity flow. But if it is much more expensive for numerical computations than the standard MRT, its advantages will not be very useful. In this regard, we fully exploit all the optimization strategies that could be used with a moment approach, such as those
specified in~\cite{dhumieres02} for the cascaded MRT LBM. It is found that for the 2D lid-driven cavity flow problem, the cascaded MRT LBM takes about
$11.6\%$ longer than the standard MRT LBM, which is acceptable in view of the significant advantages in terms of numerical stability. It should be pointed out that these results pertain only to 2D problems. Additional work is required in three-dimensions to optimize the computational cost of
the cascaded MRT LBM and also to optimize its relaxation parameters by means of a linear Fourier analysis.

\section{\label{sec:conclusion}Summary and Conclusions}
Galilean invariance is one of the main physical attributes in the description of the fluid motion. This is naturally achieved by considering dynamical
changes in terms of central moments in kinetic schemes, as was done in the recently introduced cascaded LBM. Enforcing frame invariance is generally expected to have a positive influence on numerical stability as seen in some recent work with other classical schemes. The use of multiple relaxation times (MRT) in the central moment or cascaded LBM brings in the various flexibility associated with the standard MRT LBM based on raw moments. In particular, the relaxation of different central moments at different rates introduces additional dissipation as in the raw moment based approach, which can lead to enhanced stability.

In this paper, we discussed our results from systematic numerical studies on grid convergence, accuracy, and stability of the cascaded MRT LBM. We have chosen three commonly used 2D benchmark problems including the Poiseuille flow, the decaying Taylor-Green vortex flow, and the lid-driven cavity flow. In the grid convergence tests, the cascaded MRT approach has been found to be second order accurate under diffusive scaling for all the benchmark problems considered. These results are shown to hold not only for the velocity field, but also for the components of the strain rate tensors. Furthermore, comparisons of the numerical accuracy of the cascaded MRT LBM were made with other collision models and also with prior analytical
or numerical results based on the solution of the Navier-Stokes equations. These demonstrated that the cascaded MRT LBM is in excellent agreement
with the prior results for all the canonical problems considered. In particular, the detailed flow structures for the more complex lid-driven
cavity flow predicted by the cascaded MRT LBM are in very good quantitative agreement with the standard MRT LBM. In addition, the utility and
the accuracy of the use of non-equilibrium moments with the cascaded MRT LBM for the computation of the components of the strain rate tensor
is demonstrated. Finally, stability tests on a 2D lid-driven cavity flow problem was carried out, which showed substantial improvements in numerical stability of the cascaded MRT LBM, with higher threshold Reynolds numbers, when compared to other models. With the use of proper optimization strategies, the 2D cascaded MRT LBM was found to be only about $10\%$ to $20\%$ more expensive when compared to the standard MRT LBM in terms of computational time.

Future work could include further development of more optimized formulations of the three-dimensional cascaded LBM based on central moments with a view to maintain computational efficiency and their applications to unsteady multiscale problems such as turbulence. Optimization of the relaxation parameters by a linear Fourier analysis to introduce adequate additional dissipation for enhanced numerical stability while maintaining necessary physics with this approach is also desired.

\bibliography{paper_bib}

\begin{thebibliography}{39}
\expandafter\ifx\csname natexlab\endcsname\relax\def\natexlab#1{#1}\fi
\expandafter\ifx\csname bibnamefont\endcsname\relax
  \def\bibnamefont#1{#1}\fi
\expandafter\ifx\csname bibfnamefont\endcsname\relax
  \def\bibfnamefont#1{#1}\fi
\expandafter\ifx\csname citenamefont\endcsname\relax
  \def\citenamefont#1{#1}\fi
\expandafter\ifx\csname url\endcsname\relax
  \def\url#1{\texttt{#1}}\fi
\expandafter\ifx\csname urlprefix\endcsname\relax\def\urlprefix{URL }\fi
\providecommand{\bibinfo}[2]{#2}
\providecommand{\eprint}[2][]{\url{#2}}

\bibitem[{\citenamefont{Chen and Doolen}(1998)}]{chen98}
\bibinfo{author}{\bibfnamefont{S.}~\bibnamefont{Chen}} \bibnamefont{and}
  \bibinfo{author}{\bibfnamefont{G.}~\bibnamefont{Doolen}},
  \bibinfo{journal}{Ann.\ Rev.\ Fluid Mech.} \textbf{\bibinfo{volume}{8}},
  \bibinfo{pages}{2527} (\bibinfo{year}{1998}).

\bibitem[{\citenamefont{Succi}(2001)}]{succi01}
\bibinfo{author}{\bibfnamefont{S.}~\bibnamefont{Succi}},
  \emph{\bibinfo{title}{The Lattice Boltzmann Equation for Fluid Dynamics and
  Beyond}} (\bibinfo{publisher}{Clarendon Press, Oxford},
  \bibinfo{year}{2001}).

\bibitem[{\citenamefont{Luo et~al.}(2010)\citenamefont{Luo, Krafczyk, and
  Shyy}}]{luo10}
\bibinfo{author}{\bibfnamefont{L.-S.} \bibnamefont{Luo}},
  \bibinfo{author}{\bibfnamefont{M.}~\bibnamefont{Krafczyk}}, \bibnamefont{and}
  \bibinfo{author}{\bibfnamefont{W.}~\bibnamefont{Shyy}},
  \emph{\bibinfo{title}{Lattice Boltzmann Method for Computational Fluid
  Dynamics}} (\bibinfo{publisher}{Encyclopedia of Aerospace Engineering,
  Wiley}, \bibinfo{year}{2010}).

\bibitem[{\citenamefont{Ubertini et~al.}(2010)\citenamefont{Ubertini, Asinari,
  and Succi}}]{ubertini10}
\bibinfo{author}{\bibfnamefont{S.}~\bibnamefont{Ubertini}},
  \bibinfo{author}{\bibfnamefont{P.}~\bibnamefont{Asinari}}, \bibnamefont{and}
  \bibinfo{author}{\bibfnamefont{S.}~\bibnamefont{Succi}},
  \bibinfo{journal}{Phys.\ Rev.\ E} \textbf{\bibinfo{volume}{81}},
  \bibinfo{pages}{016311} (\bibinfo{year}{2010}).

\bibitem[{\citenamefont{Chen et~al.}(1992)\citenamefont{Chen, Chen, and
  Matthaeus}}]{chen92}
\bibinfo{author}{\bibfnamefont{H.}~\bibnamefont{Chen}},
  \bibinfo{author}{\bibfnamefont{S.}~\bibnamefont{Chen}}, \bibnamefont{and}
  \bibinfo{author}{\bibfnamefont{W.}~\bibnamefont{Matthaeus}},
  \bibinfo{journal}{Phys.\ Rev.~A} \textbf{\bibinfo{volume}{45}},
  \bibinfo{pages}{5339} (\bibinfo{year}{1992}).

\bibitem[{\citenamefont{Qian et~al.}(1992)\citenamefont{Qian, d'Humi\`eres, and
  Lallemand}}]{qian92}
\bibinfo{author}{\bibfnamefont{Y.}~\bibnamefont{Qian}},
  \bibinfo{author}{\bibfnamefont{D.}~\bibnamefont{d'Humi\`eres}},
  \bibnamefont{and}
  \bibinfo{author}{\bibfnamefont{P.}~\bibnamefont{Lallemand}},
  \bibinfo{journal}{Europhys.\ Lett.} \textbf{\bibinfo{volume}{17}},
  \bibinfo{pages}{479} (\bibinfo{year}{1992}).

\bibitem[{\citenamefont{Bhatnagar et~al.}(1954)\citenamefont{Bhatnagar, Gross,
  and Krook}}]{bhatnagar54}
\bibinfo{author}{\bibfnamefont{P.}~\bibnamefont{Bhatnagar}},
  \bibinfo{author}{\bibfnamefont{E.}~\bibnamefont{Gross}}, \bibnamefont{and}
  \bibinfo{author}{\bibfnamefont{M.}~\bibnamefont{Krook}},
  \bibinfo{journal}{Phys.\ Rev.} \textbf{\bibinfo{volume}{94}},
  \bibinfo{pages}{511} (\bibinfo{year}{1954}).

\bibitem[{\citenamefont{d`Humi{\`e}res}(1992)}]{dhumieres92}
\bibinfo{author}{\bibfnamefont{D.}~\bibnamefont{d`Humi{\`e}res}}, in
  \emph{\bibinfo{booktitle}{Generalized Lattice Boltzmann Equations. Progress
  in Aeronautics and Astronautics (Eds. B.D. Shigal and D.P Weaver)}}
  (\bibinfo{year}{1992}), p. \bibinfo{pages}{450}.

\bibitem[{\citenamefont{Lallemand and Luo}(2000)}]{lallemand00}
\bibinfo{author}{\bibfnamefont{P.}~\bibnamefont{Lallemand}} \bibnamefont{and}
  \bibinfo{author}{\bibfnamefont{L.-S.} \bibnamefont{Luo}},
  \bibinfo{journal}{Phys.\ Rev.\ E} \textbf{\bibinfo{volume}{61}},
  \bibinfo{pages}{6546} (\bibinfo{year}{2000}).

\bibitem[{\citenamefont{Higuera and Jim\'enez}(1989)}]{higuera89a}
\bibinfo{author}{\bibfnamefont{F.}~\bibnamefont{Higuera}} \bibnamefont{and}
  \bibinfo{author}{\bibfnamefont{J.}~\bibnamefont{Jim\'enez}},
  \bibinfo{journal}{Europhys.\ Lett.} \textbf{\bibinfo{volume}{9}},
  \bibinfo{pages}{663} (\bibinfo{year}{1989}).

\bibitem[{\citenamefont{Higuera et~al.}(1989)\citenamefont{Higuera, Succi, and
  Benzi}}]{higuera89b}
\bibinfo{author}{\bibfnamefont{F.}~\bibnamefont{Higuera}},
  \bibinfo{author}{\bibfnamefont{S.}~\bibnamefont{Succi}}, \bibnamefont{and}
  \bibinfo{author}{\bibfnamefont{R.}~\bibnamefont{Benzi}},
  \bibinfo{journal}{Europhys.\ Lett.} \textbf{\bibinfo{volume}{9}},
  \bibinfo{pages}{345} (\bibinfo{year}{1989}).

\bibitem[{\citenamefont{Luo et~al.}(2011)\citenamefont{Luo, Liao, Chen, Peng,
  and Zhang}}]{luo11}
\bibinfo{author}{\bibfnamefont{L.-S.} \bibnamefont{Luo}},
  \bibinfo{author}{\bibfnamefont{W.}~\bibnamefont{Liao}},
  \bibinfo{author}{\bibfnamefont{X.}~\bibnamefont{Chen}},
  \bibinfo{author}{\bibfnamefont{Y.}~\bibnamefont{Peng}}, \bibnamefont{and}
  \bibinfo{author}{\bibfnamefont{W.}~\bibnamefont{Zhang}},
  \bibinfo{journal}{Phys.\ Rev.\ E} \textbf{\bibinfo{volume}{83}},
  \bibinfo{pages}{056710} (\bibinfo{year}{2011}).

\bibitem[{\citenamefont{Ginzburg}(2005)}]{ginzburg05}
\bibinfo{author}{\bibfnamefont{I.}~\bibnamefont{Ginzburg}},
  \bibinfo{journal}{Adv.\ Water Res.} \textbf{\bibinfo{volume}{28}},
  \bibinfo{pages}{1171} (\bibinfo{year}{2005}).

\bibitem[{\citenamefont{Karlin et~al.}(1999)\citenamefont{Karlin, Ferrente, and
  Ottinger}}]{karlin99}
\bibinfo{author}{\bibfnamefont{I.}~\bibnamefont{Karlin}},
  \bibinfo{author}{\bibfnamefont{A.}~\bibnamefont{Ferrente}}, \bibnamefont{and}
  \bibinfo{author}{\bibfnamefont{H.}~\bibnamefont{Ottinger}},
  \bibinfo{journal}{Eur.\ Phys.\ Lett.} \textbf{\bibinfo{volume}{47}},
  \bibinfo{pages}{182} (\bibinfo{year}{1999}).

\bibitem[{\citenamefont{Asinari and Karlin}(2009)}]{asinari09}
\bibinfo{author}{\bibfnamefont{P.}~\bibnamefont{Asinari}} \bibnamefont{and}
  \bibinfo{author}{\bibfnamefont{I.}~\bibnamefont{Karlin}},
  \bibinfo{journal}{Phys.\ Rev.\ E} \textbf{\bibinfo{volume}{79}},
  \bibinfo{pages}{036703} (\bibinfo{year}{2009}).

\bibitem[{\citenamefont{Karlin et~al.}(2011)\citenamefont{Karlin, Asinari, and
  Succi}}]{karlin11}
\bibinfo{author}{\bibfnamefont{I.}~\bibnamefont{Karlin}},
  \bibinfo{author}{\bibfnamefont{P.}~\bibnamefont{Asinari}}, \bibnamefont{and}
  \bibinfo{author}{\bibfnamefont{S.}~\bibnamefont{Succi}},
  \bibinfo{journal}{Phil.\ Trans.\ Roy.\ Soc.\ A}
  \textbf{\bibinfo{volume}{369}}, \bibinfo{pages}{2202} (\bibinfo{year}{2011}).

\bibitem[{\citenamefont{Pope}(2000)}]{pope00}
\bibinfo{author}{\bibfnamefont{S.}~\bibnamefont{Pope}},
  \emph{\bibinfo{title}{Turbulent Flows}} (\bibinfo{publisher}{Cambridge
  University Press, New York}, \bibinfo{year}{2000}).

\bibitem[{\citenamefont{Scovazzi}(2007{\natexlab{a}})}]{scovazzi07a}
\bibinfo{author}{\bibfnamefont{G.}~\bibnamefont{Scovazzi}},
  \bibinfo{journal}{Comp.\ Methods Appl.\ Mech.\ Engg.}
  \textbf{\bibinfo{volume}{196}}, \bibinfo{pages}{1108}
  (\bibinfo{year}{2007}{\natexlab{a}}).

\bibitem[{\citenamefont{Scovazzi}(2007{\natexlab{b}})}]{scovazzi07b}
\bibinfo{author}{\bibfnamefont{G.}~\bibnamefont{Scovazzi}},
  \bibinfo{journal}{Int.\ J.\ Num.\ Meth.\ Fluids}
  \textbf{\bibinfo{volume}{54}}, \bibinfo{pages}{757}
  (\bibinfo{year}{2007}{\natexlab{b}}).

\bibitem[{\citenamefont{Hughes et~al.}(2010)\citenamefont{Hughes, Scovazzi, and
  Tezduyar}}]{hughes10}
\bibinfo{author}{\bibfnamefont{T.}~\bibnamefont{Hughes}},
  \bibinfo{author}{\bibfnamefont{G.}~\bibnamefont{Scovazzi}}, \bibnamefont{and}
  \bibinfo{author}{\bibfnamefont{T.}~\bibnamefont{Tezduyar}},
  \bibinfo{journal}{J.\ Sci.\ Comp.} \textbf{\bibinfo{volume}{43}},
  \bibinfo{pages}{343} (\bibinfo{year}{2010}).

\bibitem[{\citenamefont{Geier et~al.}(2006)\citenamefont{Geier, Greiner, and
  Korvink}}]{geier06}
\bibinfo{author}{\bibfnamefont{M.}~\bibnamefont{Geier}},
  \bibinfo{author}{\bibfnamefont{A.}~\bibnamefont{Greiner}}, \bibnamefont{and}
  \bibinfo{author}{\bibfnamefont{J.}~\bibnamefont{Korvink}},
  \bibinfo{journal}{Phys.\ Rev.\ E} \textbf{\bibinfo{volume}{73}},
  \bibinfo{pages}{066705} (\bibinfo{year}{2006}).

\bibitem[{\citenamefont{Asinari}(2008)}]{asinari08}
\bibinfo{author}{\bibfnamefont{P.}~\bibnamefont{Asinari}},
  \bibinfo{journal}{Phys.\ Rev.\ E} \textbf{\bibinfo{volume}{78}},
  \bibinfo{pages}{016701} (\bibinfo{year}{2008}).

\bibitem[{\citenamefont{Premnath and Banerjee}(2009)}]{premnath09b}
\bibinfo{author}{\bibfnamefont{K.~N.} \bibnamefont{Premnath}} \bibnamefont{and}
  \bibinfo{author}{\bibfnamefont{S.}~\bibnamefont{Banerjee}},
  \bibinfo{journal}{Phys.\ Rev.\ E} \textbf{\bibinfo{volume}{80}},
  \bibinfo{pages}{036702} (\bibinfo{year}{2009}).

\bibitem[{\citenamefont{Premnath and Banerjee}(2011)}]{premnath11a}
\bibinfo{author}{\bibfnamefont{K.~N.} \bibnamefont{Premnath}} \bibnamefont{and}
  \bibinfo{author}{\bibfnamefont{S.}~\bibnamefont{Banerjee}},
  \bibinfo{journal}{J.\ Stat.\ Phys.} \textbf{\bibinfo{volume}{143}},
  \bibinfo{pages}{747} (\bibinfo{year}{2011}).

\bibitem[{\citenamefont{Premnath and Ning}(2012)}]{premnath12}
\bibinfo{author}{\bibfnamefont{K.~N.} \bibnamefont{Premnath}} \bibnamefont{and}
  \bibinfo{author}{\bibfnamefont{Y.}~\bibnamefont{Ning}},
  \bibinfo{journal}{Submitted}  (\bibinfo{year}{2012}).

\bibitem[{\citenamefont{Premnath and Banerjee}(2012)}]{premnath11b}
\bibinfo{author}{\bibfnamefont{K.~N.} \bibnamefont{Premnath}} \bibnamefont{and}
  \bibinfo{author}{\bibfnamefont{S.}~\bibnamefont{Banerjee}},
  \bibinfo{journal}{Comp.\ Phys.\ Comm., in press}  (\bibinfo{year}{2012}).

\bibitem[{\citenamefont{Chapman and Cowling}(1964)}]{chapman64}
\bibinfo{author}{\bibfnamefont{S.}~\bibnamefont{Chapman}} \bibnamefont{and}
  \bibinfo{author}{\bibfnamefont{T.}~\bibnamefont{Cowling}},
  \emph{\bibinfo{title}{Mathematical Theory of Non-Uniform Gases}}
  (\bibinfo{publisher}{Cambridge University Press, London},
  \bibinfo{year}{1964}).

\bibitem[{\citenamefont{Junk et~al.}(2005)\citenamefont{Junk, Klar, and
  Luo}}]{junk05}
\bibinfo{author}{\bibfnamefont{M.}~\bibnamefont{Junk}},
  \bibinfo{author}{\bibfnamefont{A.}~\bibnamefont{Klar}}, \bibnamefont{and}
  \bibinfo{author}{\bibfnamefont{L.-S.} \bibnamefont{Luo}},
  \bibinfo{journal}{J.\ Comput.\ Phys.} \textbf{\bibinfo{volume}{210}},
  \bibinfo{pages}{676} (\bibinfo{year}{2005}).

\bibitem[{\citenamefont{Taylor}(1923)}]{taylor23}
\bibinfo{author}{\bibfnamefont{G.}~\bibnamefont{Taylor}},
  \bibinfo{journal}{Phil.\ Mag.} \textbf{\bibinfo{volume}{46}},
  \bibinfo{pages}{671} (\bibinfo{year}{1923}).

\bibitem[{\citenamefont{Kruger et~al.}(2010)\citenamefont{Kruger, Varnki, and
  Raabe}}]{kruger10}
\bibinfo{author}{\bibfnamefont{T.}~\bibnamefont{Kruger}},
  \bibinfo{author}{\bibfnamefont{F.}~\bibnamefont{Varnki}}, \bibnamefont{and}
  \bibinfo{author}{\bibfnamefont{D.}~\bibnamefont{Raabe}},
  \bibinfo{journal}{Phys.\ Rev.\ E} \textbf{\bibinfo{volume}{82}},
  \bibinfo{pages}{025701(R)} (\bibinfo{year}{2010}).

\bibitem[{\citenamefont{Ghia et~al.}(1982)\citenamefont{Ghia, Ghia, and
  Shin}}]{ghia82}
\bibinfo{author}{\bibfnamefont{U.}~\bibnamefont{Ghia}},
  \bibinfo{author}{\bibfnamefont{K.}~\bibnamefont{Ghia}}, \bibnamefont{and}
  \bibinfo{author}{\bibfnamefont{C.}~\bibnamefont{Shin}}, \bibinfo{journal}{J.\
  Comput.\ Phys.} \textbf{\bibinfo{volume}{48}}, \bibinfo{pages}{387}
  (\bibinfo{year}{1982}).

\bibitem[{\citenamefont{Schreiber and Keller}(1983)}]{schreiber83}
\bibinfo{author}{\bibfnamefont{R.}~\bibnamefont{Schreiber}} \bibnamefont{and}
  \bibinfo{author}{\bibfnamefont{H.}~\bibnamefont{Keller}},
  \bibinfo{journal}{J.\ Comput.\ Phys.} \textbf{\bibinfo{volume}{49}},
  \bibinfo{pages}{310} (\bibinfo{year}{1983}).

\bibitem[{\citenamefont{Vanka}(1986)}]{vanka86}
\bibinfo{author}{\bibfnamefont{S.}~\bibnamefont{Vanka}}, \bibinfo{journal}{J.\
  Comput.\ Phys.} \textbf{\bibinfo{volume}{65}}, \bibinfo{pages}{138}
  (\bibinfo{year}{1986}).

\bibitem[{\citenamefont{Erturk et~al.}(2005)\citenamefont{Erturk, Corke, and
  Gokcol}}]{erturk05}
\bibinfo{author}{\bibfnamefont{E.}~\bibnamefont{Erturk}},
  \bibinfo{author}{\bibfnamefont{T.}~\bibnamefont{Corke}}, \bibnamefont{and}
  \bibinfo{author}{\bibfnamefont{C.}~\bibnamefont{Gokcol}},
  \bibinfo{journal}{Int.\ J.\ Num.\ Methods Fluids}
  \textbf{\bibinfo{volume}{48}}, \bibinfo{pages}{747} (\bibinfo{year}{2005}).

\bibitem[{\citenamefont{Bruneau and Saad}(2006)}]{bruneau06}
\bibinfo{author}{\bibfnamefont{C.-H.} \bibnamefont{Bruneau}} \bibnamefont{and}
  \bibinfo{author}{\bibfnamefont{M.}~\bibnamefont{Saad}},
  \bibinfo{journal}{Comp.\ Fluids} \textbf{\bibinfo{volume}{35}},
  \bibinfo{pages}{326} (\bibinfo{year}{2006}).

\bibitem[{\citenamefont{Lallemand and Luo}(2003)}]{lallemand03}
\bibinfo{author}{\bibfnamefont{P.}~\bibnamefont{Lallemand}} \bibnamefont{and}
  \bibinfo{author}{\bibfnamefont{L.-S.} \bibnamefont{Luo}},
  \bibinfo{journal}{J.\ Comput.\ Phys.} \textbf{\bibinfo{volume}{184}},
  \bibinfo{pages}{406} (\bibinfo{year}{2003}).

\bibitem[{\citenamefont{Premnath et~al.}(2009)\citenamefont{Premnath, Pattison,
  and Banerjee}}]{premnath09a}
\bibinfo{author}{\bibfnamefont{K.~N.} \bibnamefont{Premnath}},
  \bibinfo{author}{\bibfnamefont{M.~J.} \bibnamefont{Pattison}},
  \bibnamefont{and} \bibinfo{author}{\bibfnamefont{S.}~\bibnamefont{Banerjee}},
  \bibinfo{journal}{Phys.\ Rev.\ E} \textbf{\bibinfo{volume}{79}},
  \bibinfo{pages}{026703} (\bibinfo{year}{2009}).

\bibitem[{\citenamefont{Wesseling}(2000)}]{wesseling00}
\bibinfo{author}{\bibfnamefont{P.}~\bibnamefont{Wesseling}},
  \emph{\bibinfo{title}{Principles of Computational Fluid Dynamics}}
  (\bibinfo{publisher}{Springer, New York}, \bibinfo{year}{2000}).

\bibitem[{\citenamefont{d`Humi{\`e}res
  et~al.}(2002)\citenamefont{d`Humi{\`e}res, Ginzburg, Krafczyk, Lallemand, and
  Luo}}]{dhumieres02}
\bibinfo{author}{\bibfnamefont{D.}~\bibnamefont{d`Humi{\`e}res}},
  \bibinfo{author}{\bibfnamefont{I.}~\bibnamefont{Ginzburg}},
  \bibinfo{author}{\bibfnamefont{M.}~\bibnamefont{Krafczyk}},
  \bibinfo{author}{\bibfnamefont{P.}~\bibnamefont{Lallemand}},
  \bibnamefont{and} \bibinfo{author}{\bibfnamefont{L.-S.} \bibnamefont{Luo}},
  \bibinfo{journal}{Phil.\ Trans.\ R. Soc.\ Lond.\ A}
  \textbf{\bibinfo{volume}{360}}, \bibinfo{pages}{437} (\bibinfo{year}{2002}).

\end{thebibliography}

\end{document}